\documentclass[
 reprint,
superscriptaddress,
 amsmath,amssymb,
 aps,
 pra,onecolumn,
]{revtex4-2}

\usepackage{graphicx}% Include figure files
\usepackage{dcolumn}% Align table columns on decimal point
\usepackage{bm}% bold math
\usepackage{soul}
\usepackage[colorlinks]{hyperref}

\usepackage[english]{babel}
\usepackage{color}
\usepackage{mathtools}
\usepackage{cleveref}
\usepackage{verbatim}
\usepackage{breqn}

\makeatletter
\newcommand{\customlabel}[2]{%
   \protected@write \@auxout {}{\string \newlabel {#1}{{#2}{\thepage}{#2}{#1}{}} }%
   \hypertarget{#1}{}
}
\makeatother

\begin{document}

\title{Organization of spatially localized structures near a codimension-three cusp-Turing bifurcation}

\author{P. Parra-Rivas}
\affiliation{
Dipartimento  di  Ingegneria  dell’Informazione $,$ Elettronica  e  Telecomunicazioni$,$
Sapienza  Universit{\'a}  di  Roma$,$ via  Eudossiana  18$,$ 00184  Rome $,$ Italy}

\author{A. R. Champneys}
\affiliation{Department of Engineering Mathematics$,$ University of Bristol$,$ BS8 1TW$,$ UK}

\author{F. Al-Sahadi}
\affiliation{Department of Systems Engineering $,$ Military Technological College$,$ Muscat$,$ Oman}

\author{D. Gomila}
\affiliation{IFISC (CSIC-UIB)$,$ Instituto de F\'{\i}sica Interdisciplinar y Sistemas Complejos$,$  E-07122 Palma de Mallorca$,$  Spain}

\author{E. Knobloch}
\affiliation{Department of Physics$,$ University of California$,$ Berkeley$,$ CA 94720$,$ USA}
\date{\today}
\begin{abstract}
A wide variety of stationary or moving spatially localized structures is present in evolution problems on unbounded domains, governed by higher-than-second-order reversible spatial interactions.  This work provides a generic unfolding in one spatial dimension of a certain codimension-three singularity that explains the organization of bifurcation diagrams of such localized states in a variety of contexts, ranging from nonlinear optics to fluid mechanics, mathematical biology and beyond. The singularity occurs when a cusp bifurcation associated with the onset of bistability between homogeneous steady states encounters a pattern-forming, or Turing, bifurcation. The latter corresponds to a Hamiltonian-Hopf point of the corresponding spatial dynamics problem. Such codimension-three points are sometimes called Lifshitz points in the physics literature. In the simplest case where the spatial system conserves a first integral, the system is described by a canonical fourth order scalar system. The problem contains three small parameters, two that unfold the cusp bifurcation and one that unfolds the Turing bifurcation. Several cases are revealed, depending on open conditions on the signs of the lowest-order nonlinear terms. Taking the case in which the Turing bifurcation is subcritical, various parameter regimes are considered and the bifurcation diagrams of localized structures are elucidated. A rich bifurcation structure is revealed, which involves transitions between regions of localized periodic patterns generated  by homoclinic snaking, and mesa-like patterns with uniform cores. The theory is shown to unify previous numerical results obtained in models arising in nonlinear optics, fluid mechanics, and excitable media more generally. 
\end{abstract}

\maketitle
    
\section{Introduction}
Spatial patterns emerge in a wide variety of real-life physical problems such as weather forecasting, fluid or granular flows, flame propagation, fracture dynamics, and a wealth of new optical, biological, ecological and chemical phenomena \cite{cross_pattern_1993,hoyle_pattern_2006,murray_mathematical_2003}. Such patterns often form spontaneously through a process called {\it self-organization}, where anisotropic structures form due to spatial interactions among the components of the system without any external entity engineering the process. This typically occurs in dissipative systems driven far from thermodynamic equilibrium. In this context, a variety of states may emerge as the driving increases, which can either extend over the entire spatial domain ({\it extended patterns}) or be spatially localized ({\it localized  structures}). The former typically arise via the so-called {\it Turing} instability \cite{turing_chemical_1952}, whereby a spatially extended pattern with a single well-defined length scale emerges spontaneously from a spatially uniform state. Emergent structures of this type were named {\it dissipative structures} by the Nobel Laureate Ilya Prigogine in 1977, and since then have been found in a wide range of applications in the natural sciences including, but not limited to optics, hydrodynamics and systems of reaction-diffusion equations. 

Spontaneous pattern formation possesses universal features that are independent of the details of the underlying system. Such general properties can be captured by prototypical models that include the simplest nonlinear terms and lowest spatial derivatives respecting the symmetries of the original problem \cite{cross_pattern_1993,swift_swift-hohenberg_2008,knobloch_spatial_2015}. Thus, for instance, the Swift-Hohenberg equation (SHE) is a prototypical model that describes systems undergoing a stationary pattern-forming (or Turing) instability. Similarly, the real Ginzburg-Landau equation describes systems undergoing a pitchfork bifurcation to a pair of nontrivial homogeneous states, while the complex Ginzburg-Landau equation describes systems undergoing a Hopf bifurcation to an oscillating homogeneous state. Each model captures the general spatio-temporal dynamics associated to the corresponding codimension-one bifurcation. More generally, a system can display two or more such bifurcations in different parameter regions and, quite possibly, a crossover between the corresponding behaviors when  these bifurcations meet in a codimension-two point. The general scenarios emerging from these higher-codimension points in spatial systems have only been partially studied, but are known to involve complex spatio-temporal dynamics such as that arising in the Turing-Hopf scenario \cite{walgraef_book}. In this work we discuss the general bifurcation structure of spatially extended patterns and the associated spatially localized states in the vicinity of a codimension-three point where a system simultaneously undergoes a cusp and a Turing bifurcation.

We are interested in classes of partial differential equations (PDEs) describing the evolution of a scalar field $u(x,t)$ whose stationary patterns $u(x)$ are governed by a fourth-order scalar operator. As we shall see in Section~\ref{sec:2} below, steady states of such systems can typically be formulated in terms of an ordinary differential equation (ODE) posed on a long domain. We specifically consider a model of Swift-Hohenberg type that can be written in the form 
\begin{equation}\label{general_model1}
	\partial_t u= \eta+\mu u+\alpha u^2-\beta u^3 -(\partial_x^4+\nu\partial_x^2)u.
\end{equation}
Here, we take $(\eta,\mu,\nu)$ to be small unfolding parameters
whereas $\alpha$ and $\beta$ are $O(1)$ constants that can be used to distinguish between different cases. Through simple scaling we can choose $\alpha^2+\beta^2=1$, so that all cases can be parameterised by a single parameter $\theta$ with $\alpha=\sin\theta$, $\beta=\cos\theta$. Equation \eqref{general_model1} is a dissipative dynamical system, with the parameter $\eta$ representing the driving or, alternatively, $\mu$ representing gain. 

We seek to understand how both periodic and localized patterns are organised close to the codimension-three point $(\eta,\mu,\nu)=0$. As we shall see in Section~\ref{sec:dissipative}, near this point Eq.~\eqref{general_model1} is a particular case of the normal form of a codimension-three singularity that arises at a so-called Lifshitz point where a cusp and Turing point meet in a dissipative pattern-forming system. As we show explicitly in Section~\ref{sec:2}, this special case allows one to write \eqref{general_model1} as a variational, or gradient system.  Since we are interested in the organization of different spatial structures, i.e., steady states of \eqref{general_model1}, in parameter space rather than their stability, the stability properties of these states in time, including the transient dynamics of \eqref{general_model1}, are not of main concern. Indeed, related fourth-order ODEs arise in other pattern formation contexts and may, for example, describe the profile of a traveling wave in the wave frame, as mentioned in Section~\ref{sec:context} below, where the temporal stability properties are quite different.

Most of what follows shall be devoted to the case $\alpha=0$, $\beta=1$, since $\alpha\ne0$ plays a similar symmetry-breaking role as $\eta\ne0$. It should be noted, however, that the universal unfolding of the pitchfork bifurcation does require nonzero, albeit small, values of $\alpha$ in order to capture all possible behavior near $(\alpha,\beta)=(0,0)$ \cite{benjamin_bifurcation_1978,benjamin_bifurcation_1978b}.

%Golubitsky
\subsection{Dissipative localized structures}
\label{sec:dissipative}

\begin{figure}[!b]
	\centering
	\includegraphics[scale=1]{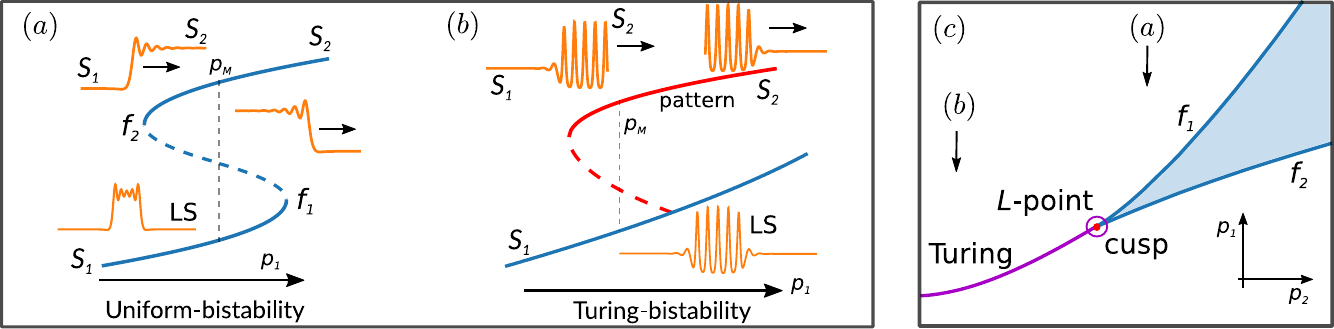}
	\caption{(a) Uniform-bistable scenario. (b) Turing-bistable scenario. (c) Lifshitz critical point scenario.}
	\label{Fig1}
\end{figure}

Spatially localized structures (LSs) arise due to a balance between nonlinearity and spatial coupling on the one hand, and a balance between energy dissipation and gain on the other. In general, the key ingredients for LS formation are {\it bistability} and {\it front pinning}. Bistability means that two different but potentially stable states of the system, $S_1$ and $S_2$, say, coexist over a range of parameter values (see Figs.~\ref{Fig1}(a) and (b), both with bifurcation parameter $p_1$). There are two bistable scenarios that generally lead to spatial localization, where either (i) $S_1$ and $S_2$ are uniform in space [see Fig.~\ref{Fig1}(a)], or (ii) one state is uniform while the other is a nonuniform or patterned state [see Fig.~\ref{Fig1}(b)].

The first configuration may appear through a {\it cusp} bifurcation, leading to a hysteresis bifurcation [see Fig.~\ref{Fig1}(c)]. Here, by changing a second parameter $p_2$, the system with a uniform state [see, e.g., the blue curve in Fig.~\ref{Fig1}(b)] changes to a situation where three distinct homogeneous states coexist [see Fig.~\ref{Fig1}(a)]. Indeed, the cusp bifurcation takes place when the two {\it folds}, $f_{1}$ and $f_2$ in Fig.~\ref{Fig1}(a), meet in a single point of the $(p_1,p_2)$ parameter space, i.e., it is a codimension-two bifurcation. In what follows, we refer to this scenario as {\it uniform-bistability}. This scenario arises in many systems including nonlinear optics, chemistry and mathematical biology 
\cite{cross_pattern_1993,hoyle_pattern_2006,murray_mathematical_2003}.

In the second scenario [Fig.~\ref{Fig1}(b)], bistability appears when a uniform state undergoes a {\it subcritical Turing bifurcation}, creating a nonuniform spatially periodic state that emerges subcritically but subsequently turns around towards larger values of $p_1$. Here the Turing bifurcation is a codimension-1 point but one is interested in the transition from supercritical behavior to subcritical behavior, and therefore also a codimension-two problem.
The resulting scenario will be referred to as {\it  Turing-bistability}.

In each of these two situations, oscillatory {\it fronts} connecting $S_1$ and $S_2$ may form as shown in Figs.~\ref{Fig1}(a) and (b). Normally, these fronts move in such a way that either $S_1$ overtakes $S_2$, or vice-versa. However, typically there is a range of values $p_1$ around the so-called Maxwell point $p_1=p_M$ of the system, where the fronts lock to one another, leading to the formation of a large variety of stationary LSs. Thus, a LS can be seen as a portion of one state embedded in the other and stationary LSs are found near $p_1=p_M$ [see Figs.~\ref{Fig1}(a) and (b)].

These two bistable scenarios may coexist in the same system, and even in the same parameter regime, leading to {\it tristability}. Two uniform states then coexist with a subcritical Turing pattern, which leads to an extraordinarily great variety of new LS configurations and complex bifurcation schemes \cite{zelnik_implications_2018}.

The bifurcation structure associated with LSs emerging in this context is related to the occurrence and unfolding of higher codimension bifurcations, specifically around a codimension-three point where the Turing bifurcation and the cusp meet and interact \cite{kozyreff_interaction_2003,kozyreff_nonvariational_2007}. In the physics literature, this point is generally referred to as the {\it Lifshitz critical point} (hereafter $L$-point), in analogy with the triple point present in phase transitions of helicoidal ferromagnetic states \cite{hornreich_lifshitz_1980}. In the following we adopt this terminology as a useful shorthand. In fact the interest in the $L$-point goes far beyond the pattern formation context, as it also appears in phase transitions in particle physics and even cosmology \cite{pisarski_pedagogical_2019}.

As shown in \cite{kozyreff_interaction_2003,kozyreff_optical_2004,kozyreff_nonvariational_2007}, near such an $L$-point, a general PDE can be reduced to the generic normal form    
\begin{equation}\label{general_model}
	\partial_t u= \eta+\mu u+\alpha u^2-\beta u^3 -(\partial_x^4+\nu\partial_x^2)u+\gamma_1u\partial_x^2u+\gamma_2|\partial_x^2u|^2. 
\end{equation}
This equation suffices to capture all essential dynamics of the system near such a point. Here, $u$ is a real-valued scalar field, and $p\equiv(\eta,\mu,\nu,\alpha,\beta)$ is the set of real parameters controlling the dynamics of the system. This model has been derived from PDEs describing pattern formation in several different areas including nonlinear optics, biology and ecology \cite{kozyreff_interaction_2003,kozyreff_nonvariational_2007}.

Despite the fact that many authors have used this model to
characterize different types of dynamics, so far a complete
understanding of its bifurcation structure is lacking. 
Equation (\ref{general_model}) can exhibit a broad range of patterned states including (but not limited to)
LSs \cite{bordeu_rodlike_2015-1,vladimirov_relative_2011-1}, oscillons \cite{burke_localized_2012,clerc_bouncing_2005} and spatially localized chaotic dynamics \cite{clerc_quasiperiodicity_2013}. We consider three scenarios or cases associated with different versions
of Eq.~(\ref{general_model}). In Case I $\gamma_{1,2}=0$ and
Eq.~(\ref{general_model}) is variational, that is, exhibits gradient dynamics, and therefore time-dependent attractors (so-called permanent dynamics) are excluded
\cite{cross_pattern_1993}. This variational structure is preserved in Case II ($\gamma_1=\gamma_2/2$), despite increased complexity. Finally, Case III focuses on the fully nonvariational scenario that appears when $\gamma_1\neq\gamma_2/2$. In this case, the system can undergo permanent temporal dynamics.

Our overarching aim is to fully understand case III. However, due to the complex nature of its behavior, the present work is restricted to case I, with Cases II and III left to future investigation.

\subsection{Other contexts}
\label{sec:context}

Models for pattern formation with similar characteristics to \eqref{general_model} arise in a wide variety of other contexts. The key ingredients are long domains, fourth-order (or two coupled second-order) spatial operators, and the presence of competing nonlinear terms.

In water wave theory, the so-called Kawahara equation, or 5th-order Korteweg-de Vries equation, has traveling wave solutions whose shape is governed by an equation whose form is similar to the right-hand side of \eqref{general_model}; see, e.g., \cite{Kawahara,Groves,haragus_local_2011} and references therein. This model has been shown rigorously to be a normal form of the dynamics of the full water wave problem with surface tension near a critical point in Bond and Froude number parameter space \cite{Toland}, and the multiplicity of LSs in the model equation persists in the full problem. Of particular relevance to the present study is the work of Iooss \cite{iooss_codimension_1995} who studies the two-parameter singularity which is equivalent to the right-hand side of \eqref{general_model1} with $\eta=\beta=0$.

Other models known to feature solitary waves and localized structures governed by fourth-order nonlinear operators include the so-called extended Fisher-Kolmogorov equation \cite{Troy} and nonlinear beam equations, modeling for example the motion of the deck of a suspension bridge \cite{McKenna}.  Another key area for such complex structures includes elastic buckling problems of struts, elasticae, twisted rods and cylindrical shells. There, in the presence of a mechanically subcritical instability, localized buckle patterns have been shown to be energetically favorable compared with their spatially periodic counterparts; see e.g.  \cite{Happy} and references therein.

Localized patterns also emerge in activator-inhibitor systems where patterns arise through the interaction between two competing fields. These models are commonly used in mathematical biology at length scales ranging from a single cell all the way to entire ecosystems; see, e.g. \cite{MeronIssue,Meron}.  It was in this context, in fact, that the original concept of a Turing bifurcation first arose
\cite{turing_chemical_1952,Dawes}. Canonical models are of Gray-Scott and Schnakenberg type; indeed, recent work \cite{al_saadi_unified_2021} shows how many of the bifurcation structures equivalent to those found in the present work can be found in a three-parameter model system that interpolates between these two types of models.

In this sense the unfolding of the localized pattern states of (\ref{general_model1}) is expected to have a far wider applicability as a normal form than as a model of dissipative structures of fourth-order pattern-forming systems.

\subsection{Outline}

The rest of this paper is organised as follows. Section~\ref{sec:2}
contains preliminary material and further motivation for the form of Eq.~(\ref{general_model}) we study.  The simplest stationary states of this equation, the uniform or homogeneous steady states (HSSs), are studied in Section~\ref{sec:3} together with their linear stability in space and time. Section~\ref{sec:5} presents multiple-scale perturbation results that approximate LSs in the weakly nonlinear limit; some results are relegated to various appendices. The next two sections of the paper contain numerical bifurcation results for opposite signs of the parameter $\nu$: Section~\ref{sec:6} is devoted to the case $\nu>0$ while Section \ref{sec:9} provides a brief summary of the differences for $\nu<0$. Section \ref{sec:11} briefly considers the case of nonzero $\alpha$, while Section \ref{sec:12} draws conclusions and suggests avenues for future work.

\section{Preliminaries}\label{sec:2}

With $\gamma_{1,2}=0$, Eq.~(\ref{general_model}) reduces to the form \eqref{general_model1}. In nonlinear optics, Eq.~\eqref{general_model1} was derived from the Maxwell-Bloch equations in the nascent optical bistability regime, and used to establish the existence of spatially localized stationary patterns \cite{tlidi_1994}.
The equation is invariant under the transformation $(\eta,u)\rightarrow -(\eta,u)$, and reversible under the transformation $u(x)\rightarrow u(-x)$. The equation can be further simplified by removing the quadratic $\alpha u^2$ through the transformation $u\rightarrow u+a$, $a\in\mathsf{R}$. However, we do not do so here since the coefficient $\alpha$ has distinct physical significance.

Equation~(\ref{general_model1}) has gradient dynamics with Lyapunov functional
\begin{equation}
	F[u]=\int_{-\infty}^{\infty}\left[\frac{1}{2}\left(\partial_x^2+\frac{\nu}{2}\right)^2u^2-\eta u-\frac{1}{2}\left(\mu+\frac{\nu^2}{4}\right)u^2-\frac{1}{3}\alpha u^3+\frac{1}{4}\beta u^4\right]dx,
\end{equation}
such that
\begin{equation}
	\partial_t u =-\frac{\delta F}{\delta u},
\end{equation}
where $\frac{\delta}{\delta u}$ represents a functional derivative. A consequence of this gradient structure is that time-dependent limit states (such as temporal oscillations and (spatio)-temporal chaos) are excluded, and the evolution of (\ref{general_model1}) settles into a time-independent state at every location. Such steady states obey the ODE
\begin{equation}
	u_{xxxx}+\nu u_{xx}-\eta-\mu u-\alpha u^2+\beta u^3 = 0.
	\label{general_model_sta}
\end{equation}
This equation forms the basis for the study in the rest of this paper. Furthermore, if we ignore boundary conditions, and consider long, formally infinite, spatial domains, we can study \eqref{general_model_sta} as if $x$ were a time-like variable, i.e.~reformulating the problem as a {\em spatial dynamics} problem \cite{haragus_local_2011}. 
%{\color{red}In this context, the existence of the gradient structure %in the PDE leads to the spatial ODE with a conserved first integral 
%\begin{equation}\label{eq:Hu}
%H(u)=\frac{1}{2}\left(d_x^2+\frac{\nu}{2}\right)^2u^2-\eta %u-\frac{1}{2}\left(\mu+\frac{\nu^2}{4}\right)u^2-\frac{1}{3}\alpha %u^3+\frac{1}{4}\beta u^4.
%\end{equation}
%A change of variables then puts the ODE into Hamiltonian form %\cite{glebsky} although we shall not exploit this structure explicitly here.}
%
%{\color{blue} The stationary equation is conservative and possesses a %Lagrangian 
%\begin{equation}
%    L=\frac{\nu}{2}(u_x)^2+\frac{1}{2}(u_{xx})^2+\eta u+\frac{\mu}{2}u^2+\frac{\alpha}{3}u^3-\frac{\beta}{4}u^4.
%\end{equation}
%The stationary equation is recovered from the Euler-Lagrange equation %
%\begin{equation}
%    \frac{d^2}{dx^2}\left(\frac{\partial L}{\partial %u_{xx}}\right)+\frac{d^2}{dx}\left(\frac{\partial L}{\partial %u_{x}}\right)-\frac{\partial L}{\partial u}=0.
%\end{equation}
%The Hamiltonian of the system $H$ is defined through the Legendre %transform
%\begin{equation}
%    H\equiv -L+\frac{\partial L}{\partial u_x}u_x+\frac{\partial %L}{\partial u_{xx}}u_{xx},
%\end{equation}
%which yields}
The steady-state equation is, moreover, Hamiltonian in space \cite{glebsky_small_1995}, with the conserved Hamiltonian
\begin{equation}
	H=-\frac{1}{2}(u_{xx})^2+u_x u_{xxx}+\frac{\nu}{2}(u_x)^2-\eta u -\frac{\mu}{2}u^2-\frac{\alpha}{3}u^3+\frac{\beta}{4}u^4, \label{eq:Hu}
\end{equation}
a property that is useful in understanding stationary fronts as homoclinic or heteroclinic orbits of the resulting spatial dynamical system.

\begin{figure}[!t]
	\centering
	\includegraphics[scale=1]{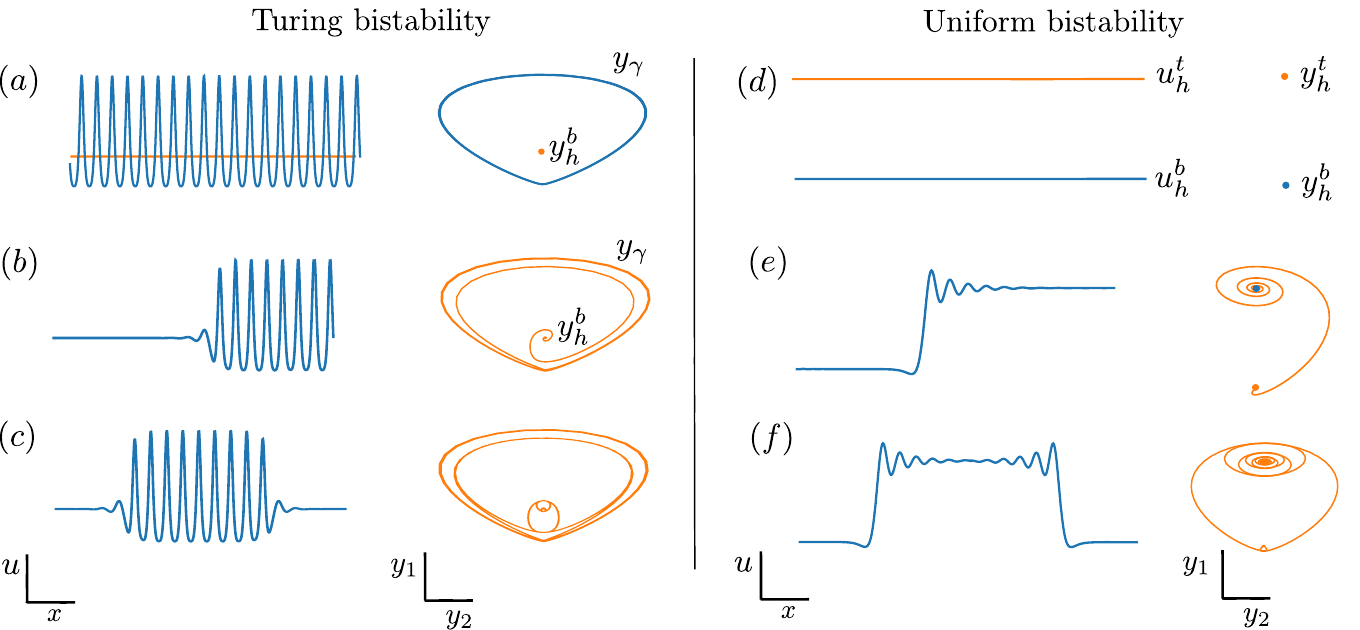}
	\caption{Analogy between the stationary solutions of Eq.~(\ref{general_model1}) and the solution of the spatial dynamical system (\ref{SD}) in the $\nu>0$ (left) and $\nu<0$ (right) regimes. For Eq.~(\ref{general_model1}) we plot $u(x)$, while in the spatial dynamics counterpart we show its projection on the $(y_2,y_1)$ phase plane. See text for details.}
	\label{duality}
\end{figure}

To understand better the different types of steady states of (\ref{general_model1}) using spatial dynamics, it is convenient to recast the steady-state ODE (\ref{general_model_sta}) as the four-dimensional dynamical system
\begin{align}\label{SD}
	\frac{dy}{dx}=\mathcal{A}(\mu,\nu)y+N(y;\alpha,\beta,\eta), && y=(y_1,y_2,y_3,y_4)^T\equiv(u,u_x,u_{xx},u_{xxx})^T, 
\end{align}
where the linear and nonlinear operators are given by 
\begin{align}
	\mathcal{A}(\mu,\nu)\equiv\left[\begin{array}{cccc}0&1&0&0\\
		0&0&1&0\\
		0&0&0&1\\
		\mu&0&-\nu&0\\
	\end{array}\right], && N(y;\alpha,\beta,\eta)\equiv\left[\begin{array}{c}0\\
		0\\
		0\\
		\alpha y_1^2-\beta y_1^3+\eta\\
	\end{array}\right].
\end{align}
The question of steady states is thus reduced to studying the dynamics of \eqref{SD} in the four-dimensional phase space. Such a study is aided by the invariance of \eqref{SD} under the involution 
\begin{equation}
	R: (x,y_1,y_2,y_3,y_4)\mapsto(-x,y_1,-y_2,y_3,-y_4), 
\end{equation}
i.e., \eqref{SD} is said to be {\it spatially reversible}. The analogy between solutions in the phase space of \eqref{SD} and steady states governed by the fourth-order ODE is illustrated qualitatively in Fig.~\ref{duality}. Here a distinction can be drawn between the kind of states observed for $\nu>0$ and for $\nu<0$.

The left side of Fig.~\ref{duality} illustrates typical states for $\nu>0$. Here, the periodic pattern corresponds to a limit cycle $y_\gamma$, while the HSS $u_h^b$ corresponds to a fixed point labeled $y_h^b$. When these states coexist patterned fronts like that shown in Fig.~\ref{duality}(b) emerge. These fronts correspond to {\it heteroclinic orbits} connecting $y_h^b$ and $y_\gamma$. These
connections form as a result of a transverse intersection between the unstable manifold of $y_h^b$ ($W^u[y_h^b]$) and the stable manifold of $y_\gamma$ ($W^s[y_\gamma]$); the robustness of this intersection is in turn a consequence of the dimensions of these manifolds, as further explained in \cite{knobloch_spatial_2015}. Furthermore, spatial reversibility implies a similar intersection between $W^u[y_\gamma]$ and $W^s[y_h^b]$, and hence the presence of a {\it heteroclinic cycle}; homoclinic orbits in $W^s[y_h^b]\cap W^u[y_h^b]$ accumulate on this cycle. An example of such an orbit is shown in Fig.~\ref{duality}(c), where the trajectory rotates several times around $y_\gamma$ before returning to $y_h^b$. Solutions of this type correspond to {\it localized patterns} (LPs) containing a long plateau where the solution resembles the spatially periodic pattern shown in Fig.~\ref{duality}(a). Each rotation around $y_\gamma$ generates an additional peak in the profile of the LP. These orbits approach or leave $u_h^b$ in an oscillatory manner, leading to the appearance of oscillatory tails in the LP profile. LSs of this type correspond to {\it Shilnikov} or {\it wild} homoclinic orbits
\cite{champneys_when_2007,homburg_homoclinic_2010}. In contrast, orbits where the behavior around the fixed point is monotonic are known as {\it tame} homoclinic orbits, and correspond to {\it spikes}
\cite{verschueren_model_2017}.

For $\nu<0$ the scenario is very different as $u_h^b$ and $u_h^t$ can coexist in a stable way [see Fig.~\ref{duality}(d)]. As a result, heteroclinic orbits can arise from the intersection between $W^u[y_h^b]$ and $W^s[y_h^t]$, forming the front shown in Fig.~\ref{duality}(e). In this regime, spatial reversibility is responsible for the formation of a variety of homoclinic orbits as well. One example of such an orbit and the corresponding LS is shown in Fig.~\ref{duality}(f).

In what follows, we shall explore the bifurcation structure of the solutions of the spatial problem using numerical computation.  All such computations are carried out within the software AUTO-07p and solving the boundary value problem defined by Eq.~(\ref{SD}) on a long domain $x\in (-L/2,L/2)$ with $L\gg  1$. Unless otherwise stated, due to the spatial reversibility of the system, we perform such computations on the half-domain using Neumann boundary conditions: 
%{\bf how do you exclude the trivial solution $u=const$ or force the solution to asymptote to $u=0$ for large $x$? {\color{blue} I do not understand the question ..}}
\begin{align}
	u_{x}(0)=u_{xxx}(0)=0, && u_x(L/2)=u_{xxx}(L/2)=0.
	\label{eq:BCs}
\end{align}
When computing bifurcation diagrams of nontrivial solutions, we often choose signed $L_1$ norm
\begin{equation}
	\|u\|_{L_1}\equiv\frac{1}{L}\int_{-L/2}^{L/2}\,u(x)dx,
	\label{L1}
\end{equation}
which, of course, can take positive or negative values. Temporal stability of the steady states we find shall not be of primary concern in this paper but can in principle be easily determined using (numerical) spectral methods. That is, on finding a steady state $u(x,t)=u_s(x)$, its temporal stability can be investigated by solving the eigenvalue problem
\begin{align}
	\mathcal{L}(u_s)\psi(x)=\sigma\psi(x), && \mathcal{L}(u_s)\equiv\mu+2\alpha u_s-3\beta u_s^2 -(\partial_x^4+\nu\partial_x^2),
\end{align}
where $\psi$ and $\sigma$ are the corresponding eigenfunction and eigenvalue associated with $\mathcal{L}$. Thus, a given state $u(x)$ is linearly stable if all eigenvalues satisfy $\mbox{Re}[\sigma]<0$, and unstable otherwise.

%--------------------------------------------------------------------------------------------------

\section{Homogeneous steady states}\label{sec:3}

The simplest steady state solutions are
the homogeneous (or uniform) steady states (HSSs) $u_h$ obtained by setting the $x$ derivatives in Eq.~(\ref{general_model_sta}) to zero. These states satisfy the cubic algebraic equation
\begin{equation}\label{HSS_cubic}
	\eta+\mu u_h+\alpha u_h^2-\beta u_h^3=0.	
\end{equation}
Since $u_h(\eta,\mu)$ is given implicitly by \eqref{HSS_cubic}, we sometimes use $u_h$ as a control parameter instead of $\eta$. 
The HSS undergo two folds or turning points at: 
\begin{equation}
	u_h=u_{f}^{l,r}\equiv\frac{\alpha\pm\sqrt{\alpha^2+3\beta\mu}}{3\beta},
\end{equation}
such that $u_h$ is single-valued if $\alpha^2+3\beta\mu<0$ and multivalued if $\alpha^2+3\beta\mu>0$. The transition between these two situations in the ($\eta$, $\mu$) parameter space occurs at the cusp bifurcation point
\begin{equation}
	(\eta_C,\mu_C)\equiv\left(-\frac{\alpha^2}{3\beta},-\frac{\alpha}{3\beta}\right).
\end{equation}

\begin{figure}[!t]
	\centering
	\includegraphics[scale=0.9]{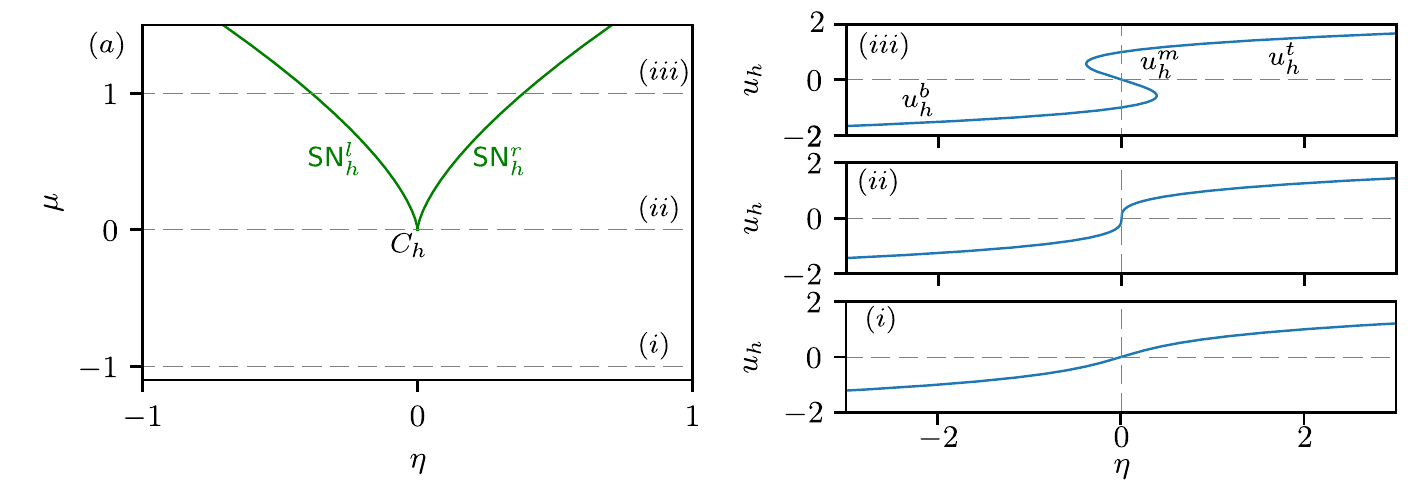}
	\includegraphics[scale=0.9]{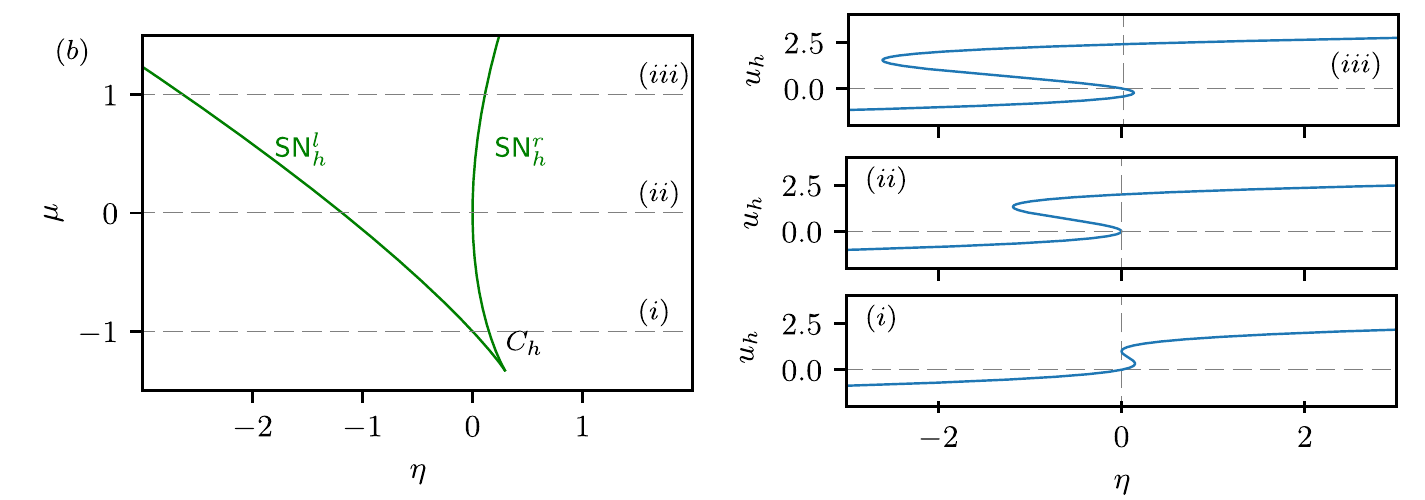}
	\caption{Homogeneous steady state ({\rm HSS}) solutions for (a) $\alpha=0$, $\beta=1$ and (b) $\alpha=2$, $\beta=1$. Left columns show the fold lines of the {\rm HSS} $u_h$ in the $(\eta,\mu)$ parameter space. Right columns show the {\rm HSS} solutions as a function of $\eta$ for the three values of $\mu$ indicated by the dashed horizontal lines in the left panels.}
	\label{hom_I_II}
\end{figure}

Figure~\ref{hom_I_II}(a) shows the cusp point $C_h$ and the fold lines $\eta_f^{l,r}(\mu)=\eta(u^{l,r}_f;\mu)$, labeled as SN$^{l,r}_h$ respectively, for $\alpha=0$. The horizontal dashed lines indicate the values of $\mu$ for which the HSS branches are shown in the right panels (i)-(iii). For $\mu=-1$ [panel (i) in Fig.~\ref{hom_I_II}(a)], the HSS is single-valued. Increasing $\mu$, the cusp occurs at $\mu=0$ [panel (ii)], and for $\mu>0$ the system shows three coexisting states that we label $u_h^b$, $u_h^m$, and $u_h^t$, as shown in panel (iii). These configurations are symmetric with respect to the transformation $(\eta,u)\rightarrow -(\eta,u)$. For $\alpha\neq0$, however, the symmetry $(\eta,u)\rightarrow -(\eta,u)$ is broken, resulting in the tilted scenario shown in Fig.~\ref{hom_I_II}(b).  

\subsection{Linear stability properties of HSSs}
To determine the linear stability of these states against spatiotemporal perturbations we suppose that $u(x,t)=u_h+\epsilon\phi(x,t)$, with $\epsilon\ll1$ and $\phi$ a perturbation. At $\mathcal{O}(\epsilon)$ we obtain the linear equation 
\begin{equation}
	\partial_t\phi=-\partial_x^4\phi-\nu\partial_x^2\phi+(\mu+2\alpha u_h-3\beta u_h^2)\phi,
\end{equation}
for the spatiotemporal evolution of the perturbation $\phi$.

To solve this equation we consider perturbations of the form $\phi(x,t)\sim e^{\sigma t}e^{ikx}$, where $\sigma$ is the growth rate of the Fourier mode with wavenumber $k$, satisfying the dispersion relation
\begin{equation}
	\sigma(k)=-k^4+\nu k^2+\mu+2\alpha u_h-3\beta u_h^2.
\end{equation}
Thus, the HSS $u_h$ is unstable against perturbations with wavenumber $k$ if $\sigma(k)>0$, and stable otherwise. The limiting situation, i.e. $\sigma(k)=0$, yields to the {\it marginal instability curve} (MIC):
\begin{equation}\label{Marginal_1}
	u_h=\frac{\alpha\pm\sqrt{\alpha^2+3\beta(\mu-k^4+\nu k^2)}}{3\beta},
\end{equation} 
defining for a given set of parameters the unstable wavenumbers, and determining the dominant spatiotemporal instabilities of the system. Indeed, the threshold  and critical wavenumber of such instabilities can be determined from the extrema of the MIC, which satisfy simultaneously the conditions $\sigma(k)=0$ and $d\sigma(k)/dk=0$. The critical wavenumber is given by $k_c^2=k_T^2=\nu/2$ if $\nu >0$, and $k_c=0$ otherwise. The threshold $u_c$ is given by Eq.~(\ref{Marginal_1}) evaluated at $k=k_c$.

\begin{figure*}[!t]
	\centering
	\includegraphics[scale=1]{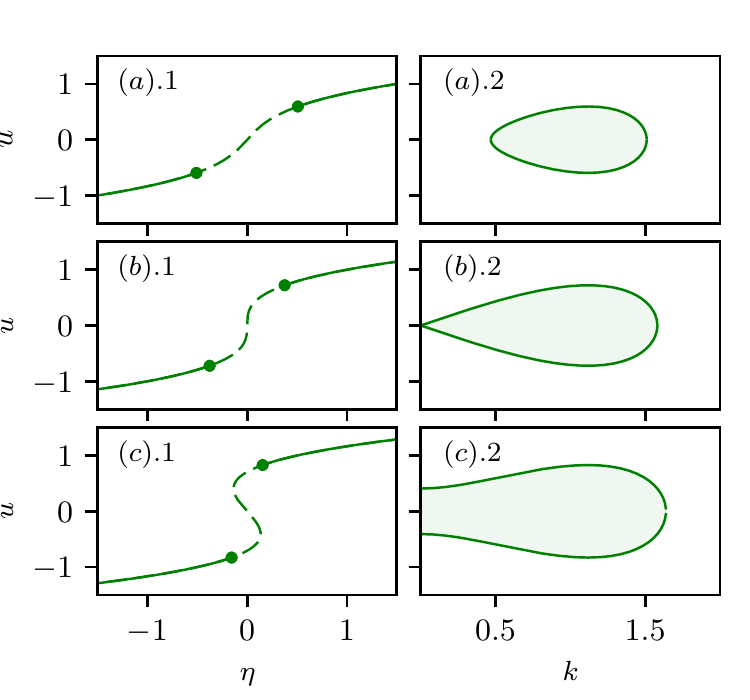}
	\includegraphics[scale=1]{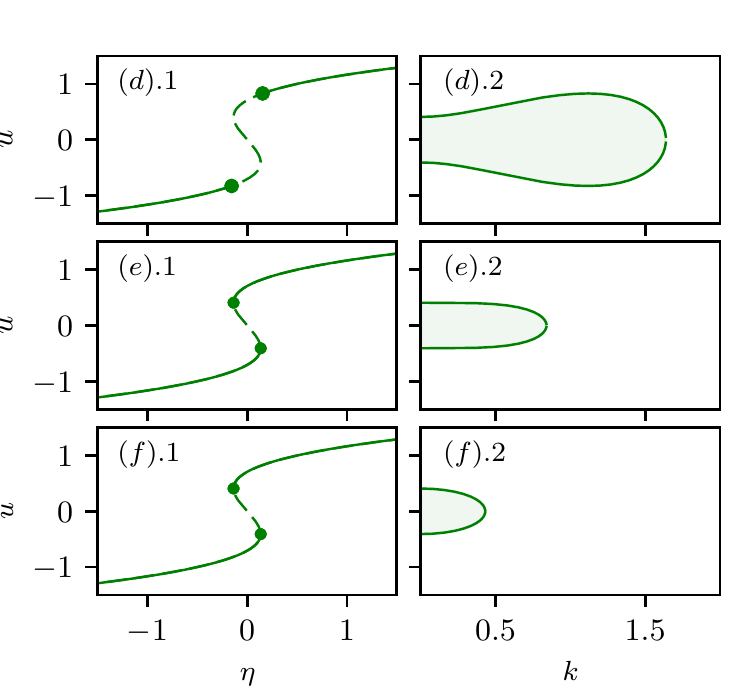}
	\caption{Panels (a)-(c) show the linear stability of the {\rm HSSs} $u_h$ for $\nu=2.5$ and different values of $\mu$. Left columns show the {\rm HSSs} as a function of $\eta$ for $\mu=-0.5$ in (a).1, $\mu=0$ in (b).1 and $\mu=0.5$ in (c).1, while panels (a).2, (b).2  and (c).2 show the corresponding marginal instability curves. Solid (dashed) lines correspond to stable (unstable) solutions. Panels (d)-(f) show the same as panels (a)-(c) for $\mu=0.5$ and different values of $\nu$. Left columns show the {\rm HSSs} as a function of $\eta$ for (d) $\nu=2.5$, (e) $\nu=0$, and (f) $\nu=-2.5$, while the right column shows the corresponding marginal instability curves. The region inside these curves corresponds to unstable {\rm HSSs}. 
		%{\bf Pedro, can you please check this caption - it does not make sense to me. Note that I also changed the text since it did not agree with the caption! Add values of $\alpha,\beta$.}
		Here $\alpha=0$, $\beta=1$.}
	\label{Marginal_caseI}
\end{figure*}

The type of instability undergone by the system thus depends on the value of $\nu$. If $\nu > 0$ the system undergoes first a Turing instability (TI), where $u_h$ becomes unstable to perturbations with wavenumber $k_c$. The onset of the TI is given by
\begin{equation}\label{uc}
	u_c=u_T^{t,b}=\frac{\alpha}{3\beta}\pm\frac{1}{3\beta}\sqrt{\alpha^2+3\beta\left(\mu+\frac{\nu^2}{4}\right)}.
\end{equation} 
Note that TI requires that $\alpha^2+3\beta(\mu+\nu^2/4)>0$, or equivalently $\mu>-\alpha^2/3\beta-\nu^2/4$. In contrast, if $\nu \leq 0$ the system undergoes an uniform stationary instability, where $u_h$ becomes unstable to uniform perturbations. The onset of the instability is then given by $u_c=u_f$, corresponding to the folds, or saddle-node bifurcations SN$_h^{l,r}$, of the HSSs.

Figure~\ref{Marginal_caseI} shows the stability of the HSSs as a function of $\eta$ for different values of $\nu$ and $\mu$. Figures~\ref{Marginal_caseI}(a)-(c) are drawn for $\nu=2.5$ and three different values of $\mu$, namely $\mu=-0.5$ in (a), $\mu=0$ in (b) and $\mu=0.5$ in (c). Left panels show the HSS while the right panels show the associated MIC. For a fixed value of $k=\tilde{k}$, $u_h$ is unstable to perturbations with wavenumber $\tilde{k}$ if it lies inside the area delimited by the MIC (shaded light green area), and stable otherwise. The resulting stability of the HSS is indicated using solid (dashed) lines for stable (unstable) states [see Figs.~\ref{Marginal_caseI}(a).1-(c).1].
In Figs.~\ref{Marginal_caseI}(a).1,2 $u_h$ is single-valued and with increasing $\eta$ it first becomes unstable against spatiotemporal perturbations with $k_c=k_T$ between the two TI points (green dots). As the parameter $\mu$ approaches zero, the MIC expands towards smaller $k$ touching $k=0$ at $\mu=0$. This point corresponds to the cusp bifurcation. This is the situation shown in Figs.~\ref{Marginal_caseI}(b).1,2. For $\mu>0$, $u_h$ becomes triple-valued and the three solution branches $u_h^b$, $u_h^m$ and $u_h^t$ appear as shown in Figs.~\ref{Marginal_caseI}(c).1,2 for $\mu=0.5$. The middle branch $u^m_h$ is unstable against any type of perturbation including those with $k=0$. The branches $u^b_h$ and $u^t_h$, however, are stable against $k=0$ perturbations, but unstable to perturbation with $k\neq0$ between the two TI points.

So far we have studied the stability and configuration of the HSS $u_h$ as a function of $\mu$ at a fixed value of $\nu$. However, one could also wonder what happens when $\mu$ is fixed and $\nu$ varies. Figures~\ref{Marginal_caseI}(d)-(f) show this scenario for $\mu=0.5$ and different values of $\nu$. The situation shown in Figs.~\ref{Marginal_caseI}(d).1,2 for $\nu=2.5$ is the same as that depicted in Figs.~\ref{Marginal_caseI}(c).1,2. As $\nu$ decreases towards zero, the instability onset moves towards $k=0$ ($k_T\rightarrow 0$), and the TI points move towards the folds $u_f^{l,r}$. For $\nu=0$, the TI collapses onto the folds and the wavenumber of the critical pattern arising from TI becomes $k_T=0$. This is the situation shown in Figs.~\ref{Marginal_caseI}(e).1,2. 
For $\nu<0$ (see Figs.~\ref{Marginal_caseI}(f).1,2 for $\nu=-2.5$) $u_h^t$ and $u_h^b$ are both stable against any spatiotemporal perturbation and no spatially periodic states emerge from them.

%%%%%%%%%%%%%%%%%%%%%%%%%%%%%%%%%%%%%%%%%%%%%%%%%%%%%%%%%
% Edgar comment
%%%%%%%%%%%%%%%%%%%%%%%%%%%%%%%%%%%%%%%%%%%%%%%%%%%%%%%%%

%For $\nu>0$, bistability between the subcritical periodic Turing state
%$P$ and $u_h^b$ [see Fig.~\ref{duality}(a)] appears.
%{\bf you have said nothing about the Turing state being subcritical or supercritical. Should this result be included here? {\color{blue} Yes, indeed. I didn't included here because I still don't have that information, which come later in the current order of the paper. For that we have to wait to the weakly nonlinear analysis.}}

%\begin{figure}[!t]
%\centering
%\includegraphics[scale=1]{duality.pdf}
%\caption{Analogy between the stationary solutions of the Eq.~(\ref{general_model1}) and the solution of the spatial dynamical system (\ref{SD}) in the $\nu>0$ and $\nu<0$ regimes. For the solution Eq.~(\ref{general_model1}) we plot $u(x)$, while in the spatial dynamics counterpart we consider its projection on the $(y_2,y_1)$ phase-plane.}
%\label{duality}
%\end{figure}	

\subsection{Linearized spatial dynamics around HSSs}

Linearization of the spatial dynamical system (\ref{SD}) around its fixed points, and the study of its eigenspectrum yield very relevant information regarding the potential emergence of different type of LSs. The spatial eigenvalues $\lambda$ of the Jacobian matrix associated with (\ref{SD}) satisfy the biquadratic equation
\begin{equation}\label{bi-q}
	\lambda^4+\nu\lambda^2+3\beta u_h^2-2\alpha u_h-\mu=0,
\end{equation}
with solutions
\begin{equation}\label{spa_eigen}
	\lambda^2=\frac{1}{2}\left(-\nu\pm\sqrt{\nu^2+4(3\beta u_h^2-2\alpha u_h-\mu)}\right).
\end{equation}
\begin{figure}[!t]
	\centering
	\includegraphics[scale=0.95]{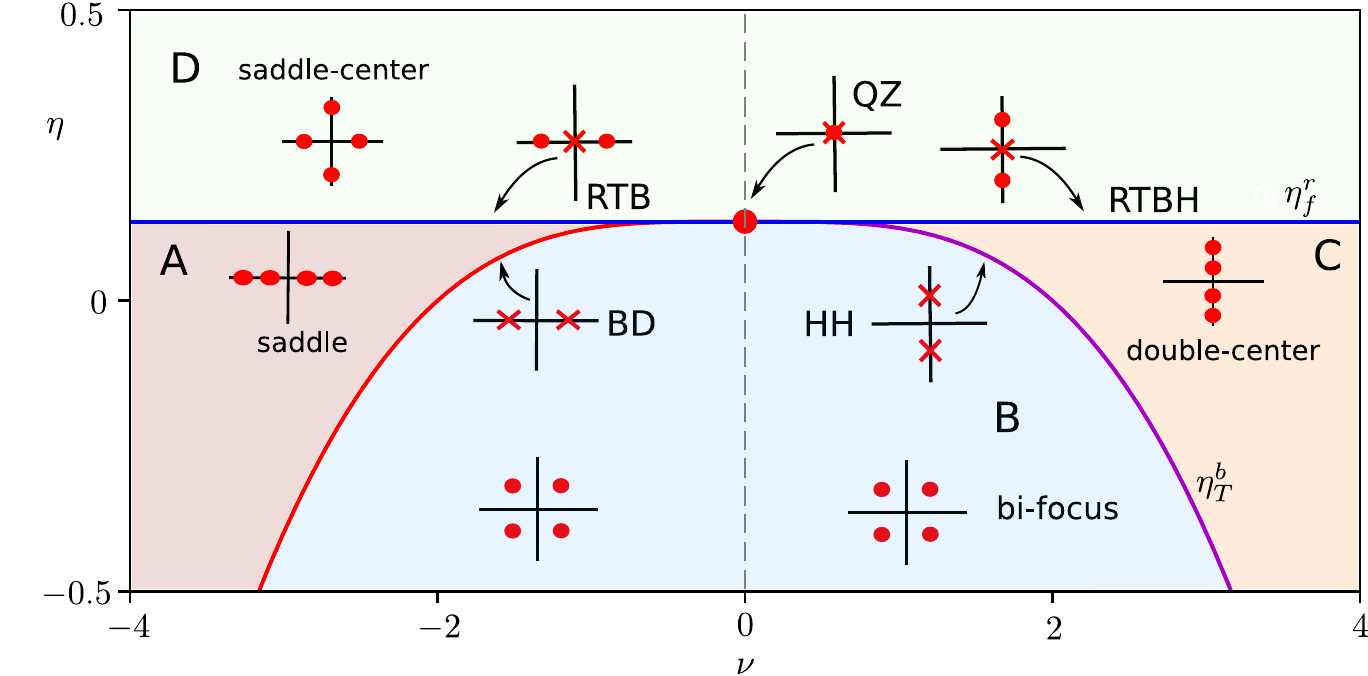}
	\caption{Schematic unfolding of the {\rm QZ} point in the $(\eta,\nu)$ parameter space. For $\nu<0$, the fold $\eta_f^r$ corresponds to a {\rm RTB} bifurcation and the BD transition corresponds to the red line. At the {\rm QZ} point ($\nu=0$), these bifurcations collide, and for $\nu>0$, $\eta_f^r$ becomes a {\rm RTBH} bifurcation and {\rm BD} turns into a {\rm HH} bifurcation (purple line). These four lines organize the different type of equilibria of the system. The terminology is explained in Table~\ref{tab}.}
	\label{eigenvalue_regions}
\end{figure}
This equation is invariant under $\lambda\rightarrow-\lambda$ and $\lambda\rightarrow\bar{\lambda}$ and leads to eigenvalue configurations symmetric with 
respect to both axes as depicted in Figure~\ref{eigenvalue_regions}.
The form of this equation is a consequence of spatial reversibility \cite{devaney_reversible_1976,homburg_homoclinic_2010,knobloch_spatial_2015}.

According to the values of the control parameters, one can identify four qualitatively different eigenvalue configurations:
\begin{enumerate}
	\item[$\mathsf{A}$] The eigenvalues are real: $\lambda_{1,2}=\pm a_1$, $\lambda_{3,4}=\pm a_2$
	\item[$\mathsf{B}$] There is a quartet of complex eigenvalues: $\lambda_{1,2,3,4}=\pm a \pm ib$
	\item[$\mathsf{C}$] The eigenvalues are imaginary: $\lambda_{1,2}=\pm ib_1$, $\lambda_{3,4}=\pm ib_2$
	\item[$\mathsf{D}$] Two eigenvalues are real and two imaginary: $\lambda_{1,2}=\pm a$, $\lambda_{3,4}=\pm ib$
\end{enumerate}

The transitions between these possible eigenvalue configurations are shown in Figure~\ref{eigenvalue_regions}, where the main spatial bifurcations and transitions of the system around SN$^r_h$ are shown for a fixed value of $\mu$ (the same scenario appears when considering SN$^l_t$). Here we adopt the nomenclature  for these transitions introduced in our earlier work \cite{parra-rivas_origin_2016,parra-rivas_parametric_2020}, which is summarized in Table~\ref{tab} and described below, although other conventions are possible:
\begin{itemize}
	\item[$\mathsf{BD}$] The transition from region $\mathsf{A}$ to region
	$\mathsf{B}$ is through a Belyakov-Devaney (BD)
	\cite{devaney_reversible_1976,champneys_homoclinic_1998} line with
	eigenvalues $\lambda_{1,2}=\pm b_0$, $\lambda_{3,4}=\pm b_0$. At
	this transition, a {\it broom } global homoclinic bifurcation takes
	place \cite{champneys_homoclinic_1998,homburg_homoclinic_2010}, such
	that the spatial period of a spatially periodic state becomes
	infinite and the periodic orbit becomes a tame homoclinic orbit.
	\item[$\mathsf{HH}$] The transition from region $\mathsf{B}$ to region
	$\mathsf{C}$ occurs through a Hamiltonian-Hopf (HH) bifurcation
	\cite{iooss_perturbed_1993,haragus_local_2011}. At this bifurcation
	the spatial eigenspectrum reads $\lambda_{1,2}=\pm ib_c$,
	$\lambda_{3,4}=\pm ib_c$. This spatial bifurcation corresponds to a
	temporal Turing instability creating a spatially periodic state with
	wavelength $2\pi/b_c$.
	\item[$\mathsf{RTB}$] The transition from region $\mathsf{A}$ to
	region $\mathsf{D}$ takes place via a reversible Takens-Bogdanov
	(RTB) bifurcation with eigenvalues $\lambda_{1,2}=\pm q_0$,
	$\lambda_{3}=\lambda_{4}=0$ \cite{champneys_homoclinic_1998}. Normal
	form theory predicts that near this point tame homoclinic orbits of
	the form $C_1{\rm sech}(C_2x)$ exist. This transition can also be
	referred to as a Hamiltonian (or reversible) fold bifurcation.
	\item[$\mathsf{RTBH}$] The transition from region $\mathsf{C}$ to
	region $\mathsf{D}$ occurs via a reversible Takens-Bogdanov-Hopf
	(RTBH) bifurcation characterized by eigenvalues $\lambda_{1,2}=\pm
	ib$, $\lambda_{3}=\lambda_{4}=0$
	\cite{champneys_homoclinic_1998}. This transition can also be thought of
	as different case of the Hamiltonian (or reversible) fold, here in the presence of a pair of purely imaginary eigenvalues instead of real eigenvalues.
\end{itemize}
The transition lines in the parameter plane emanate from a codimension-two quadruple zero (QZ) point characterized by the eigenvalues $\lambda_1=\lambda_2=\lambda_{3}=\lambda_{4}=0$
\cite{champneys_homoclinic_1998,haragus_local_2011}. Unfolding of this bifurcation is evidently essential for understanding the nonlinear behavior associated with the transitions shown in Fig.~\ref{eigenvalue_regions}. The first attempt at a systematic study of this point was carried out by Iooss \cite{iooss_codimension_1995} but a complete understanding of the possible behavior near this point is essential for understanding the $L$-point.

%In regions $\mathsf{A}$ and $\mathsf{B}$ of the $(\nu,\eta)$-phase diagram shown in  Figure~\ref{eigenvalue_regions}, $u_h$ is hyperbolic, i.e. Re$[\lambda]\neq0$, and therefore homoclinic orbits bi-asymptotic to $u_h$ are of codimension zero, and therefore they must persist under small reversible perturbations.

Let us now analyze the spatial eigenvalues defined by Eq.~(\ref{spa_eigen}) along the different transition curves in Fig.~\ref{eigenvalue_regions}. The red and purple curves shown in the figure are defined by the equations
\begin{align}
	\eta_c=-\mu u_c-\alpha u_c^2-\beta u_c^3, && u_c=\frac{\alpha}{3\beta}\pm\frac{1}{3\beta}\sqrt{\alpha^2+3\beta\left(\mu+\frac{\nu^2}{4}\right)}.
\end{align}
Along these lines the spatial eigenvalues reduce to  
\begin{equation}
	\lambda_{1,2,3,4}=\left(\pm \sqrt{\frac{-\nu}{2}},\pm \sqrt{\frac{-\nu}{2}}\right),
\end{equation}
i.e., two pairs of eigenvalues, each of double multiplicity.  This expression leads to two different situations depending on the sign of $\nu$. When $\nu<0$ the previous expression becomes 

\begin{equation}
	\lambda_{1,2,3,4}=\left(\pm\sqrt{\frac{|\nu|}{2}},\pm\sqrt{\frac{|\nu|}{2}}\right),
\end{equation}
and $\eta_c$ then corresponds to a Belyakov-Devaney (BD) transition, shown with a red line in Fig.~\ref{eigenvalue_regions}. In contrast, when $\nu>0$
\begin{equation}
	\lambda_{1,2,3,4}=\left(\pm i\sqrt{\frac{|\nu|}{2}},\pm i\sqrt{\frac{|\nu|}{2}}\right),
\end{equation}
and $\eta_c$ then corresponds to a HH spatial bifurcation or, equivalently, the Turing bifurcation TI (purple line).

The other relevant points of this scenario correspond to the two saddle-node bifurcations of $u_h$, namely $u_{f}^{l,r}$. At these points, the spatial eigenvalues become
\begin{equation}
	\lambda_{1,2,3,4}=\left(0,0,\pm\sqrt{-\nu}\right). 
\end{equation}
In Fig.~\ref{eigenvalue_regions} we only plot $\eta_f^r$, although the situation is analogous for $\eta_f^l$.

\begin{table}[t]
	\centering
	\begin{tabular}{|l|c|c|c|c|}
		%	\hline
		\hline
		Cod & $(\lambda_{1,2,3,4})$  &  Name  & Label  \\
		\hline
		Zero & $(\pm q_0\pm i k_0)$  & Bifocus & BiF \\
		\hline
		Zero & $(\pm q_1,\pm q_2)$  & Saddle & S \\
		\hline
		Zero & $(\pm ik_1,\pm ik_2)$  & Double-Center & DC  \\
		\hline
		Zero & $(\pm q_0,\pm ik_0)$   & Saddle-Center & SC  \\
		\hline
		One & $(\pm q_0,0,0)$  & Rev.Takens-Bogdanov & RTB  \\
		\hline
		One & $(\pm ik_0,0,0)$ & Rev.Takens-Bogdanov-Hopf & RTBH   \\
		\hline
		One & $(\pm q_0,\pm q_0)$  &  Belyakov-Devaney & BD \\
		\hline
		One & $(\pm ik_c,\pm ik_c)$  & Hamiltonian-Hopf & HH\\
		\hline
		Two & $(0,0,0,0)$  & Quadruple Zero & QZ \\
		\hline
		%\hline
	\end{tabular}
	\caption{Nomenclature used to refer to different transitions in the spatial eigenspectrum.}
	\label{tab}
\end{table}

As in the previous case, two different cases appear depending on the sign of $\nu$. For $\nu<0$, the previous expression reduces to 
\begin{equation}
	\lambda_{1,2,3,4}=\left(0,0,\pm\sqrt{|\nu|}\right),
\end{equation}
and therefore $\eta_f^r$ corresponds to a reversible Takens-Bogdanov (RTB) bifurcation. In contrast, when $\nu>0$ the spatial eigenvalues read
\begin{equation}
	\lambda_{1,2,3,4}=\left(0,0,\pm i\sqrt{|\nu|}\right),
\end{equation}
and the line $\eta_f^r$ corresponds to a reversible Takens-Bogdanov-Hopf (RTBH) bifurcation.

A classification of the different types of fixed points and bifurcations is summarized in Table~\ref{tab}, where the corresponding codimension, name and label are given. In the following, regions $\mathsf{A}$ and $\mathsf{B}$ of the $(\nu,\eta)$ phase diagram shown in  Figure~\ref{eigenvalue_regions} will be of greatest interest. In these regions $u_h$ is hyperbolic, i.e. Re$[\lambda]\neq0$, and therefore homoclinic orbits biasymptotic to $u_h$ are of codimension zero and therefore persist under small reversible perturbations.

\begin{figure}[!t]
	\centering
	\includegraphics[scale=1]{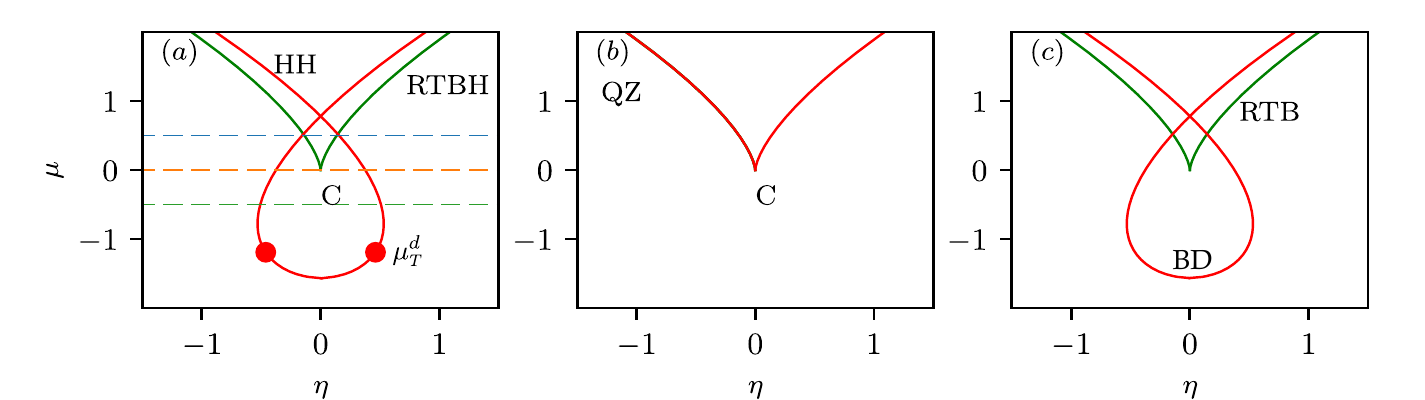}
	\includegraphics[scale=1]{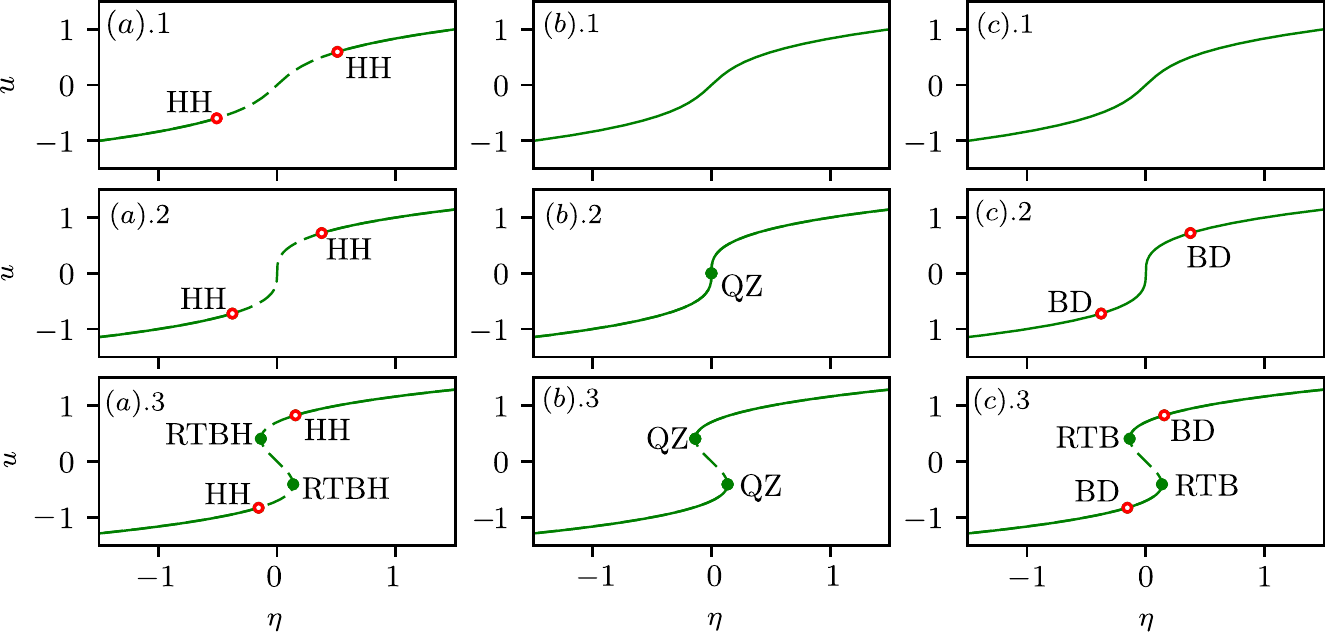}
	\caption{Spatial bifurcations in the $(\eta,\mu)$ parameter space for $\alpha=0$, $\beta=1$, and three values of $\nu$: (a) $\nu=2.5$, (b) $\nu=0$, and (c) $\nu=-2.5$. The green and red lines are defined by $\eta_f=\eta(u_f)$ and $\eta_T=\eta(u_T)$. For $\nu>0$, panel (a), $\eta_f$ and $\eta_T$ correspond to {\rm RTBH} and {\rm HH}, respectively; 
		{\rm HH} generates a supercritical (subcritical) periodic orbit when $\mu<\mu_T^d$ ($\mu>\mu_T^d$). The location of $\mu_T^d$ is marked by the red dots on the HH line.
		In contrast, for $\nu<0$, panel (c), $\eta_f$ and $\eta_T$ become
		{\rm RTB} and {\rm BD}, respectively. 
		The case $\nu=0$, panel (b), shows that the transition between these two scenarios occurs through a degenerate {\rm QZ} bifurcation. 
		The panels below show the homogeneous steady state solutions as a function of $\eta$ for three constant values of $\mu$, namely $\mu=-0.5$ (column {\rm .1}), $\mu=0$ (column {\rm .2}) and $\mu=0.5$ (column {\rm .3}), indicated by horizontal dashed lines in panel (a). The green dots signal the crossing of one of the transition lines. }
	\label{spatial_phase}
\end{figure}

Figure~\ref{spatial_phase} shows the different spatial bifurcations and transitions in the $(\eta,\mu)$ parameter space for three values of $\nu$. For $\nu>0$, the red line is a {\rm HH} bifurcation while the folds correspond to RTBH bifurcations. Three representative slices of this diagram are shown in Figs.~\ref{spatial_phase}(a).1-3. Approaching $\nu=0$, the loop of the HH line shrinks and the red line overlaps with the fold lines, yielding a degenerate QZ line [see Figs.~\ref{spatial_phase}(b) and (b).1-3]. For $\nu<0$ the fold lines become RTB bifurcations and the red line reappears as a BD transition. In the following we focus on the bifurcation scenario shown in Figs.~\ref{spatial_phase}(a) and \ref{spatial_phase}(c).

\section{Weakly nonlinear analysis}\label{sec:5}

%\begin{itemize}
%\item \btext{Divide up the parameter space into regions where different regimes apply (and shade them in appropriate two-parameter slices) }
%\item \btext{Only include analyses that are really novel, perhaps with the details in the appendices}
%\end{itemize}

Linear theory only tells us what happens before or after a given instability or bifurcation, but fails to describe the behavior at the bifurcation itself. The analysis of the system around a local bifurcation reveals the existence of periodic patterns, fronts, and LSs of different types. To perform such analysis, one can either use formal normal form theory \cite{haragus_local_2011,godey_bifurcation_2017} or follow a multiscale perturbation approach as done in \cite{burke_classification_2008}. Here, we follow the latter approach. 

First, we decompose the solution into a homogeneous and a space-dependent part, $u(x)=u_h+\phi(x)$, such that the problem splits in two, yielding an equation for the homogeneous part:
\begin{subequations}
	\begin{equation}\label{gen_hom_eq}
		\beta u_h^3-\alpha u_h^2-\mu u_h-\eta=0,
	\end{equation}
	and a space-dependent equation:
	\begin{equation}\label{gen_nhom_eq}
		-\partial_x^4\phi-\nu\partial_x^2\phi+(\mu-3\beta u_h^2+2\alpha u_h)\phi+(\alpha-3\beta u_h)\phi^2-\beta\phi^3=0.
	\end{equation}
\end{subequations}
Taking $\eta$ as the control parameter of the system and the other parameters as fixed, we expand $\eta=\eta(u_h)$ about the bifurcation point $u_h=u_c$,
\begin{equation}
	\eta=\eta_c+\epsilon^2\delta,
\end{equation}
with $\epsilon$ being a small expansion parameter measuring the distance from the bifurcation point.

In the neighborhood of a local bifurcation, the leading order contribution to a weakly nonlinear state near $u_h$ is in general captured by the ansatz
\begin{equation}
	\phi(x,X)\sim \epsilon^m (A(X)e^{ik_cx}+c.c.),
\end{equation}
where $k_c$ is the characteristic wavenumber of the marginal mode at the bifurcation (i.e. $k_c=0$ for the fold bifurcation, and $k_c=k_T$ for the TI) and $A$ is its  amplitude or envelope describing spatial modulations occurring on a larger spatial scale $X=\epsilon^lx$. The selection of the exponents $m$ and $l$ depends on the problem. For bifurcations in which the amplitude of the nonlinear solution grows as the square root of the distance from the threshold, such as those considered in this work, $m=1$.   

%We have to point out that the temporal stability of such asymptotic states can be estimated analytically as done in \cite{kolokolnikov_existence_2005,van_heijster_pulse_2008,chen_stability_2011}. However, here, the temporal stability is calculated numerically (see Sec.~\ref{sec:5}).

From here on, the calculation depends on the type of bifurcation and the solution that we are looking for. We first compute the weakly nonlinear steady states arising from SN$_h^{l,r}$ [see Sec.~\ref{sec:5.1}], followed by the periodic and weakly nonlinear LSs near TI [see Sec.~\ref{sec:5.2}].

\subsection{Weakly nonlinear solutions near the fold bifurcations}\label{sec:5.1}

We suppose that the saddle-node bifurcations SN$_h^{l,r}$ occur at $u_h=u_f$, and consider the expansions
\begin{align}\label{expansions_fold}
	u_h=u_f+\epsilon u_1+\epsilon^2 u_2+\cdots,&& \phi=\epsilon \phi_1+\epsilon^2 \phi_2+\cdots,
\end{align}
where $\phi(X)$ depends on the long scale $X=\sqrt{\epsilon} x$ only, $\eta=\eta_f+\epsilon^2\delta$ and $\epsilon\propto u_h-u_f$.
\begin{figure}[!t]
	\centering\includegraphics[scale=1]{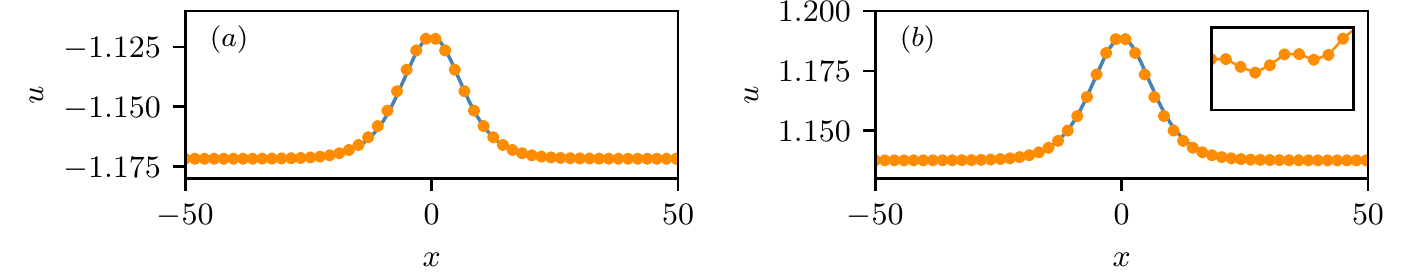}
	\caption{Panel (a) shows the asymptotic approximation (\ref{LSs_fold}) in blue and the exact numerical solution (orange dots) close to {\rm SN}$_h^b$ when $(\nu,\mu)=(-2,4)$ and $\eta=\eta_b-0.001$. We observe very good agreement between the approximate and exact solutions.
		In this case, {\rm SN}$^b_h$ corresponds to a {\rm RTB} spatial bifurcation, and the {\rm LS} arising from it corresponds to a tame homoclinic orbit. Panel (b) shows a similar result for {\rm SN}$^t_h$ when $(\nu,\mu)=(2,4)$ and $\eta=\eta_t+0.001$. In this case, {\rm SN}$^t_h$ corresponds to a {\rm RTBH} spatial bifurcation, and the {\rm LS} arising from it correspond to a generalized homoclinic orbit. The inset shows a close-up view of the tails.}
	\label{agree1}
\end{figure}	
Keeping the leading order terms, the weakly nonlinear LS solutions take the form
\begin{equation}
	u(x)=u_f+\epsilon \left(u_1+\phi_1(x)\right)+\mathcal{O}(\epsilon^2),
\end{equation}
where $u_1=\pm\sqrt{\delta/(3\beta u_f-\alpha)}$, $\phi_1(x)=u_1A(X)$ and $A(X)$ satisfies the second order equation
\begin{align}\label{normal_fold}
	c_0A_{XX}+2A+A^2=0,&& c_0=\frac{\nu}{\sqrt{\delta (3\beta u_f-\alpha)}}.
\end{align}
This amplitude equation supports solutions of the form 
\begin{equation}
	A(X)=-3{\rm sech}^2\left(\frac{1}{2}\sqrt{-\frac{2}{\nu}\sqrt{\delta(3\beta u_f-\alpha)}}X\right),
\end{equation}
leading to the 
weakly nonlinear LS
\begin{equation}\label{LSs_fold}
	u(x)= u_f+ \sqrt{\frac{\eta-\eta_f}{3\beta u_f-\alpha}}\left[1-3{\rm sech}^2\left(\frac{1}{2}\sqrt{-\frac{2}{\nu}\sqrt{(3\beta u_f-\alpha)(\eta-\eta_f)}}x\right)\right]+\mathcal{O}(\eta-\eta_f).
\end{equation}
This expression holds whenever $\nu<0$, i.e., when the saddle-node bifurcations at $u_f$ corresponds to a RTB spatial bifurcation. We label this family of solutions as $\bar\Sigma$. When $u_f$ is instead a RTBH bifurcation the situation is rather more delicate \cite{haragus_local_2011}, and new states, commonly known as {\it generalized solitary waves}, may arise. In contrast to the former, the latter states are biasymptotic to a spatially periodic state of constant but arbitrarily small amplitude. A proper computation of these states requires the application of a careful normal form approach to Eq.~(\ref{SD}) \cite{haragus_local_2011,godey_bifurcation_2017}. However, as found by Gandhi {\it et al.}\cite{gandhi_spatially_2018}, the weakly nonlinear solution (\ref{LSs_fold}), obtained through formal multiscale perturbation analysis that ignores the center eigenvalues, may provide a good approximation to such states provided one replaces $-\nu$ in Eq.~(\ref{LSs_fold}) by $|-\nu|$. The details of this calculation are presented in Appendix~\ref{Appendix:1}.

Figure~\ref{agree1}(a) shows an excellent agreement between the weakly nonlinear solution (\ref{LSs_fold}) (blue line) and the numerically exact solution computed through a Newton-Raphson solver (orange dots) when $u_f$ is of RTB type. Figure~\ref{agree1}(b) shows the corresponding results near RTBH. The inset shows a detail of the small constant amplitude oscillations in the background.

\subsection{Weakly nonlinear solutions near the Turing bifurcation}\label{sec:5.2}

Near the Turing bifurcation, we write
\begin{align}\label{expansion_Turing}
	u_h=u_T+\epsilon^2u_2+\cdots,&& \phi=\epsilon\phi_1(x,X)+\epsilon^2\phi_2(x,X)+\epsilon^3\phi_3(x,X)\cdots\,.
\end{align}
In what follows we allow $\phi$ to depend on both the short scale $x$ and the long scale $X=\epsilon x$, where $\epsilon$ is  defined through the relation $\eta=\eta_T+\epsilon^2\delta$, with $\delta<0$ for $\eta_T^b$ and $\delta>0$ for $\eta_T^t$.
%Introducing these expansion into Eqs.~(\ref{gen_hom_eq}) and (\ref{gen_nhom_eq}) we can solve the problem at each order in $\epsilon$. 

At leading order in $\epsilon$, the weakly nonlinear steady states around TI are described by 
\begin{equation}
	u(x)=u_T+\epsilon \phi_1(x)+\epsilon^2u_2,
\end{equation}
where $u_2=-\delta/\mathcal{M}_T$ with $\mathcal{M}_T\equiv \mu+2\alpha u_T-3\beta u_T^2=-\nu^2/4$, and 
\begin{equation}
	\phi_1(x,X)=A(X)e^{ik_cx}+c.c.
\end{equation}
Here $k_c=k_T$, and the complex amplitude can be written as $A(X)=B(X)e^{i\varphi}$ where $B(X)$ is real and satisfies the equation
\begin{equation}\label{amp_eq_HH}
	\delta B-c_1B_{XX}+c_3B^3=0.
\end{equation}
Here $c_1$ and $c_3$ are two coefficients that depend on the parameters of the system [see Appendix~\ref{Appendix:2}]. Equation~(\ref{amp_eq_HH}) admits uniform solutions satisfying
\begin{equation}
	B(\delta+c_3B^2)=0,
\end{equation}
namely $B=0$ and $B=\sqrt{-\delta/c_3}$. The latter corresponds to the spatially periodic pattern
\begin{equation}\label{HH_pattern}
	u(x)=u_T-\mathcal{M}_T^{-1}(\eta-\eta_T)+2\sqrt{\frac{\eta-\eta_T}{-c_3}}{\rm cos}(k_Tx+\varphi).
\end{equation}
This pattern appears through a pitchfork bifurcation at $\delta=0$ and is supercritical when $c_3>0$ and subcritical when $c_3<0$. The transition between these two scenarios occurs in a degenerate TI (a condimension-two bifurcation) at $(\eta_T,\mu^d_T)$ where $\mu^d_T$ satisfies $c_3|_{\mu^d_T}=0$. This condition yields 
\begin{figure}[!t]
	\centering\includegraphics[scale=1]{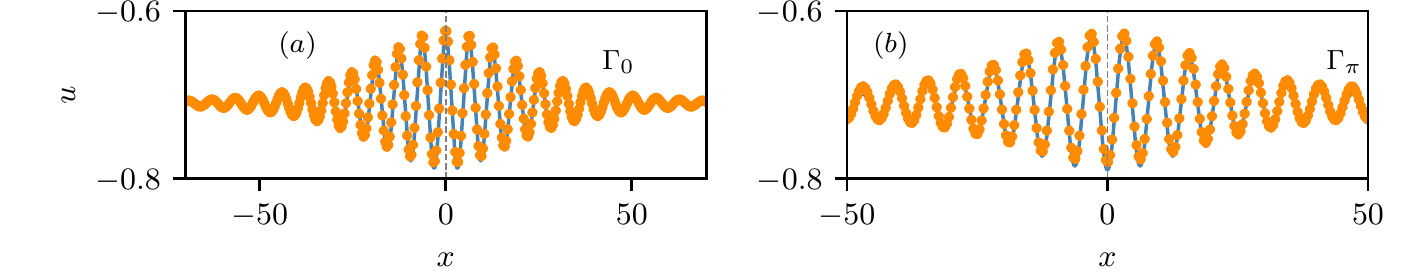}
	\caption{Panel (a) shows the agreement between the asymptotic approximation (\ref{HH_LS}) in blue and the numerically exact solution (orange dots) close to $u_T^b$ for $(\nu,\mu)=(2,0.5)$, $\eta=-0.003$ and $\varphi=0$. Panel (b) shows the agreement between the weakly nonlinear solution (\ref{HH_LS}) and the numerically exact solution for $\varphi=\pi$.}
	\label{agree2}
\end{figure}
\begin{equation}
	\mu_T^d\equiv-\frac{1}{3\beta}\left(\frac{87}{152}\beta\nu^2+\alpha^2\right).
\end{equation}
In the subcritical regime, Eq.~(\ref{amp_eq_HH}) also admits pulse solutions of the form 
\begin{equation}
	B(X)=\sqrt{\frac{2\delta}{-c_3}}{\rm sech}\left(\sqrt{\frac{\delta}{c_1}}X\right), 
\end{equation}
leading to the LS solution
\begin{equation}\label{HH_LS}
	u(x)=u_T-\mathcal{M}_T^{-1}(\eta-\eta_T)+2\sqrt{\frac{\eta-\eta_T}{-c_3/2}}{\rm sech}\left(\sqrt{\frac{\eta-\eta_T}{c_1}}x\right){\rm cos}(k_Tx+\varphi).
\end{equation}	
The spatial phase $\varphi$ of the spatially periodic states is arbitrary and reflects invariance under translations. However, this is no longer so for the LS, where a beyond all orders calculation predicts two specific values of $\varphi$, $\varphi=0,\pi$, both preserving the spatial reversibility symmetry $(x,u)\rightarrow(-x,u)$ of Eq.~(\ref{general_model1}) \cite{melbourne_derivation_1998,burkeknobloch2006,kozyreff_asymptotics_2006,chapman_exponential_2009,kozyreff_localized_2012}. Hence, there are two types of localized weakly nonlinear solutions, one with a maximum at the center of the domain ($x=0$), corresponding to $\varphi=0$, and another with a minimum at $x=0$, associated with $\varphi=\pi$. In the following we label such families of solutions as $\Gamma_0$, and $\Gamma_\pi$, respectively. Continuing these families away from HH they can develop two distinct bifurcation structures depending on the parameters, namely standard homoclinic snaking, discussed in Sec.~\ref{sec:6.1}, or collapsed snaking, discussed in Sec.~\ref{sec:6.2}. In the latter case we will refer to $\Gamma_{0,\pi}$ as $\Gamma'_{0,\pi}$ to distinguish the two cases.

In the present scenario ($\alpha=0$, $\beta=1$), the weakly nonlinear spatially periodic patterns emerging from $u_T^b$ (i.e., the bottom HH bifurcation) read
\begin{align}
	u(x)=u_T^b+\frac{\eta-\eta^b_T}{\nu^2/4}+2\sqrt{\frac{\eta-\eta_T^b}{-c_3^b}}{\rm cos}(k_Tx+\varphi), && c_3=\frac{\sqrt{3}(29\nu^2+152\mu)}{36\sqrt{\nu^2+4\mu}},
\end{align}
%where 
%\begin{equation}
%C_3=\frac{\sqrt{3}(29\nu^2+152\mu)}{36\sqrt{\nu^2+4\mu}}.
%\end{equation}
and the degenerate HH point occurs at 
\begin{equation}
	\mu_T^d\equiv-\frac{29\nu^2}{152}.
\end{equation}
This point is marked with ${\color{red}\bullet}$ in Fig.~\ref{spatial_phase}(a). 
Thus, the spatially periodic states arise subcritically whenever $\mu>\mu^d_T$, and supercritically otherwise.

Furthermore, for $\mu>\mu^d_T$ small amplitude homoclinic orbits of the form  
\begin{equation}\label{simplified_LS}
	u(x)=u_T^b+\frac{\eta-\eta^b_T}{\nu^2/4}+\sqrt{\frac{\eta-\eta^b_T}{-c_3^b/2}}{\rm sech}\left(\sqrt{\frac{\eta-\eta_T^b}{c_1^b}}x\right){\rm cos}(k_Tx+\varphi)
\end{equation}
bifurcate together with the periodic pattern and in the same direction ($\eta>\eta_T$). Here
\begin{equation}
	c_1^{b,t}=\frac{-\nu^3}{2f^{b,t}_0}=\frac{\nu^3}{12u_T^{b,t}}.
\end{equation}
Figure~\ref{agree2}(a) and (b) shows an excellent agreement between the weakly nonlinear solution (\ref{simplified_LS}) (blue line) and the numerically exact solution computed through a Newton-Raphson solver (orange dots) for $\varphi=0$ and $\varphi=\pi$, respectively.  
These approximate solutions are valid near the bifurcation point HH, but fail far from it. However, these families of states can be tracked far from HH using a numerical path-continuation algorithm \cite{doedel_numerical_1991,doedel_numerical_1991-1} resulting in their full bifurcation structure and region of existence.

\section{Two-parameter bifurcation diagrams for $\nu>0$}
\label{sec:6}

In this section we present a detailed study of the bifurcation structure and stability of the family of LSs emanating from the HH bifurcation when $\nu>0$. Here and in Section \ref{sec:9} we focus on the symmetric scenario corresponding to $\alpha=0$. In this case, the scaling $\hat{u}= u/\sqrt{\beta}$, $\hat{\eta} = \eta/\sqrt{\beta}$ allows us to assume $\beta=1$ without loss of generality. The main differences arising when $\alpha\neq0$ are reported in Section~\ref{sec:11}.
\begin{figure}[!t]
	\centering
	\includegraphics[scale=1]{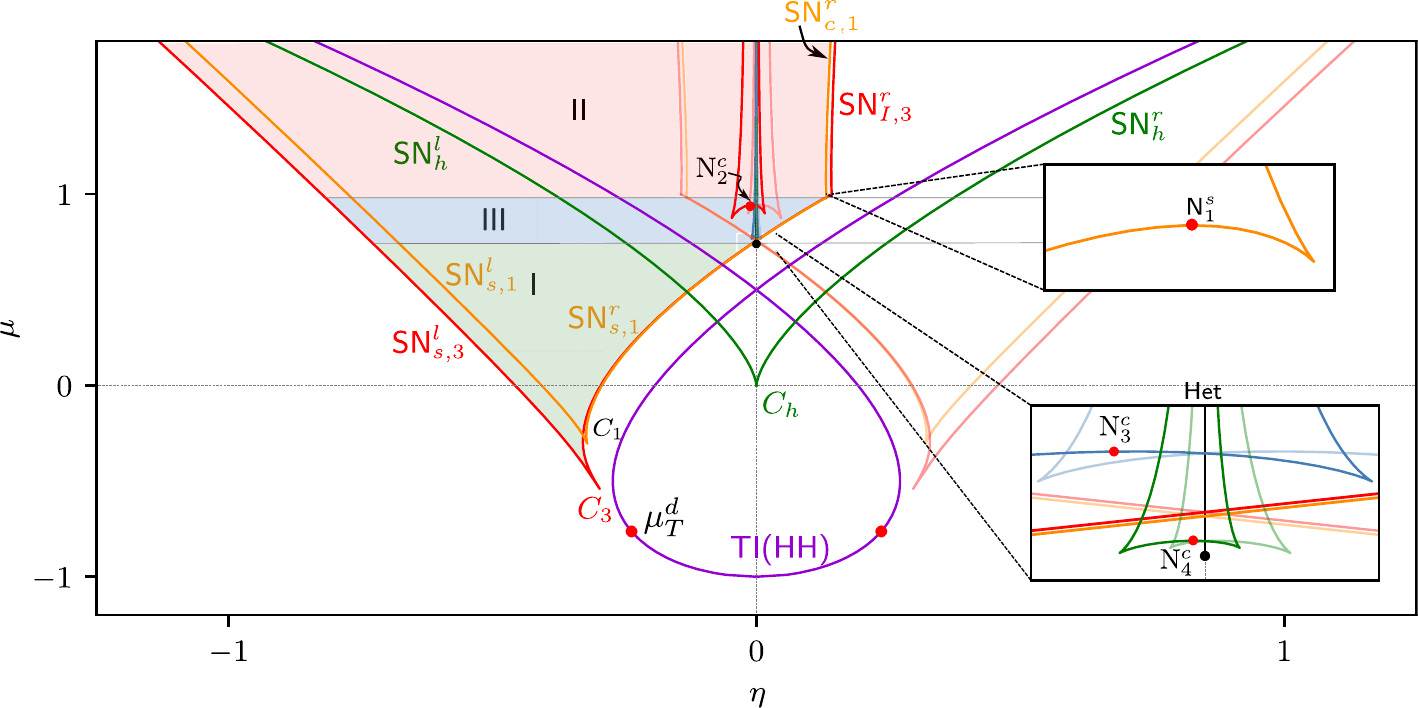}
	\caption{The phase diagram in the $(\eta,\mu)$ parameter space for $\alpha=0$, $\beta=1$ and $\nu=2$, showing the main bifurcation curves for bright {\rm LSs}: ${\rm HH}$ (violet), ${\rm SN}_h^{l,r}$ (green), ${\rm SN}_{s,1}^{l,r}$ and ${\rm SN}^{r}_{c,1}$ (orange), {\rm SN}$_{s,3}^l$ and {\rm SN}$_{I,3}^r$ (red). The cusp bifurcation $C_h$ of the uniform state, the points $C_1$, $C_3$ and the degenerate point $\mu_d$ are also shown. There are two main regions: {\rm I} with standard homoclinic snaking, and {\rm II} with collapsed snaking. The bifurcation lines corresponding to the dark states are related by symmetry to those for the bright states, and are shown in lighter colors. }
	\label{phase_diagram_nupos}
\end{figure}

The type of bifurcation structure that these states undergo far from HH depends on how the stability of the spatially periodic solution and of the HSSs changes with the parameters of the system. Depending on these features three different scenarios appear:
\begin{itemize}
	\item[I.] Standard homoclinic snaking
	\item[II.] Collapsed homoclinic snaking 
	\item[III.] Transition region between the two kinds of homoclinic snaking.
\end{itemize}

Figure~\ref{phase_diagram_nupos} provides a
summary of the bifurcation structures of LSs in the $(\eta,\mu)$ plane. Scenario I corresponds to the light green region below the lower gray dot-dashed line. Scenario II corresponds to the region above the upper gray line. Finally, there is a crossover region between these two scenarios we label as Scenario III.

The green lines in Fig.~\ref{phase_diagram_nupos} correspond to SN$_h^{l,r}$, the violet line represents TI or HH, and the red dot is the degenerate HH. Due to symmetry under the transformation $(u,\eta)\rightarrow(-u,-\eta)$, a solution $u(x;-\eta)$ implies the presence of a solution $-u(x;\eta)$; the corresponding bifurcation curves are related by the same transformation. This leads to the symmetry of the $(\eta,\mu)$ phase diagram with respect to $\eta\rightarrow-\eta$, and one can therefore limit attention to the bifurcation lines for one sign of $\eta$ only. Here we plot lines corresponding to the LSs sitting on $u_h^b$, hereafter bright LSs. In the following section we elucidate the nature of each of these lines.
\begin{figure}[!t]
	\centering
	\includegraphics[scale=1]{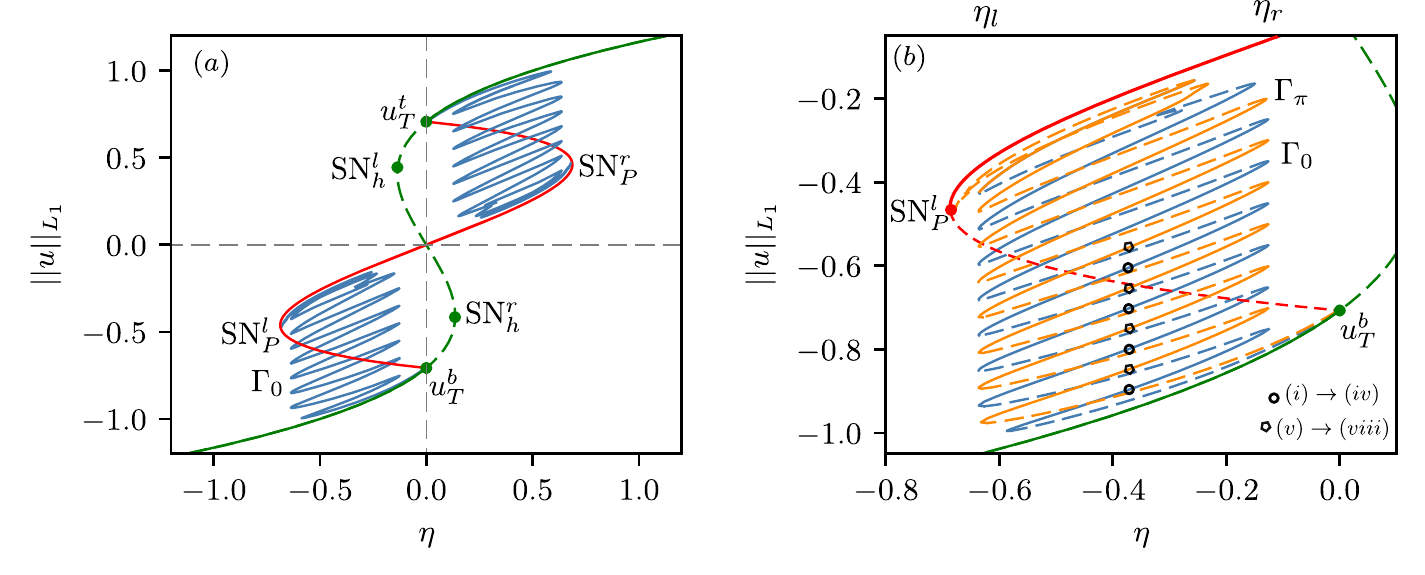}
	\includegraphics[scale=1]{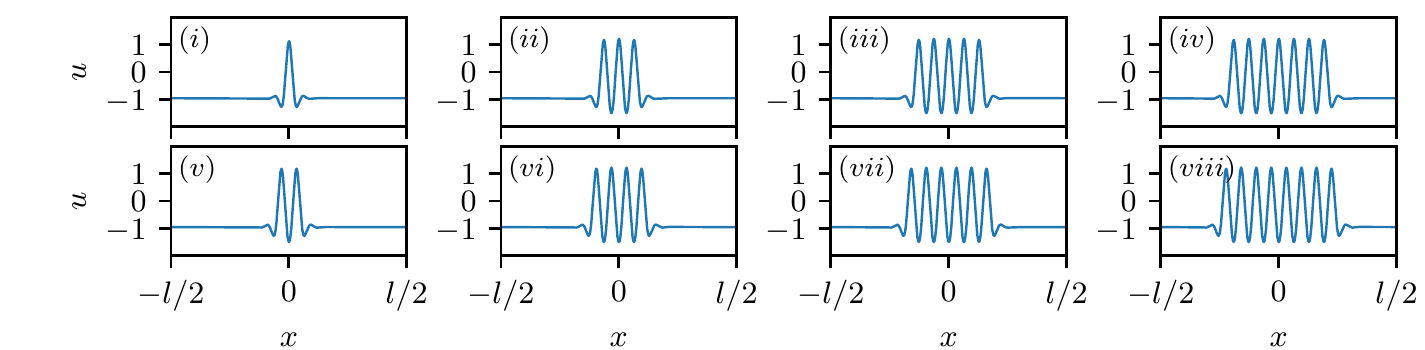}
	\caption{(a) Bifurcation diagram for $\nu=2$ and $\mu=0.5$, showing {\rm HSSs} (green), periodic patterns arising from the {\rm HH} points (red) and the branches of $\Gamma_0$ {\rm LSs} for both bright and dark states (blue). (b) Close-up view of panel (a) focusing on homoclinic snaking of the bright states with $\Gamma_0$ (blue) and $\Gamma_\pi$ (orange). Sample bright profiles along $\Gamma_0$ are shown in panels (i)-(iv) and along $\Gamma_{\pi}$ in panels (v)-(viii). The snaking or pinning region lies between $\eta_l$ and $\eta_r$.  }
	\label{dia_mu05}
\end{figure}

\subsection{Standard homoclinic snaking: Scenario I}\label{sec:6.1}
In this section, we characterize the bifurcation structure of the LSs within the light green region of the phase diagram shown in Fig.~\ref{phase_diagram_nupos}.

Figure~\ref{dia_mu05} shows the bifurcation diagram of LSs as a function of $\eta$ for $\mu=0.5$ in terms of the norm \eqref{L1}. Note that this norm may be negative. This diagram corresponds to a slice at constant $\mu$ of the $(\eta,\mu)$ phase diagram. The stability of the different homogeneous (green line) and LS (blue and orange lines) solutions is indicated using solid (dashed) lines for stable (unstable) states. The dark red line in this diagram corresponds to the spatially periodic states P (stability not indicated). The latter
states arise subcritically from HH at $u_T^b$, and stabilize at the saddle-node SN$_P^l$ at $\eta=\eta_P^l$. After crossing this fold, the P states remain stable until SN$_P^r$ ($\eta=\eta_P^r$) where the P branch folds back and terminates at a second HH point at $u_T^t$. In this way the stable pattern P and the HSS branches $u_h^b$
and $u_h^t$ define two Turing bistability regions, namely
$\eta_P^l<\eta<\eta_T^b$ and $\eta_T^t<\eta<\eta_P^r$. In Fig.~\ref{dia_mu05}(a), due to the choice of parameters, $\eta_T^b = \eta_T^t=0$.

\subsubsection{Snakes-and-ladders structure}
The LSs within each bistability region fall on one of two subcritical branches $\Gamma_{0,\pi}$ as shown in Fig.~\ref{dia_mu05}(a). The bottom left snaking structure corresponds to bright LSs while the top right snaking structure corresponds to dark LSs. Owing to invariance under the transformation $(\eta,u)\mapsto (-\eta,-u)$, the diagram is symmetric with respect to the point $(\eta,u_h)=(0,0)$. For this reason we focus in the following on the bottom left snaking structure. This structure is shown in close-up in Fig.~\ref{dia_mu05}(b), showing the solution curve $\Gamma_0$ in blue and $\Gamma_\pi$ in orange. Some representative examples of $\Gamma_0$ and $\Gamma_\pi$ profiles are shown in Fig.~\ref{dia_mu05}(i)-(iv) and Fig.~\ref{dia_mu05}(v)-(viii), respectively. These states can be seen as a slug of the spatially periodic pattern emerging from HH embedded in a $u_h^b$ background, and will be referred to in the following as {\it localized patterns} (LPs). Near HH, the $\Gamma_0$ and $\Gamma_\pi$ states are well described by the asymptotic expressions (\ref{simplified_LS}). Both arise subcritically from HH and are therefore initially unstable, but as $\eta$ varies they enter the snaking or pinning region $\eta_l<\eta<\eta_r$. Within this region, $\Gamma_{0,\pi}$ both undergo standard homoclinic snaking, i.e., a sequence of back-and-forth oscillations in $\eta$ around the Maxwell point $\eta_M^p$ where $F[u_p]=F[u_h]$. The oscillations in amplitude reflect the successive addition of pattern peaks at either side of the structure each time the solution passes a right saddle-node SN$_i^{r}$ as one proceeds up the diagram. Here the subindex $i$ indicates the number of peaks of the solution. Furthermore, temporal stability of $\Gamma_{0,\pi}$ changes at each SN$_i^{l,r}$ leading to intervals of stable (solid line) and unstable (dashed line) solutions as one proceeds up each snaking branch. These bifurcation points converge monotonically and exponentially to $\eta_l$ and $\eta_r$, respectively, as the number of peaks within the LP increases. For this reason the folds SN$_3^{l,r}$ in Fig.~\ref{phase_diagram_nupos} are already very close to $\eta_l$ and $\eta_r$ (not shown in the figure).
In finite domains, the pairwise peak nucleation process must terminate once the spatial domain is filled. As a result, the $\Gamma_0$ and $\Gamma_\pi$ branches connect with one of the many periodic states arising from $u_h$ between the HH bifurcations at $u_T^b$ and  $u_T^t$.

The formation of LPs and their organization in a homoclinic snaking structure can be understood from a geometrical perspective in terms of a heteroclinic tangle that occurs within $\eta_l<\eta<\eta_r$. Within this range, the unstable manifold of $u_h^b$ ($W_u[u_h^b]$) intersects transversally the stable manifold of a given spatially periodic pattern ($W_s[\gamma]$) as $\eta$ varies. The first tangency $W_u[u_h^b]\cap W_s[\gamma]$ at $\eta_l$ corresponds to the birth of Shilnikov-type homoclinic orbits, which are biasymptotic to the bifocus equilibrium $u_h^b$. The last tangency at $\eta_r$ corresponds to their destruction.% \cite{}.

In addition, the $\Gamma_0$ and $\Gamma_\pi$ branches are connected by a sequence of rung states (not shown) consisting of asymmetric states arising from symmetry-breaking pitchfork bifurcations that occur very close to SN$_i^{l,r}$. Because of its gradient structure these states are stationary, albeit unstable solutions of (\ref{general_model1}). With this set of rung branches, the standard homoclinic snaking takes the form of what is known as a {\it snakes-and-ladders} structure \cite{burke_snakes_2007}.

\subsubsection{Persistence}
The phase diagram of Fig.~\ref{phase_diagram_nupos} shows the changes in the homoclinic snaking region as $\mu$ varies. Decreasing $\mu$, the snaking region shrinks, and the different LP states disappear in a sequence of cusp bifurcations C$_i$ where SN$_i^r$ and SN$_i^l$ annihilate one another. The cusp bifurcations C$_{i}$ with $i=1,3$ associated with the LPs of 1 and 3 peaks are shown in Fig.~\ref{phase_diagram_nupos}. With increasing $\mu$, however, the snaking region expands, and the saddle-node bifurcations SN$_i^r$ are destroyed in a sequence of {\it necking bifurcations} where the snakes-and-ladders structure breaks into separate isolas. The first of such bifurcations, N$^s_1$, occurs at $(\eta,\mu)\approx(0.13299,
0.98085)$ (see inset in Fig.~\ref{phase_diagram_nupos}) and marks an upper limit for the standard snaking region. After that point the homoclinic snaking starts to break up as its branches reconnect with other states. We explain in detail this transition in Section~\ref{sec:6.3}.

%----------------------------------------------------------------------------------- 

The dark LPs follow the same type of structure [Fig.~\ref{dia_mu05}(a)] related to that shown in Fig.~\ref{dia_mu05}(b) through the symmetry $(\eta,u)\rightarrow (-\eta,-u)$. The corresponding bifurcation curves are thus related to those in the $(\eta,\mu)$ phase diagram of Fig.~\ref{phase_diagram_nupos} through the transformation $\eta\rightarrow-\eta$. For simplicity we do not show these additional curves here.

%------------------------------------------------------------------------------------------------------	

%Increasing $\mu$ the region of existence of LPs enlarge [see Fig.~\ref{phase_diagram_nupos}(b)], and eventually the saddle-nodes close to $\eta_r$, i.e., SN$_i^r$
%are destroyed in a sequence of {\it necking bifurcations} [see Fig.~\ref{phase_diagram_nupos}(c)], 
%The various necking bifurcations along the snaking diagrams occur at nearly identical set $(\eta,\mu)=(\eta_N,\mu_N)\approx(0.142774,0.997775)$. In such a way this critical point defines an upper limit of the snaking region.  

\subsection{Collapsed homoclinic snaking: Scenario II}\label{sec:6.2}
\begin{figure}[!t]
	\centering
	\includegraphics[scale=1]{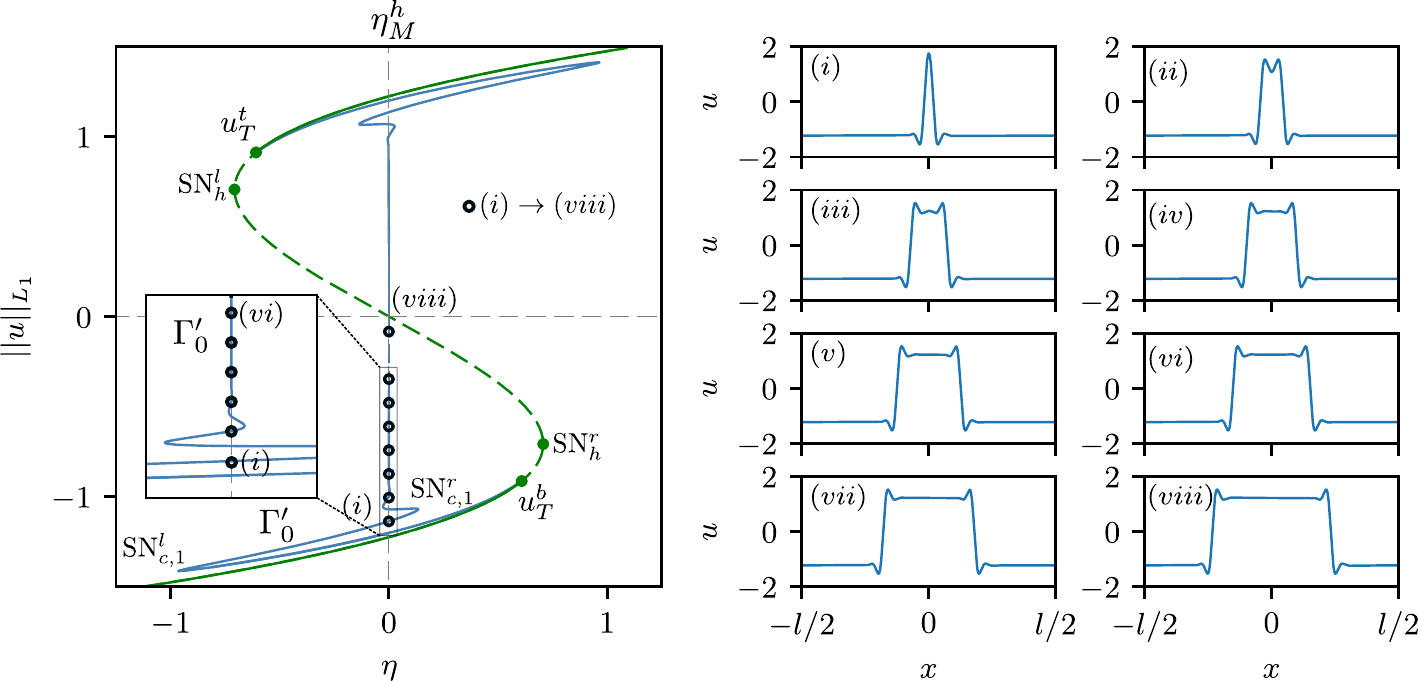}
	\caption{(a) Bifurcation diagram for $\nu=2$ and $\mu=1.5$, showing {\rm HSSs} (green) and the collapsed snaking branch of {\rm LSs} (blue). The inset shows a close-up view of the bright states in the lower portion of diagram. The vertical dashed line $\eta_M^h$, at $\eta=0$ indicates the Maxwell point between the two homogeneous solutions. The green dots indicate the folds of the top and bottom homogeneous solutions, $u_t$ and $u_b$, and the respective Turing instabilities $u_T^{t,b}$.
		(b) Sample bright solution profiles at successive locations indicated by the open dots in (a).}
	\label{dia_collapsed1}
\end{figure}

Above N$_s^1$, the states $\Gamma'_{0,\pi}$ arising from HH organize in a different bifurcation structure. One example of such a scenario is depicted in Fig.~\ref{dia_collapsed1} for $\mu=1.5$ and $\varphi=0$. The vertical dashed line at $\eta=\eta^h_M\equiv0$ represents the HSS Maxwell point, where $F[u_h^b]=F[u_h^t]$. Within the parameter interval $\eta_T^t<\eta<\eta_T^b$, $u_h^b$ and $u_h^t$ coexist and are stable. In the following we refer to this interval as the uniform-bistability region. Within this region, stationary fronts connecting $u_h^b$ and $u_h^t$ ($u_h^{b\rightarrow t}$), and vice versa ($u_h^{t\rightarrow b}$), may be present.

These heteroclinic orbits may approach and leave the HSS $u_h^{b,t}$ either monotonically or in an oscillatory manner, depending on the spatial stability of such equilibria. In the parameter range studied here ($\eta\in[\eta_T^t,\eta_T^b]$), $u_h^{b,t}$ are of bifocus type, and therefore the spatial eigenvalues are $\lambda_{1,2,3,4}={\pm a\pm i b}$, with $a,b\in \mathsf{R}^+$. Hence, any orbit approaching such an equilibrium does so in an oscillatory fashion described by $u-u_h^{b,t}\sim e^{\pm ax}{\rm cos}(bx)$ whenever $|u-u_h^{b,t}|\ll1$. In the following, we use the term {\it oscillatory tail} to refer to the part of a front far from its core, the part described by the preceding approximation. In the presence of such oscillatory tails, two back-to-back fronts, $u_h^{b\rightarrow t}$ and $u_h^{t\rightarrow b}$, may interact and lock, leading to a large number of different LSs close to $\eta^h_M$. These states consist of a {\it plateau} of one of the HSSs embedded in the other.  

%-----------------------------------------------------------

\subsubsection{Collapsed snaking structure}
Before introducing the front locking mechanism, let us discuss the bifurcation structure shown in Fig.~\ref{dia_collapsed1}. Here the $\Gamma'_0$ states again arise subcritically from HH and hence are unstable, as in the scenario described in Sec.~\ref{sec:6.1}. The single peak LSs persist to high amplitude as $\eta$ decreases to SN$_{s,1}^l$, where it folds back to approach the front-locking region close to $\eta_M^h$ (see inset of Fig.~\ref{dia_collapsed1}). Beyond SN$_{s,1}^l$ this state is stable and remains so until it reaches SN$^r_{c,1}$. This LS, shown in Fig.~\ref{dia_collapsed1}(i), resembles that in Fig.~\ref{dia_mu05}(i) and indeed one can pass between them by continuously changing the parameters from region I to region II. Proceeding up the bifurcation diagram, $\Gamma'_0$ starts to oscillate around $\eta_M^h$, approaching it asymptotically as $||u||_{L_1}\rightarrow 0$. All along $\Gamma'_0$, the LSs undergo successive saddle-node bifurcations SN$^{l,r}_{c,i} (i=1,2,3,\cdots)$ lying, respectively, to the left/right of $\eta_M^h$. Because of the exponentially decreasing amplitude of these oscillations the bifurcation structure in Fig.~\ref{dia_collapsed1} is commonly known as {\it collapsed homoclinic snaking} \cite{knobloch_homoclinic_2005,burke_classification_2008,parra-rivas_dark_2016}. At each SN$_{c,i}^{r}$ on $\Gamma'_0$, an additional spatial oscillation is nucleated around $x=0$, leading to a continuous growth in the LS width as the inner part of the LS fills with $u_h^t$. This broadening can be observed in the series of LS profiles shown in Fig.~\ref{dia_collapsed1}(i)-(viii). In the broadest states [e.g., profiles (iv)-(viii)] one can easily see the connection of $u_h^b$ with $u_h^t$ and back again to $u_h^b$, a result of the spatial locking of the heteroclinic orbits $u_h^{b\rightarrow t}$ and $u_h^{t\rightarrow b}$.

Owing to the finite size of the domain, the continued broadening of the LS must terminate, as in the case of the LPs in Sec.~\ref{sec:6.1}. Here, however, as $||u||_{L_1}$ becomes positive, the extent of the plateau around $u_h^t$ becomes longer than the interval occupied by $u_h^b$, and the bright LSs turn smoothly into dark LSs, with $u_h^b$ now embedded in a $u_h^t$ background. These dark LSs also undergo collapsed homoclinic snaking as can be seen in the top part of the diagram depicted in Fig.~\ref{dia_collapsed1}. Near HH on $u_T^t$ the dark LSs are again well described by (\ref{simplified_LS}) before their disappearance at $u_T^t$.

\begin{figure}[!t]
	\centering
	\includegraphics[scale=1]{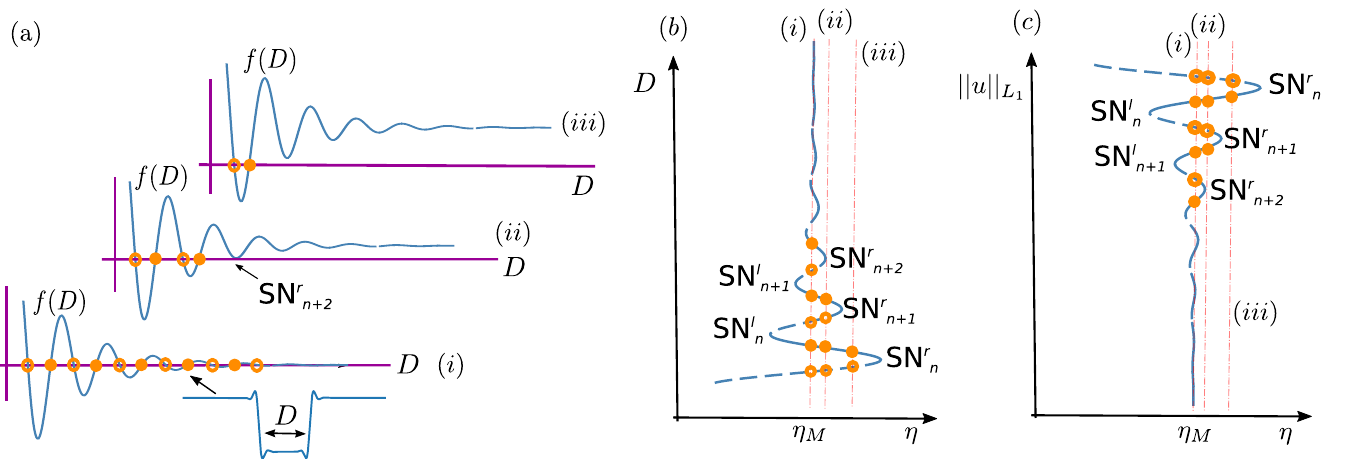}
	\caption{Schematic representation of the interaction and locking of domains walls. (a) Equilibria of Eq.~(\ref{frontinteraction}) for different values of $\eta$: $\eta=0$ at location (i), and $\eta_{1,2}$ with $\eta_2>\eta_1$ at locations (ii) and (iii). Stable (unstable) states are indicated by full $\bullet$ (empty $\circ$) circles. (b) Bifurcation diagram showing the equilibrium front separation $D$ from Eq.~(\ref{frontinteraction}) as a function of $\eta$ and the three locations (i), (ii) and (iii). (c) Same as (b) but using the norm $||u||_{L_1}$ instead of $D$. This last diagram is a schematic picture of the collapsed snaking shown in Fig.~\ref{dia_collapsed1}.}
	\label{inter_esquema}
\end{figure}

The front locking mechanism responsible for collapsed snaking can be understood phenomenologically via the model equation \cite{coullet_nature_1987,coullet_localized_2002} 
\begin{equation}\label{frontinteraction}
	\partial_t D=\varrho e^{-aD}{\rm cos}(bD)+\Theta\equiv f(D),
\end{equation} 
describing the temporal evolution of the separation $D$ between two fronts. Here, $\Theta$ is proportional to the distance from the Maxwell point $\eta_M^h$ (i.e., $\Theta\sim \eta-\eta^h_M$), $\varrho$ depends on the parameters of the system and $a$ ($b$) is the real (imaginary) part of the leading spatial eigenvalue associated with the bifoci $u_h^{b,t}$. In special cases this equation can be derived by a systematic procedure \cite{clerc_analytical_2010,escaff_non-local_2011,tlidi_localized_2015}. Its equilibria and their stability are summarized in Fig.~\ref{inter_esquema} and reproduce the collapsed snaking diagram in Fig.~\ref{dia_collapsed1}.

In summary, when the fronts are close to one another the interaction between them is strong and the pinning interval is broad; when the fronts are far apart the interaction is exponentially weak and the distance between successive saddle-nodes shrinks exponentially.

\subsubsection{Other states and persistence}

The collapsed snaking structure exists in the pink shaded area (II) shown in the $(\eta,\mu)$ phase diagram of Fig.~\ref{phase_diagram_nupos}. As $\mu$ increases the collapsed snaking persists, and the region of existence of the LSs expands. Decreasing $\mu$, however, leads to the break up of collapsed snaking at the necking bifurcation N$_s^1$ as region I is approached. Below this point, isolas of LSs are present but such isolas disappear again in another sequence of necking bifurcations near $\eta_M^h$, labeled N$^c_i$, as $\mu$ decreases.  The extent of these isolas is determined by the {\it forked tongues} shown in Fig.~\ref{phase_diagram_nupos}, Fig.~\ref{isolas_collapsed} and Fig.~\ref{phase_transit}.  We study this transition in the next section.

%#-------------------------------------------------------------

\subsection{The transition region: Scenario III}
\label{sec:6.3}
\begin{figure}[!t]
	\centering
	\includegraphics[scale=1]{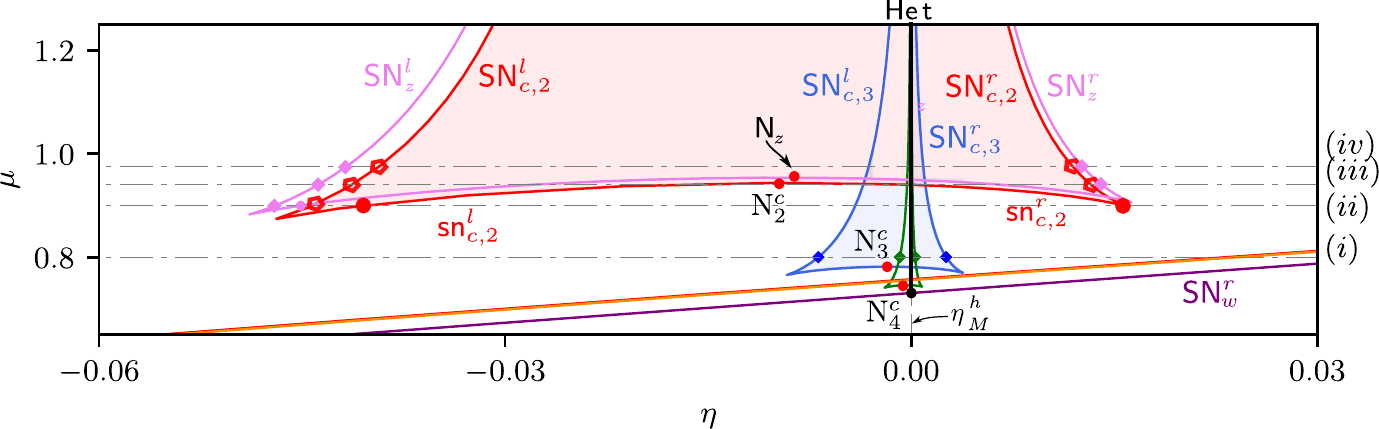}
	\includegraphics[scale=1]{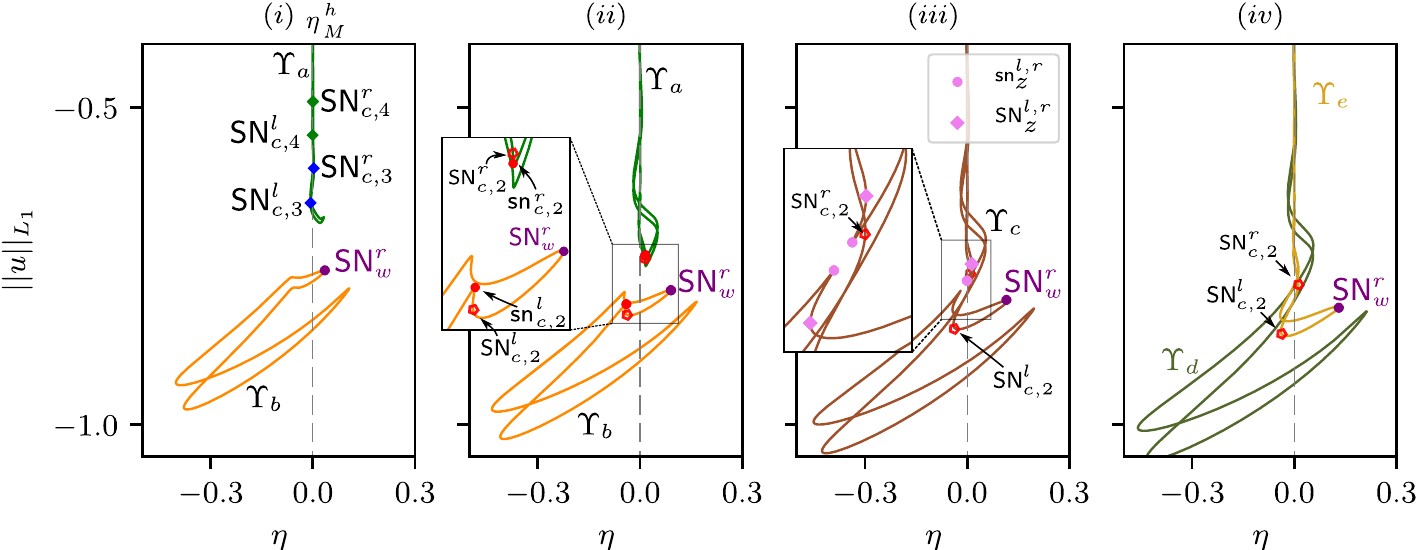}
	\caption{Top panel: a close-up view of the $(\eta,\mu)$ phase diagram in Fig.~\ref{phase_diagram_nupos}, showing the details of the transition region between regions {\rm I} and {\rm II} around $\eta_M^h$. Here we plot the main saddle-node, necking and cusp bifurcations of the system. See text for additional details.
		Labels (i)-(iv) refer to the constant-$\mu$ slices depicted in the remaining panels, corresponding to $\mu=0.8,0.9,0.95$ and $0.975$, respectively.}
	\label{isolas_collapsed}
\end{figure}

\begin{figure}[!t]
	\centering
	\includegraphics[scale=1]{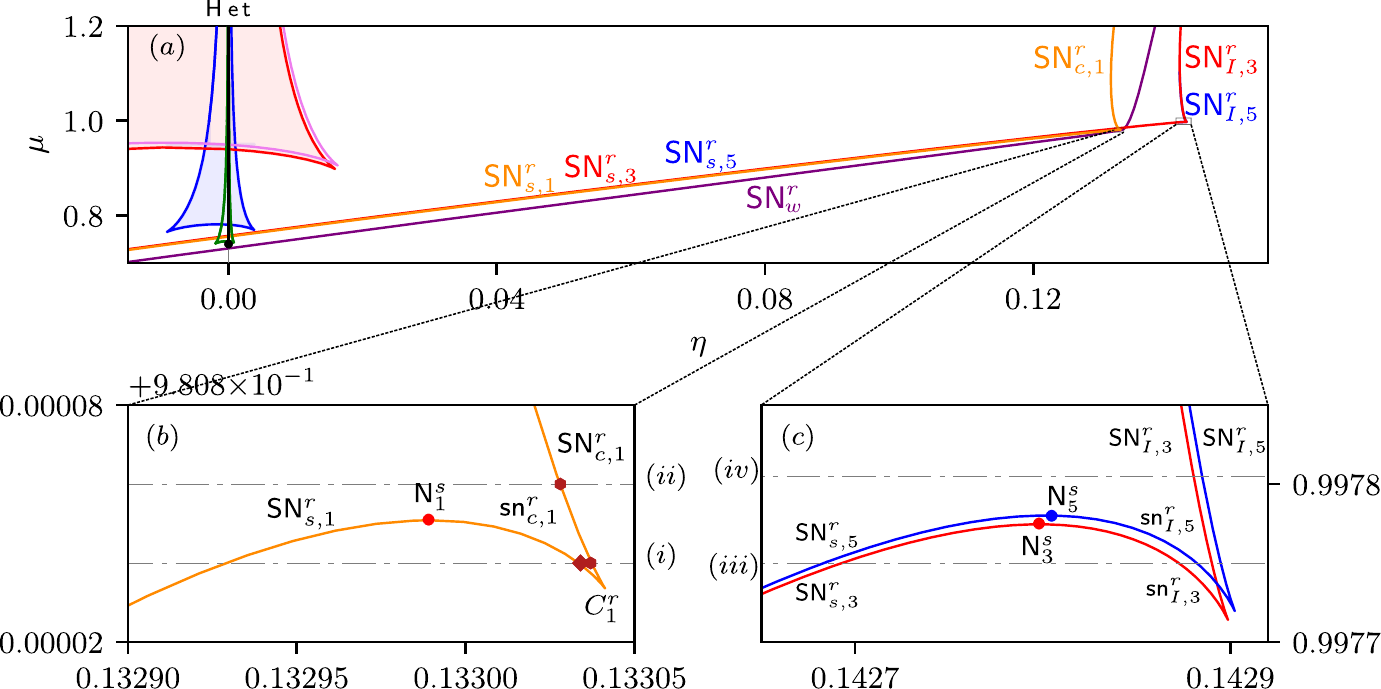}
	\caption{(a) Close-up view of the phase diagram shown in Fig.~\ref{phase_diagram_nupos} near the standard-to-collapsed homoclinic snaking transition. The horizontal point-dashed lines (i)-(iv) correspond to the diagrams shown in Fig.~\ref{isolas_collapsed}.  Panels (b) and (c) show close-up views of panel (a) around the necking bifurcations {\rm N}$_1^s$ and {\rm N}$_{3,5}^s$, respectively. The forked tongues around $\eta_M^h$ are related to the {\rm LSs} formed through the locking of fronts connecting the two {\rm HSSs}, and exhibit a cascade of necking bifurcations labeled {\rm N}$_c^i$. The horizontal dashed lines (i)-(ii) in panel (b) correspond to the diagrams shown in Fig.~\ref{necking12}; those in (c) correspond to the diagrams shown in Fig.~\ref{necking34a}.}
	\label{phase_transit}
\end{figure}

In the previous sections we have analyzed the bifurcation structure associated with LSs arising from HH in region I, where they undergo standard homoclinic snaking, and region II where they exhibit collapsed homoclinic snaking. The transition between these two scenarios is quite complex, and is mediated by a sequence of necking bifurcations. As a result, different segments of the previous bifurcation curves merge with a number of isolas. The transition region associated with this process is depicted in two close-up views of Fig.~\ref{phase_diagram_nupos}, shown in Figs.~\ref{isolas_collapsed} and \ref{phase_transit}. There are two types of necking bifurcations: those associated with the destruction of standard homoclinic snaking, labeled N$_i^s$, and those related with the destruction of collapsed snaking, labeled N$_i^c$. In what follows we analyze each transition separately.

Before considering these necking transitions in detail,
one can ask what is the fundamental mechanism that forces this necking to happen. Here we can learn from the recent work in \cite{al_saadi_unified_2021} which considers a similar transition in the context of reaction-diffusion systems. The key idea is that, as one varies a second parameter ($\mu$ in our case), one-parameter diagrams undergo collapsed snaking (for sufficiently large absolute values of $\|u\|_{L1}$) whenever there exists a heteroclinic cycle, that is, a pair of symmetry-related heteroclinic connections (fronts) between the spatially homogeneous states. The transition occurs at a codimension-two point for which the heteroclinic orbits become degenerate due to a quadratic tangency between the two two-dimensional stable and unstable manifolds involved. This codimension-two bifurcation can be seen in Fig.~\ref{isolas_collapsed}, at the point marked 'Het'. The (black) curve of heteroclinic cycles emerges from this point. 

The case studied in \cite{al_saadi_unified_2021} was for systems that, while reversible, did not conserve a first integral. Here the computation of heteroclinic cycles is easier, because the cycle must exist for the value of the parameters for which the two equilibria in question have the same value of $H$ given by Eq.~(\ref{eq:Hu}). For the equilibria $u=\pm \sqrt{\mu/\beta}$, the choice $\alpha=0$, $\beta=1$ implies that $\mu>0$ and such cycles must therefore occur along the half-line $\{\eta=0, \mu>0\}$. However, numerical computations reveal that such heteroclinic cycles do not exist along this whole half-line.
Instead there is a finite $\mu$ value, $\mu_{\rm Het}(\eta_M^h)$,
for which the aforementioned heteroclinic tangency occurs, such that the heteroclinic cycle exists along $\{\eta=\eta_M^h=0, \mu>\mu_{\rm Het}\}$. This is the point marked `Het' in Fig.~\ref{isolas_collapsed}.

The transition between regular and collapsed snaking can thus be explained, in the limit of wide patterns, by the resulting change in the ultimate limit state of the snaking bifurcation curve. However, from the point of view of the evolution of one-parameter bifurcation diagrams, the process is much more complex, which we shall now describe.

\subsubsection{Transition mediated by N$_i^c$}

Let us start with a description of the process taking place through the sequence of necking bifurcations labeled N$_i^c$. These bifurcations occur below N$^s_1$, and therefore, inside region I, where patterned LSs undergo standard snaking. A close-up view of this area is depicted in the top panel of Fig.~\ref{isolas_collapsed}. The three fork-shaped tongues show the regions of existence of LSs of different extent, bounded by the saddle-node bifurcations SN$_{c,i}^{l,r}$ and sn$_{c,i}^{l,r}$ (with $i=1,2,3$). Each pair SN$_{c,i}^{l,r}$ and sn$_{c,i}^{l,r}$ collide in a cusp bifurcation labeled $C_{c,i}^{l,r}$, while sn$_{c,i}^l$ and sn$_{c,i}^r$ coalesce at the necking bifurcations N$_i^c$. The horizontal point-dashed lines in the phase diagram of Fig.~\ref{isolas_collapsed} refer to the constant-$\mu$ slices shown in Figs.~\ref{isolas_collapsed}(i)-(iv) depicting a certain number of the isolas around the Maxwell point $\eta_M^h$.

Figure~\ref{isolas_collapsed}(i) shows the $\Upsilon_a$ isola for $\mu=0.8$. The LSs between SN$_{c,3}^{l,r}$ correspond to a three-peak state like that shown in Fig.~\ref{dia_collapsed1}(iii). For this value of $\mu$, a second isola, labeled $\Upsilon_b$, is also present. The states on this isola are related to the two-peak state shown in Fig.~\ref{dia_collapsed1}(ii). As $\mu$ decreases further only isolas of the broadest LSs remain, and even all these disappear below 'Het'.
In contrast, increasing $\mu$, results in a situation like that depicted in Fig.~\ref{isolas_collapsed}(ii) for $\mu=0.9$. By this value of $\mu$, sn$_{c,2}^{l,r}$ have already formed through the cusp bifurcations $C_{c,2}^{l,r}$. A close-up view of the resulting isolas is shown in the inset, where the top part of $\Upsilon_b$ and the bottom part of $\Upsilon_a$ are depicted together with   SN$_{c,2}^{l,r}$ and sn$_{c,2}^{l,r}$.  The bifurcations SN$_w^r$, relevant for increasing $\mu$, are also labeled.

At N$_2^c$ (i.e. $\mu=\mu_{{\rm N}_2^c}$), sn$_2^l$ and sn$_2^r$ collide forming the stable branch bounded by SN$_{c,2}^l$ and SN$_{c,2}^r$. As a result, $\Upsilon_a$ and $\Upsilon_b$ merge into a single isola $\Upsilon_c$. This situation is shown in Fig.~\ref{isolas_collapsed}(iii) for $\mu=0.95$, just after N$_2^c$. The inset in that figure shows a close-up view where sn$_{z}^{l,r}$ and SN$_{c,2}^r$ can easily be appreciated. Increasing $\mu$ slightly more leads to a secondary necking bifurcation N$_T$ at  $(\eta,\mu)\approx(-0.01,0.96)$ where sn$_{z}^{l}$ and sn$_{z}^{r}$ collide. Owing to this merger two new isolas $\Upsilon_d$ and $\Upsilon_e$ arise. This new configuration is depicted in Fig.~\ref{isolas_collapsed}(iv) for $\mu=0.975$. 

In what follows, we focus on the reconnections encountered by $\Upsilon_e$ as $\mu$ increases.

\begin{figure}[!t]
	\centering
	
	\includegraphics[scale=1]{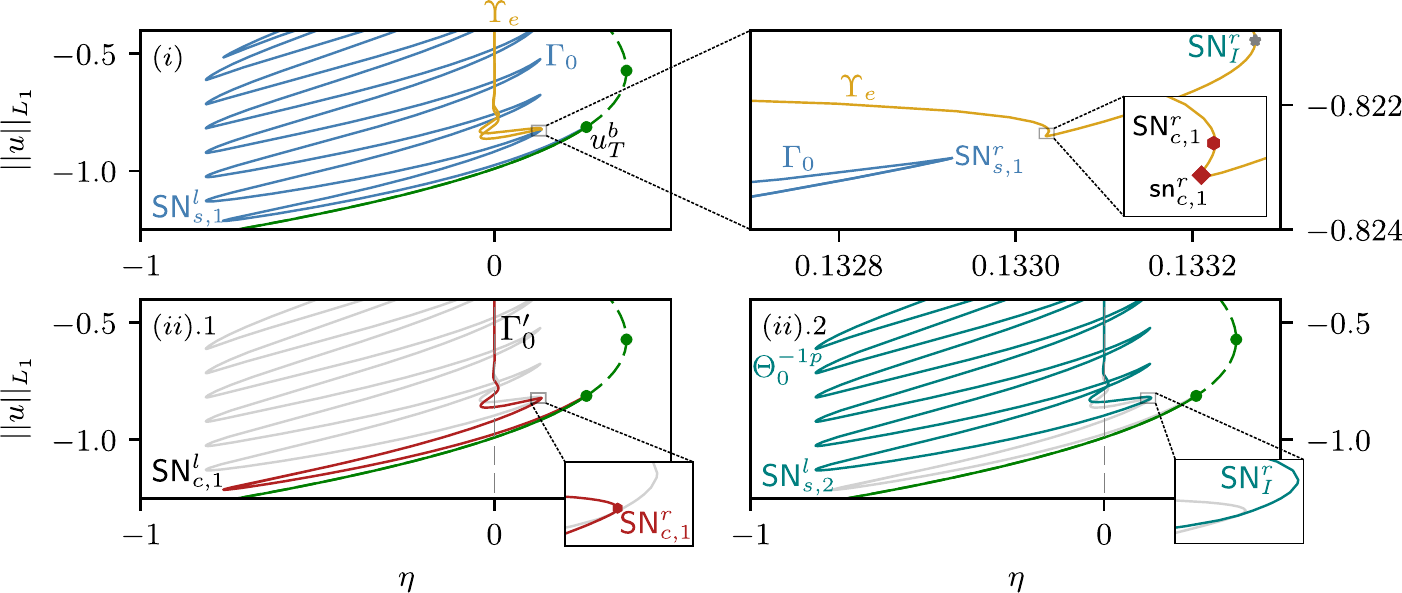}
	\caption{Destruction of standard homoclinic snaking through the necking bifurcation ${\rm N}_1^s$ occurring at $(\eta,\mu)=(0,1329,0.98085)$. In (i) we show the bifurcation scenario very close to {\rm N}$_1^s$, at $\mu=0.98084$, and a close-up view. Panels (ii) shows the reorganization of the branches for $\mu=0.980861$, just after the occurrence of ${\rm N}_1^s$.}
	\label{necking12}
\end{figure}
\begin{figure*}[!t]
	\centering
	\includegraphics[scale=1]{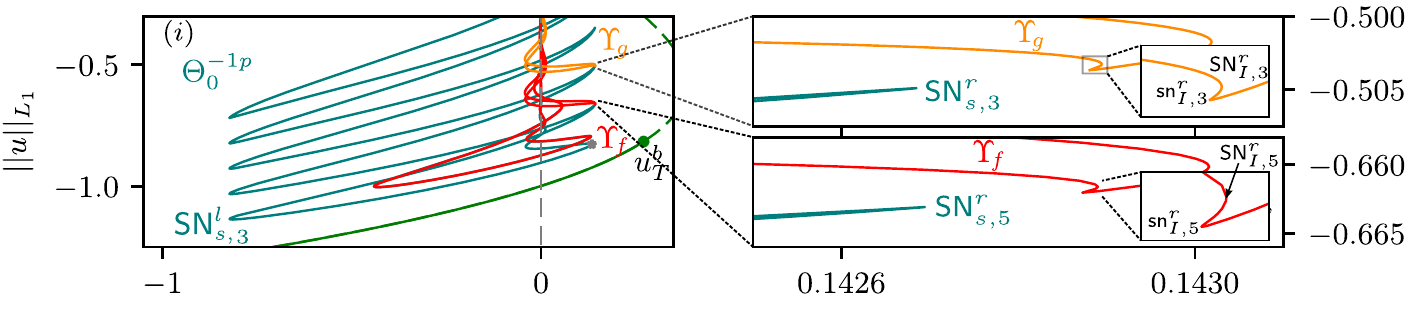}
	\includegraphics[scale=1]{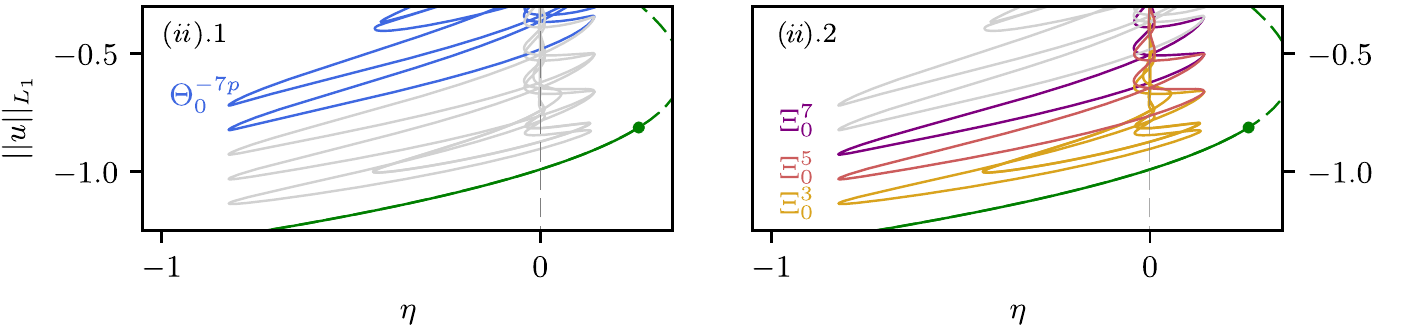}
	\caption{Destruction of the isola $\Theta_0^{-1p}$ and the formation of $\Theta_0^{-7p}$ and $\Xi_0^{3}$, $\Xi_0^{5}$ and $\Xi_0^{7}$ when crossing {\rm N}$_3^s$ and {\rm N}$_5^s$. In panel (i) $\mu=0.98084$, while panels (ii).1 and (ii).2
		show the situation for $\mu=0.997809$ after crossing {\rm N}$_3^s$ and {\rm N}$_5^s$.	
		%Destruction of the standard homoclinic snaking through a necking bifurcation {\rm N}$_1^s$ occurring at $(\eta,\mu)=(,)$. {\bf pls complete} In (i) and the close-ups to the right we show the bifurcation scenario very close to N$_a$ at $\mu=0.997751$. In (iv).1,2 we show the reorganization of the branches for $\mu=0.997809$, just after N$_3^s$.
	}
	\label{necking34a}
\end{figure*}

\begin{figure}[!t]
	\centering
	\includegraphics[scale=0.95]{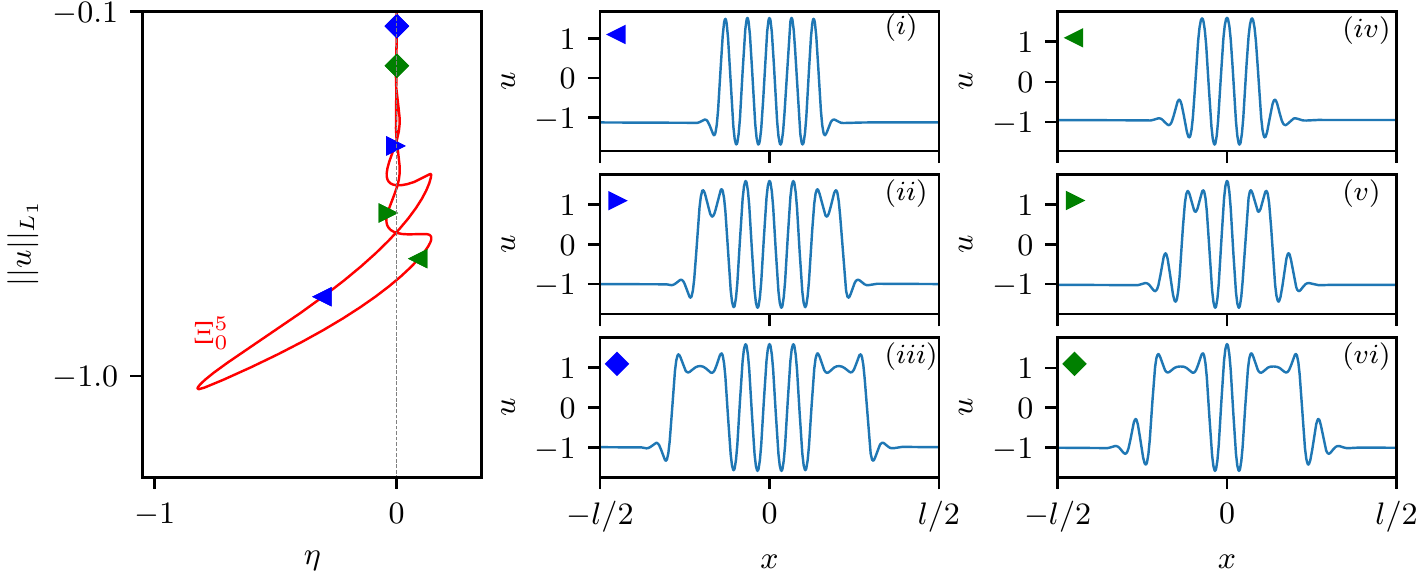}
	\caption{Hybrid states associated with $\Xi_0^5$.}
	\label{hybrid}
\end{figure}

\subsubsection{Transition mediated by N$_i^s$}
So far, we have analyzed the entanglement of isolas $\Upsilon_{a,...,e}$, through the necking bifurcations N$_i^c$ (see Fig.~\ref{isolas_collapsed}). Here, we continue that analysis and describe the process leading to the sequential destruction of standard snaking and appearance of collapsed snaking. This destruction is mediated by a new set of necking bifurcations that we label N$_i^s$ (with $i$ associated with the SN$^{l,r}_{s,i}$).

Figure~\ref{phase_transit}(a) shows a detailed view of the $(\eta,\mu)$ phase diagram of Fig.~\ref{phase_diagram_nupos}, showing the necking bifurcations N$_i^c$ and N$_i^s$ (with $i=1,3,5$). Let us first tackle the transitions mediated by N$_1^s$. An enlargement of Fig.~\ref{phase_transit}(a) around N$_1^s$ is shown in Fig.~\ref{phase_transit}(b), together with the saddle-node bifurcations SN$_{s,1}^r$, sn$_{s,1}^r$ and SN$_{c,1}^r$.
To understand this transition we show in Fig.~\ref{necking12} two slices of Fig.~\ref{phase_transit}(b) just before [see Fig.~\ref{necking12}(i) for $\mu=0.98084$] and after [see Fig.~\ref{necking12}(ii) for $\mu=0.98086$] N$_1^s$.

In Fig.~\ref{necking12}(i) we show the standard snaking related to $\Gamma_0$ and the isola $\Upsilon_e$ previously depicted in  Fig.~\ref{isolas_collapsed}(iv). At this location, SN$_{s,1}^r$ and sn$_{c,1}^r$ are very close to one another (see inset), as a consequence of the proximity of the cusp $C_{1}^r$ creating them. At N$_1^s$ [$(\eta,\mu)\approx (0.1329,0.98085)$], SN$_{s,1}^r$ and sn$_{c,1}^r$ merge, leading to the reconnection of $\Gamma_0$ and $\Upsilon_e$. In this fusion, the first two branches of $\Gamma_0$ reconnect with a part of $\Upsilon_e$ leading to the destruction of standard snaking and consequent creation of the collapsed snaking structure shown in Fig.~\ref{necking12}(ii).1. The remaining parts of $\Upsilon_e$ and the remnants of the $\Gamma_0$ tangle combine to yield the LP isola depicted in Fig.~\ref{necking12}(ii).2. In what follows, we refer to this isola as $\Theta_0^{-1p}$. As a result of this reconnection, SN$_{c,1}^r$ becomes the first saddle-node on the right of $\Gamma'_0$, while SN$_{s,1}^l$ is the first on the left. Hereafter, we rename SN$_{s,1}^l$ as SN$_{c,1}^l$.
Conversely, in $\Theta_0^{-1p}$, the solution branches corresponding to the one-peak LP are absent, a fact indicated by the superscript $-1p$. Therefore, above N$_1^s$, collapsed snaking coexists with different types of isola remnants from $\Gamma_0$.
%, where $2n+1$ in the exponent refers to the LP solution branches erased.

With increasing $\mu$, we approach N$_3^s$ and N$_5^s$ [see Fig.~\ref{phase_transit}(c)]. Figure~\ref{necking34a}(i) shows a slice of Fig.~\ref{phase_transit}(c) just below N$_3^s$, for $\mu=0.98084$, where three isolas are depicted, namely $\Theta_0^{-1p}$, $\Upsilon_f$ and $\Upsilon_g$. The panels on the right show close-up views of Fig.~\ref{necking34a}(i) around SN$_{s,3}^r$ and SN$_{s,5}^r$.
At N$^s_{3}$, SN$_{s,3}^r$ collides with sn$_{I,3}^r$, and SN$_{s,5}^r$ meets  sn$_{I,3}^r$ at N$^s_{5}$. Soon after passing both necking bifurcations [see horizontal line (iv) in Fig.~\ref{phase_transit}(c), $\mu=0.997809$], the isolas recombine forming $\Theta_0^{-7p}$ [see Fig.~\ref{necking34a}(ii).1] and the three hybrid isolas $\Xi_0^{3}$, $\Xi_0^{5}$ and $\Xi_0^{7}$ shown in Fig.~\ref{necking34a}(ii).2. Increasing $\mu$ further, this process repeats until the $\Theta_0$-isolas eventually disappear in the last necking bifurcation N$_i^s$. Once this point is passed, only the collapsed snaking branch $\Gamma'_0$ and the hybrid $\Xi_0$-isolas coexist. 

Hybrid states belonging to $\Xi_0^5$ are depicted in Fig.~\ref{hybrid}, together with the isola $\Xi_0^5$ for comparison. Profile (i) corresponds to a localized pattern of 5 peaks, previously related to standard homoclinic snaking. Following the diagram rightwards, this state encounters a first fold where it becomes unstable, while the two outermost peaks develop small dips. After stabilizing in a second fold, this state resembles that shown in Fig.~\ref{hybrid}(ii). Proceeding further in the diagram, the dip broadens via the nucleation of spatial oscillations (SOs). At this stage, the structure looks like the profile plotted in Fig.~\ref{hybrid}(iii). This state can be seen as a combination of a localized pattern of 3 peaks and a state formed through the locking to fronts connecting $u_h^b$ and $u_h^t$. 

If we now follow the diagram from the profile (i) leftwards, this state changes as illustrated in Figs.~\ref{hybrid}(iv)-(vi). In the first fold on the left, the five-peak LS becomes unstable, and the amplitude of the two outermost peaks decreases, forming the state shown in Fig.~\ref{hybrid}(iv), consisting of just 3 peaks. Proceeding up the diagram this state undergoes a morphological modification similar to that already described: the outermost peaks develop SOs leading to the formation of the states shown in Figs.~\ref{hybrid}(v) and \ref{hybrid}(vi).

\section{Two-parameter bifurcation diagrams for $\nu<0$}\label{sec:9}
In this section, we follow the procedure used in the previous section to study now the bifurcation structure of the LSs emerging from SN$_h^{l,r}$ changes when $\nu<0$. In this regime the SN$_h^{l,r}$ correspond to the spatial bifurcations RTB. Near these points small amplitude localized solutions of the form (\ref{LSs_fold}) exist but asymptote to an even smaller amplitude periodic pattern in the far field.
\begin{figure*}[!t]
	\centering
	\includegraphics[scale=1]{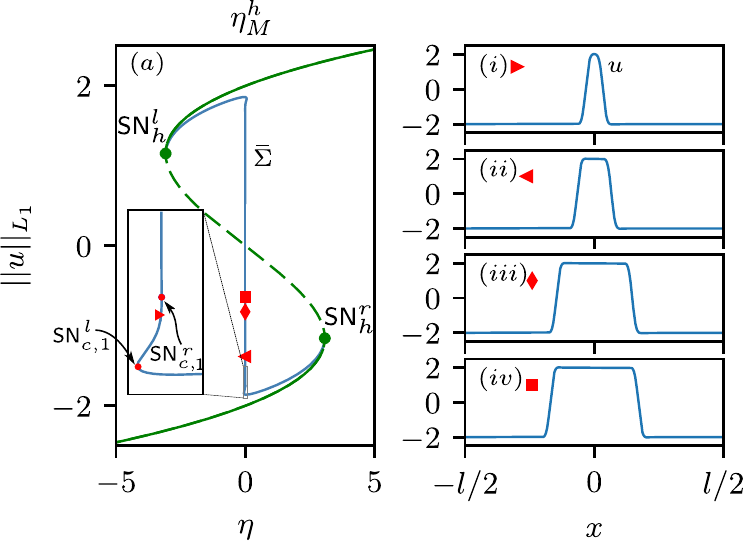}
	\includegraphics[scale=1]{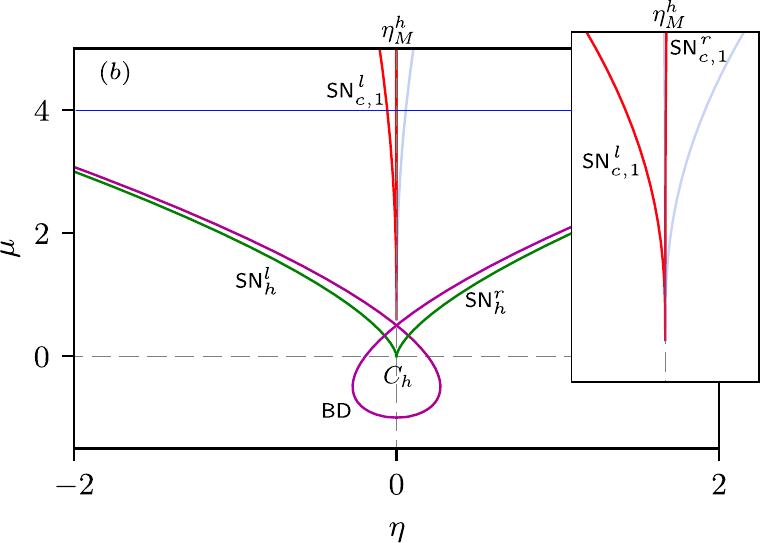}
	\caption{Panel (a) shows the collapsed snaking diagram $\bar{\Sigma}$ for $(\alpha,\beta,\nu,\mu)=(0,1,-2,4)$ for bright and dark {\rm LSs}. Labels (i)-(vi) correspond to the {\rm LSs} shown on the right. The Maxwell point of the system is indicated by the dashed vertical line at $\eta=0$. The inset represents a close-up view of the bottom part of the diagram. Panel (b) depicts the $(\eta,\mu)$ phase diagram with the main bifurcation lines of the system drawn for $\nu=-2$. The $\bar{\Sigma}$ diagram in (a) corresponds to the slice of (b) at constant $\mu=4$ (horizontal blue line).}
	\label{diagram_nu-2}
\end{figure*}
The results of numerical continuation for $\nu=-2$ and $\mu=4$ lead to the bifurcation diagram shown in Fig.~\ref{diagram_nu-2}(a). Here $u_h^b$ and $u_h^t$ are both temporally stable, and therefore HH bifurcations are absent. Thus only collapsed snaking takes place. The bottom right part of the resulting branch $\bar\Sigma$ corresponds to bright LSs while the top left part corresponds to dark states. Owing to invariance under the transformation $(\eta,u)\mapsto (-\eta,-u)$, the diagram in Fig.~\ref{diagram_nu-2}(a) is symmetric with respect to the point $(\eta,u_h)=(0,0)$. For this reason we focus in the following on the bottom right part of the collapsed snaking structure. 

The state arising from SN$_h^r$ does not exhibit oscillatory tails around $u_h^b$ and so corresponds to a tame homoclinic orbit. Decreasing $\eta$, the system undergoes a BD transition generating oscillatory tails on $u_h^{b\rightarrow t}$ and $u_h^{t\rightarrow b}$, and resulting, via front-locking, in LSs resembling those in Figs.~\ref{diagram_nu-2}(i)-(iv). As a result the bifurcation diagram shown in Fig.~\ref{diagram_nu-2}(a) corresponds to collapsed snaking of the family of solutions $\bar{\Sigma}$. Following $\bar{\Sigma}$ to the left, the amplitude of LS increases, and at SN$_{c,1}^{l}$ the state stabilizes. An example of the LS at this stage is shown in Fig.~\ref{diagram_nu-2}(i).  This state remains stable until SN$_{c,1}^{r}$. After this bifurcation, nucleation of SOs takes place as explained in Sec.~\ref{sec:6.2}, and as a result the LSs widen as one proceeds up along $\bar{\Sigma}$ [see profiles in Figs.~\ref{diagram_nu-2}(ii)-(viii)], asymptotically approaching $\eta_M^h$. The oscillatory tails in this case are highly damped, resulting in strongly confined collapsed snaking.

The bifurcation diagram in Fig.~\ref{diagram_nu-2}(a) corresponds to a slice at $\mu=4$ of the $(\eta,\mu)$ phase diagram shown in Fig.~\ref{diagram_nu-2}(b). This phase diagram shows in blue some of the main bifurcation lines of the collapsed snaking branch $\bar\Sigma$. Increasing $\mu$, the uniform-bistability region widens, and so does the region of existence of these LSs. This region is bounded by SN$_{c,1}^{l}$ and SN$_{c,1}^{r}$. Decreasing $\mu$, however, both regions shrink and the LSs disappear through the collision of the different saddle-nodes SN$_{c,i}^{l,r}$ in a sequence of cusp bifurcations (not shown).

\begin{figure}[!t]
	\centering
	\includegraphics[scale=1]{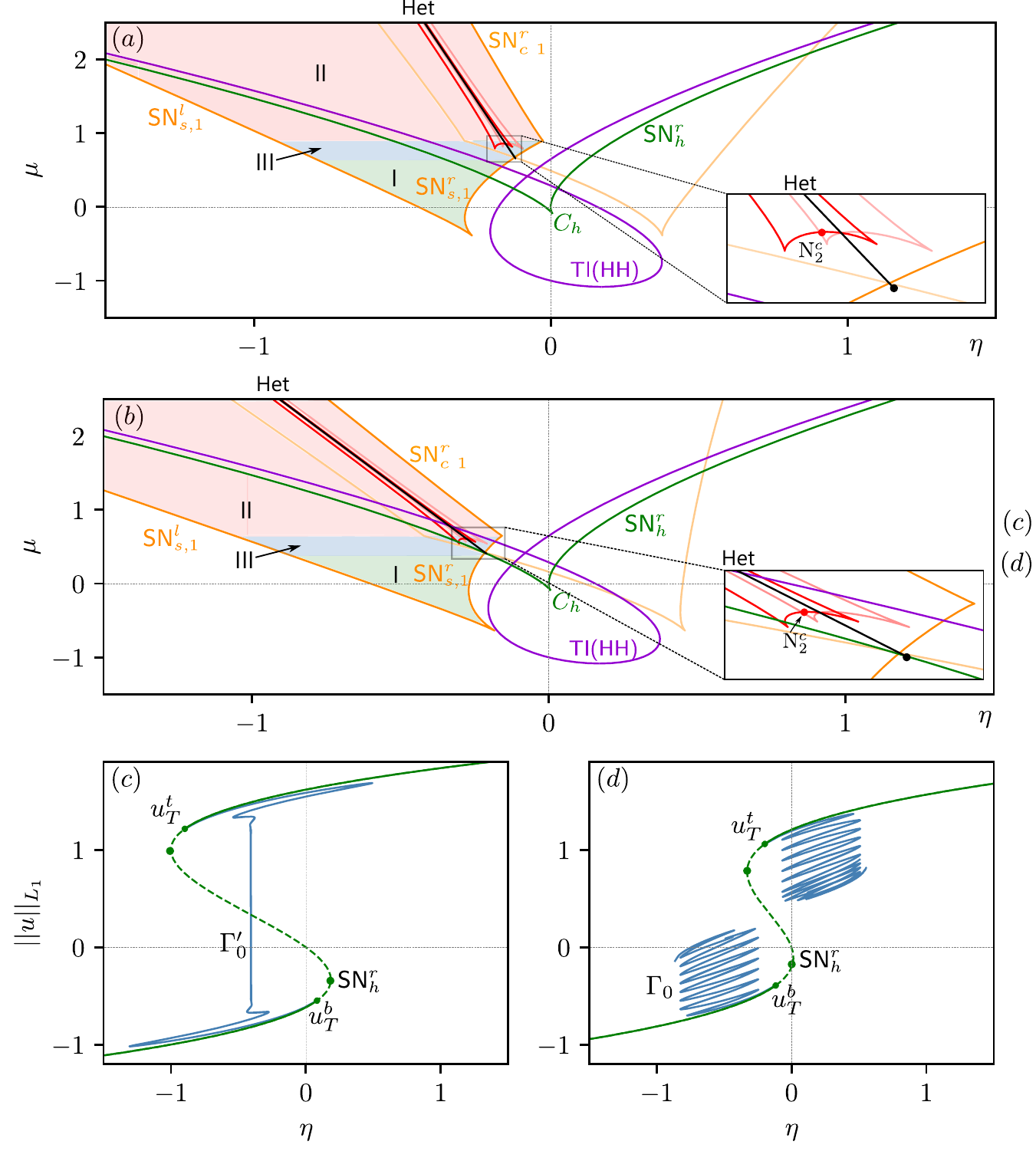}
	\caption{Asymmetric bifurcation diagrams for $\alpha\neq0$. Panel (a) shows the $(\eta,\mu)$ plane for $\alpha=0.5$, while (b) shows the same diagram for $\alpha=1$. Panels (c) and (d) show the bifurcation diagrams corresponding to two slices of (b), at $\mu=1$ and $\mu=0.25$, respectively.  }
	\label{diagrams_nu2_alphas}
\end{figure}

\section{Implication of asymmetry on the bifurcation scenario}\label{sec:11}

So far, we have focused on the simplest version of the model, where the different solutions are symmetric with respect to the transformation $(\eta,u)\rightarrow(-\eta,-u)$. However, generic systems will be asymmetric. In this section we analyze how the quadratic nonlinearity $u^2$ controlled through the parameter $\alpha$ may modify the bifurcation scenarios described previously. 

In Figs.~\ref{diagrams_nu2_alphas}(a) and  \ref{diagrams_nu2_alphas}(b) we show the $(\eta,\mu)$ plane for two values of $\alpha$. In both diagrams, we plot the main bifurcation lines: SN$_{s,1}^{r,l}$, SN$_{c,2}^{l,r}$, HH and the homogeneous folds SN$_h^{l,r}$.
For $\alpha=0.5$ [Fig.~\ref{diagrams_nu2_alphas}(a)], all the bifurcation lines tilt towards negative values of $\eta$, and the regions bounded by such lines are asymmetric. This can be easily appreciated by looking at the regions between SN$_{s,1}^{r,l}$ and SN$_{c,2}^{l,r}$. Furthermore, the lines corresponding to the Maxwell points and the heteroclinic bifurcation 'Het' are tilted with respect to the $\alpha=0$ scenario. Increasing $\alpha$ further tilts these curves further, yielding the situation depicted in Fig.~\ref{diagrams_nu2_alphas}(b) for $\alpha=1$.  

To clarify the resulting asymmetric scenario, let us consider transverse cuts of Fig.~\ref{diagrams_nu2_alphas}(b) at two constant values of $\mu$, one in the collapsed snaking regime and the other in the standard snaking regime. These one-parameter bifurcation diagrams are plotted in Figs.~\ref{diagrams_nu2_alphas}(c) and \ref{diagrams_nu2_alphas}(d). Figure~\ref{diagrams_nu2_alphas}(c) shows the collapsed snaking bifurcation diagram corresponding to $(\mu,\alpha)=(1,1)$. We see that the Maxwell point is shifted towards negative values of $\eta$, with a similar shift in the standard homoclinic snaking scenario shown in Figure~\ref{diagrams_nu2_alphas}(d) for $(\mu,\alpha)=(1.25,1)$. 

A similar configuration is obtained when $\alpha$ takes negative values, albeit with a positive tilt. Despite these changes, the bifurcation structure is completely analogous to that reported for the fully symmetric case. Examples of asymmetric scenarios where the bifurcation structure of LSs has been studied include nonlinear cavity optics \cite{parra-rivas_dark_2016,parra-rivas_coexistence_2017,arabi_localized_2020,parra-rivas_dark_2021} and ecological models \cite{al_saadi_unified_2021}.

%------------------------------------------------

\section{Conclusion}\label{sec:12}

In this work we have presented a detailed study of the bifurcation structure and stability of spatially localized states arising near a Lifshitz critical point. Around this point, the essential behavior of any system can be captured by the nonvariational normal form (\ref{general_model}) as shown in
\cite{kozyreff_nonvariational_2007}. However, owing to the complexity of this model, we have focused in this paper on the 'simplest' variational version of Eq.~(\ref{general_model}), namely Eq.~(\ref{general_model1}). We have referred to this situation as Case I. This scenario was introduced in Sec.~\ref{sec:2}, and its variational structure described. 

One of the first things to determine when dealing with nonlinear equations describing pattern-forming systems is to understand its homogeneous, or uniform, states and their temporal stability properties as a function of the parameters of the system. This study was carried out in Sec.~\ref{sec:3}, where we have determined the location of the Turing instability (TI) for two cases: the fully symmetric case $\alpha=0$, and asymmetric case $\alpha\neq0$. The results of this analysis define the parameter region with coexisting stable uniform states (i.e., the uniform-bistable regime)
as well as the region of bistability between a uniform state and a pattern state. In each case we have focused on the accompanying stationary localized steady states, and thus, on solutions of the time-independent Eq.~(\ref{general_model_sta}). This equation is equivalent to the 4D spatial dynamical system (\ref{SD}) on the real line, a reformulation that permits the use of well-known results from dynamical systems theory in the study of the spatial bifurcations generating localized coherent structures. We computed the nontrivial weakly nonlinear states in the vicinity of these bifurcation points using multiple scale perturbation methods (Sec.~\ref{sec:5}). In the case of the TI the calculation determines the parameter regime within which spatially periodic states bifurcate subcriticality, and thus the Turing bistability region.

While LSs emerging from fold bifurcations always display collapsed homoclinic snaking (Sec.~\ref{sec:9}), LSs emerging from the TI can display two distinct bifurcation structures, namely standard homoclinic snaking (Sec.~\ref{sec:6.1}) and collapsed homoclinic snaking (Sec. \ref{sec:6.2}), depending on the parameters. The former is observed in the region of coexistence between a stable periodic solution and a HSS;  the latter arises when two HSSs coexist. In applications these two LS scenarios are typically observed in different regimes. However, near the Lifshitz point these two scenarios collide in the codimension-three point studied in this work, allowing us to investigate the transition between these two scenarios within a unified framework (Sec.~\ref{sec:6.3}).

There are many aspects of this problem that have not been fully explored here. One omission is the study of the temporal stability of the LSs we have computed. Also, as stated in the Introduction, we have only considered one particular case of the normal form \eqref{general_model}, with the full problem likely to exhibit greater complexity, owing to its nonvariational structure. 

\subsection*{Generalized solitary waves}

In the present study we have focused only on LSs that are bi-asymptotic to a uniform or flat state. However, in a system like ours, LSs approaching asymptotically a spatially periodic state of finite, but arbitrary, amplitude are also possible. Such states are commonly known as {\it generalized solitary waves}; see, for example, \cite{kolossovski_multi-pulse_2002,gandhi_spatially_2018,godey_bifurcation_2017}.

These generalized homoclinic orbits emerge from RTBH spatial bifurcations with eigenvalues $\lambda_{1,2,3,4}=(0,0,\pm ia)$. Here, the RTBH point corresponds to a saddle-node bifurcation of $u_f=u_h^{b,t}$ when $\nu>0$. Close to such points, weakly nonlinear states are well approximated by the expression
\begin{equation}\label{RTBH}
	u(x)= u_f+ \sqrt{\frac{\eta-\eta_f}{3u_f}}\left[1-3{\rm sech}^2\left(\frac{1}{2}\sqrt{\frac{2}{|\nu|}\sqrt{3 u_f(\eta-\eta_f)}}x\right)\right]
\end{equation}	
even though this expression does not capture the precise nature of the oscillatory background.

\begin{figure}[!t]
	\centering
	\includegraphics[scale=1]{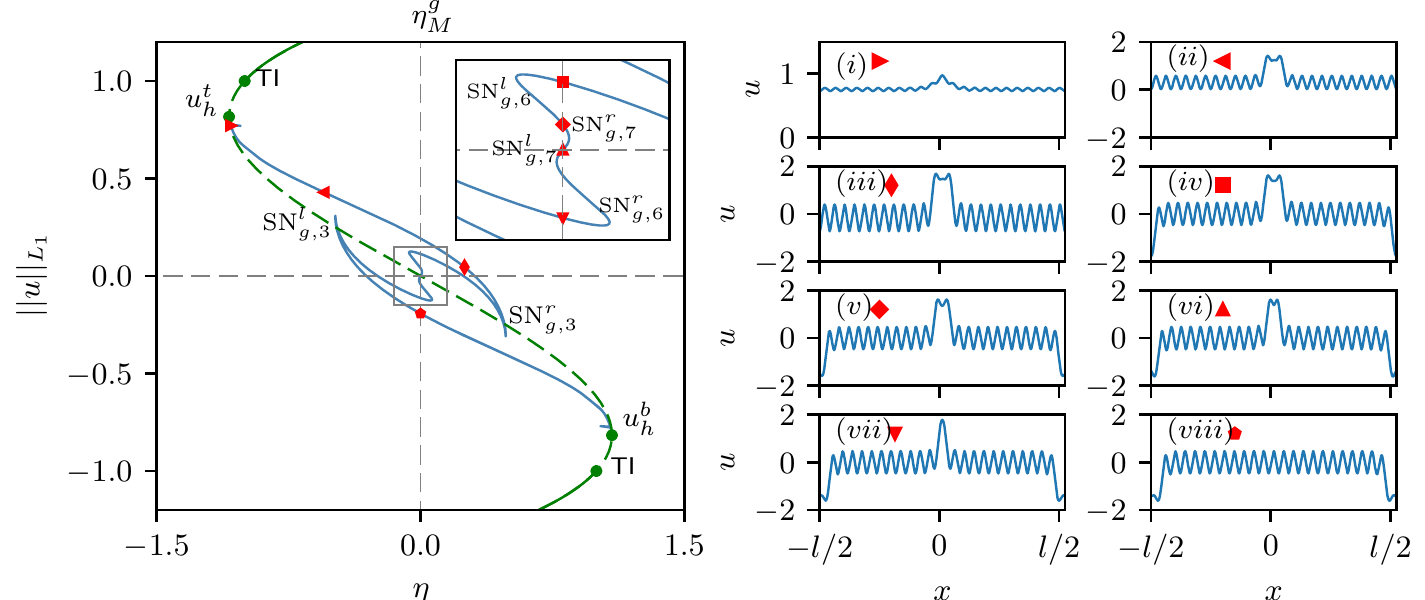}
	\caption{Bifurcation diagram (in solid blue) associated with generalized LSs at $\mu=1$ when $\nu=2$. The profiles on the right show the resulting LSs at the labeled points. This diagram corresponds to a slice of the $(\eta,\mu)$ parameter plane shown in Fig.~\ref{phase_generalized}.}
	\label{diageneral3}
\end{figure}

It should be stressed that the codimension of generalized solitary waves in reversible Hamiltonian system is different to that of the other LSs we have studied.
That is, for each value of the parameters, there will in general be a one-parameter family of generalized LSs, parameterized either by the amplitude of the periodic state in the far-field, or by the phase shift between the far-field oscillations on either side. See \cite{kolossovski_multi-pulse_2002} and references therein. Here we report some preliminary results by performing a path-continuation computation on a fixed $L$ domain with the Neumann boundary conditions \eqref{eq:BCs}, starting from the initial guess (\ref{RTBH}). Thus, we only present a single member of the expected one-parameter family at each parameter value. The result leads to the bifurcation structure shown in Fig.~\ref{diageneral3}(a) when $\mu=1$. Note that this bifurcation structure overlaps with the standard snaking studied in Sec.~\ref{sec:6.1}.

\begin{figure}[!t]
	\centering
	\includegraphics[scale=1]{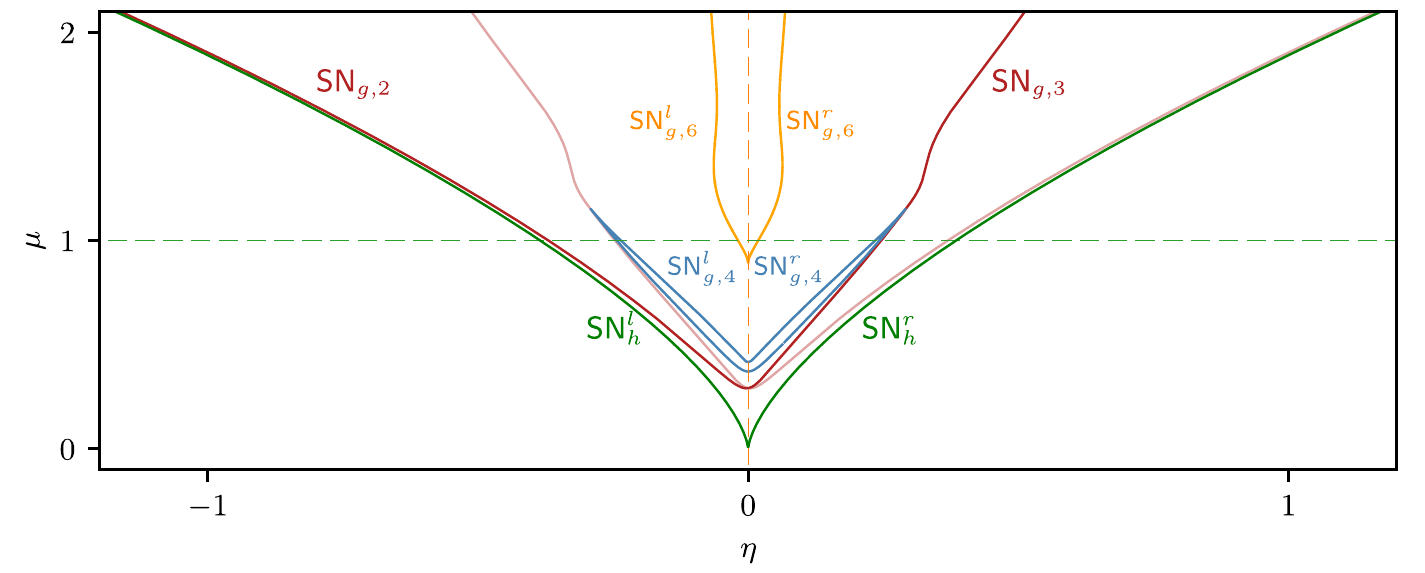}
	\caption{Phase diagram in the $(\eta,\mu)$ parameter plane for $\nu=2$ showing the main bifurcation curves associated with generalized LSs. These bifurcations are depicted in the bifurcation diagram shown in Fig.~\ref{diageneral3}, which is a slice of this diagram at $\mu=1$ (dashed line). The TI line is not plotted.}
	\label{phase_generalized}
\end{figure}

%------------------------------------------

Very close to SN$_h^l$, the LS resembles that depicted in  Fig.~\ref{diageneral3}(i), i.e. a localized pulse embedded in a spatially periodic state of small and constant amplitude. Increasing $\eta$ further the amplitude of the spatially periodic background increases, and the preceding state changes into that shown in Fig.~\ref{diageneral3}(ii). Note the emergence of several spatial oscillations at the center of the structure. This process continues until SN$^r_{g,3}$ [see Fig.~\ref{diageneral3}(iii)]. Passing SN$_{g,6}^l$, a negative peak starts to form at the boundary of the domain, i.e., at $x=\pm l/2$ [see Fig.~\ref{diageneral3}(iv)]. Proceeding down in the diagram we see that width of the central peak decreases, while that of the boundary peak increases. This occurs through the creation (destruction) of spatial oscillations around the peak at the boundary (center), while crossing the bifurcations SN$_{g,6}^{l,r}$ and SN$_{g,7}^{l,r}$. This process is depicted in Figs.~\ref{diageneral3}(v)-(vii).  Once SN$_{g,3}^l$ is crossed, the peak at the boundary undergoes the same process as that in the center, eventually disappearing at SN$_h^l$ in another RTBH bifurcation.
Our linear stability analysis reveals that all these states are unstable. This is likely related to the instability of the spatially periodic background.% In what follows we refer to this bifurcation structure as $\Pi$.

%and once SN$_{g,2}^l$ is passed, the it becomes the high amplitude single-peak LS shown in Fig.~\ref{diagenera2}(iv).

%---------------------------------------------------------------------

%Increasing $\eta$, the amplitude of the periodic background increases [see Fig.~\ref{diageneral3}(v)], and after passing SN$_{g,3}^r$, SN$_{g,4}^r$ and SN$_{g,5}^r$, a new peak with a negative amplitude starts to grow at $x=L/2$. This situation is depicted in Fig.~\ref{diagenera2}(vi). The bright and dark states reach the same amplitude at $\eta=0$. With $\eta$ taking negative values, the dark peak increases its amplitude to the detriment of the bright one, which disappears at SN$_{g,1}^l$. Once SN$_{g,3}^r$, SN$_{g,4}^r$ and SN$_{g,5}^r$ are crossed, the dark LS undergoes the same process as that for the bright LS, and eventually disappears at SN$_h^l$ in another RTBH bifurcation. 

The bifurcation diagram shown in Fig.~\ref{diageneral3} corresponds to a slice for $\mu=1$ of the $(\eta,\mu)$-phase diagram shown in Fig.~\ref{phase_generalized}. In this diagram we plot the bifurcation curves associated with those depicted in Fig.~\ref{diageneral3}, and some other bifurcations emerging for other values of $\mu$. %SN$_{g,4}^l$ and SN$_{g,5}^l$ meet in a cusp bifurcation C$_{g,45}^l$ where they disappear. Symmetrically, the same thing occurs at C$_{g,45}^r$.

\section*{Acknowledgments}
P.P.R. acknowledges support from the European Union’s Horizon 2020 Research and Innovation Programme under Marie Sklodowska-Curie grant agreement no. 101023717. The work of F. AlS. was supported by the Ministry of Higher Education, Research and Innovation and the Military Technological College, Oman. The work of E.K. was supported in part by the National Science Foundation under grant under Grant DMS-1908891.

\newpage
\appendix
\section{Weakly nonlinear solutions near HSS folds}\label{Appendix:1}
In this section we compute weakly nonlinear solutions around the HSS folds occurring at $(\eta_f,u_f)$. We proceed as explained in Sec.~\ref{sec:5} by first splitting the problem into homogeneous and space-dependent Eqs.~(\ref{gen_hom_eq}) and (\ref{gen_nhom_eq}), and employing the perturbation expansion in  Eq.~(\ref{gen_hom_eq}). In what follows we summarize the calculations for both the homogeneous and the space-dependent problem. 

\subsection{Homogeneous problem around the folds}
From Eq.~(\ref{gen_hom_eq}) and Eq.~(\ref{expansions_fold}) we obtain, order by order,
\begin{subequations}
	\begin{equation}
		\begin{array}{lcc}
			\mathcal{O}(\epsilon^0): & & \eta_r+\mu u_f+\alpha u_f^2-\beta u_f^3=0
		\end{array}
	\end{equation}
	\begin{equation}
		\begin{array}{lcc}
			\mathcal{O}(\epsilon^1): & & (\mu+2\alpha u_f-3\beta u_f^2)u_1=0
		\end{array}
	\end{equation}	
	\begin{equation}
		\begin{array}{lcc}
			\mathcal{O}(\epsilon^2): & & (\mu+2\alpha u_f-3\beta u_f^2)u_2+(\alpha-3\beta u_f)u_1^2+\delta=0.
		\end{array}
	\end{equation}
\end{subequations}
We use the subscript $f$ to denote $t$ (top fold) or $b$ (bottom fold).

The equation at $\mathcal{O}(\epsilon^1)$ has the nontrivial solution $u_1=C_f$ provided the solvability condition 
\begin{align}
	\mu+2\alpha u_f-3\beta u_f^2=0
\end{align}
is satisfied. This leads to the two folds at
\begin{equation}
	u_f=\frac{\alpha}{3\beta}\pm\frac{\sqrt{\alpha^2+3\mu\beta}}{3\beta},\qquad\beta\neq0. 
\end{equation}
At $\mathcal{O}(\epsilon^2)$, the first term of the equation vanishes and one concludes that
\begin{equation}
	C_f=\pm\sqrt{\frac{\delta}{3\beta u_f-\alpha}}.
\end{equation}
\subsection{Space-dependent problem around the folds}
Equation~(\ref{gen_nhom_eq}) can be written as
\begin{equation*}\label{compact}
	(\mathcal{L}+\mathcal{N})\phi=0,
\end{equation*}
where the linear and nonlinear operators can be expanded as
\begin{align*}
	\mathcal{L}=\mathcal{L}_0+\epsilon\mathcal{L}_1+\epsilon^2 \mathcal{L}_2+\cdots, 
	&&
	\mathcal{N}=\epsilon\mathcal{N}_1+\epsilon^2\mathcal{N}_2+\cdots, 
\end{align*}
with the order by order contributions
\begin{align*}
	\mathcal{L}_0=\mu+2\alpha u_f-3\beta u_f^2,
\end{align*}
\begin{align*}
	\mathcal{L}_1=-\nu\partial^2_X+2(\alpha-3\beta u_f)u_1, &&
	\mathcal{L}_2=-\partial^4_X+2(\alpha-3\beta u_f)u_2-3\beta u_1^2,
\end{align*}
and 
\begin{align*}
	\mathcal{N}_1=(\alpha-3\beta u_f)\phi_1,&&
	\mathcal{N}_2=2(\alpha-3\beta u_f)\phi_2-3\beta u_1\phi_1.
\end{align*}
Collecting the terms at the same order in $\epsilon$ we obtain the equations
\begin{subequations}
	\begin{equation}
		\begin{array}{lcc}
			\mathcal{O}(\epsilon^1): && \mathcal{L}_0\phi_1=0,
		\end{array}
	\end{equation}	
	\begin{equation}
		\begin{array}{lcc}
			\mathcal{O}(\epsilon^2): && \mathcal{L}_0\phi_2+(\mathcal{L}_1+\mathcal{N}_1)\phi_1=0.
		\end{array}
	\end{equation}
\end{subequations}
The equation at $\mathcal{O}(\epsilon^1)$ can be solved on assuming that $\phi_1(X)=u_1A(X)$. The amplitude $A(X)$ is determined from the solvability condition at $\mathcal{O}(\epsilon^2)$:
\begin{align}
	c_0A_{XX}+2A+A^2=0, && 	c_0=\frac{\nu}{\sqrt{\delta (3\beta u_f-\alpha)}}.
\end{align}
%which has as solutions the homoclinic orbit
%\begin{equation}
%A(X)=-\frac{3}{2}\frac{c_1}{c_2}{\rm %sech}^2\left(\frac{1}{2}\sqrt{-\frac{c_1}{c_0}}X\right).
%\end{equation} 
This amplitude equation supports small amplitude pulse solutions of the form 
\begin{equation}
	A(X)=-3{\rm sech}^2\left(\frac{1}{2}\sqrt{-\frac{2}{\nu}\sqrt{\delta(3\beta u_f-\alpha)}}X\right),
\end{equation}
provided $\nu<0$. With this solution, the weakly nonlinear homoclinic orbit in the neighborhood of the HSS fold is given by
\begin{equation}
	u(x)= u_f+ \sqrt{\frac{\eta-\eta_f}{3\beta u_f-\alpha}}\left[1-3{\rm sech}^2\left(\frac{1}{2}\sqrt{-\frac{2}{\nu}\sqrt{(3\beta u_f-\alpha)(\eta-\eta_f)}}x\right)\right]+\mathcal{O}(\eta-\eta_f).
\end{equation}

\section{Weakly nonlinear analysis near the Turing bifurcation}\label{Appendix:2}
Here we present the details of the multiscale perturbation analysis carried out around the Turing bifurcation point $(\eta_T,u_T)$. Inserting the expansions in terms of the small parameter $\epsilon$ defined in Eq.~(\ref{expansion_Turing}) into  Eqs.~(\ref{gen_hom_eq}) and (\ref{gen_nhom_eq}), we obtain the following problems. 

\subsection{Homogeneous problem around the Turing bifurcation}
Around the Turing bifurcation, the homogeneous problem decomposes into the two equations
\begin{align}
	\mathcal{O}(\epsilon^0): && \beta u_T^3-\alpha u_T^2-\mu u_T-\eta=0,&& &&
\end{align}
\begin{align}
	\mathcal{O}(\epsilon^2): && u_2=-\delta/\mathcal{M}_T,&&\mathcal{M}_T\equiv\mu+2\alpha u_T-3\beta u_T^2.
\end{align}
At $\mathcal{O}(\epsilon^0)$ we obtain the HSS solution evaluated at the Turing point.

\subsection{Space-dependent problem around the Turing bifurcation}
The space-dependent equation can be written in the form
\begin{equation*}\label{compact2}
	(\mathcal{L}+\mathcal{N})\phi=0,
\end{equation*}
where the linear and nonlinear operator expansions in $\epsilon$ now read 
\begin{align*}
	\mathcal{L}=\mathcal{L}_0+\epsilon\mathcal{L}_1+\epsilon^2 \mathcal{L}_2+\cdots,&& \mathcal{N}=\epsilon\mathcal{N}_1+\epsilon^2\mathcal{N}_2+\cdots.
\end{align*}
Here, the linear operators are given by
\begin{align*}
	\mathcal{L}_0=\mathcal{M}_T-\partial_x^4-\nu\partial_x^2, &&  \mathcal{L}_1=-2(\nu\partial_x\partial_X+2\partial_X\partial_x^3),&&	\mathcal{L}_2=-(\nu\partial^2_X+6\partial^2_x\partial_X^2),
\end{align*}
while the nonlinear ones are 
\begin{align*}
	\mathcal{N}_1=(\alpha-3\beta u_T)\phi_1,&&\mathcal{N}_2=2(\alpha-3\beta u_T)u_2+(\alpha-3\beta u_T)\phi_2-\beta\phi_1^2.
\end{align*}

Inserting these expansions in Eq.~(\ref{compact2}) we have 
\begin{equation*}
	\left[\mathcal{L}_0+\epsilon\left(\mathcal{L}_1+\mathcal{N}_1\right)+\epsilon^2\left( \mathcal{L}_2+\mathcal{N}_2\right)+\cdots  \right](\epsilon \phi_1+\epsilon^2 \phi_2+\epsilon^3 \phi_3+\cdots)=0,
\end{equation*}
leading to the hierarchy of equations
\begin{subequations}
	\begin{equation}\label{order1}
		\mathcal{O}(\epsilon): \qquad
		\mathcal{L}_0\phi_1=0,
	\end{equation}
	\begin{equation}
		\begin{array}{lcc}
			\mathcal{O}(\epsilon^2): & & \mathcal{L}_0\phi_2+(\mathcal{L}_1+\mathcal{N}_1)\phi_1=0,
		\end{array}
	\end{equation}
	\begin{equation}
		\begin{array}{lcc}
			\mathcal{O}(\epsilon^3): & &  \mathcal{L}_0\phi_3+(\mathcal{L}_1+\mathcal{N}_1)\phi_2+(\mathcal{L}_2+\mathcal{N}_2)\phi_1=0.
		\end{array}
	\end{equation}
\end{subequations}
\subsubsection{Solution at first-order in $\epsilon$}
At $\mathcal{O}(\epsilon)$ an adequate solution ansatz is
\begin{equation}
	\phi_1(x,X)=A(X)e^{ik_Tx}+c.c.
\end{equation}
Inserting this ansatz in Eq.~(\ref{order1}) we obtain 
$$\mathcal{L}_0\phi_1=\mathcal{L}_0\left(A(X)e^{ik_Tx}+c.c.\right)=\left(\mathcal{M}_T-k_T^4+\nu k_T^2\right)\left(A(X)e^{ik_Tx}+c.c.\right)=0.$$
This equation has a nontrivial solution provided the solvability condition  
\begin{equation}\label{wavenumber}
	\mathcal{M}_T-k_T^4+\nu k_T^2=0
\end{equation}
holds.
\subsubsection{Solution at second order in $\epsilon$}
At $\mathcal{O}(\epsilon^2)$, the equation reads
\begin{equation}\label{order2}
	\mathcal{L}_0\phi_2=-(\mathcal{L}_1+\mathcal{N}_1)\phi_1.  
\end{equation}
To solve this equation we first evaluate the forcing term, i.e., the right-hand side. This leads to
\begin{multline*}
	(\mathcal{L}_1+\mathcal{N}_1)\phi_1=(\mathcal{L}_1+\mathcal{N}_1)\left[A(X)e^{ik_Tx}+{\rm c.c.}\right]=2k_T(2k_T^2-\nu) i\partial_X A e^{ik_T x}+{\rm c.c.}+\\\\(\alpha-3\beta u_T)\left[A(X)e^{ik_Tx}+{\rm c.c.}\right]^2=f_0|A|^2+f_1 i\partial_X A e^{i k_T x}+f_2A^2e^{2i k_T x}+{\rm c.c.},  
\end{multline*}
with 
\begin{align*}
	f_0\equiv 2(\alpha-3\beta u_T), && f_1\equiv 2k_T(2k_T^2-\nu), && f_2\equiv f_0/2.
\end{align*}
The solvability condition at this order is obtained by projecting on the subspace defined by the null eigenvector of the self-adjoint operator. To obtain this condition we first define the scalar product
\begin{equation}
	\langle f,g\rangle\equiv\int_{-l/2}^{l/2}f^*(x)g(x)dx.
\end{equation}
With this definition $\mathcal{L}_0$ is self-adjoint and the null eigenspace is spanned by the two null eigenvectors $w = (e^{i k_T x}, e^{-i k_T x})$, such that $\mathcal{L}^\dagger_0 w=\mathcal{L}_0 w=0$.

The solvability condition then implies 
$$\langle (\mathcal{L}_1+\mathcal{N}_1)\phi,w\rangle=0,$$
which leads to $f_1$, and therefore to the relation
\begin{equation}\label{wavenumber2}
	k_T^2=\nu/2.
\end{equation}
Together with Eq.~(\ref{wavenumber}) this equation determines the location of TI.

Once this condition holds, Eq.~(\ref{order2}) can be solved adopting the ansatz
\begin{equation}
	\phi_2(x,X)=W_0|A(X)|^2+W_2A^2(X)e^{2ik_Tx}+{\rm c.c.}
\end{equation}
Applying the operator $\mathcal{L}_0$ and collecting terms with the same exponential dependence we obtain the equations 
\begin{align*}
	\mathcal{M}_T W_0=-f_0, && (\mathcal{M}_T-16 k_T^4+4\nu k_T^2)W_2=-f_0/2.
\end{align*}
The solutions of these equations are 
\begin{align}
	W_0=-f_0/\mathcal{M}_T, && W_2=\frac{-f_0/2}{\mathcal{M}_T-16 k_T^4+4\nu k_T^2} .
\end{align}

\subsubsection{Solution at third order in $\epsilon$}

At $\mathcal{O}(\epsilon^3)$ the equation we need to solve is 
\begin{equation}
	\mathcal{L}_0\phi_3=-(\mathcal{L}_1+\mathcal{N}_1)\phi_2-(\mathcal{L}_2+\mathcal{N}_2)\phi_1.
\end{equation}
The first term on the right hand side becomes
\begin{equation*}
	(\mathcal{L}_1+\mathcal{N}_1)\phi_2=g_1(X)e^{ik_T x}+g_2(X)e^{2ik_T}+g_3(X)e^{3ik_T}+{\rm c.c.},
\end{equation*}
with 
%\begin{equation}
%    g_1(X)\equiv (\alpha-3\beta u_T)(W_0+W_2)|A|^2A=g_1^c|A|^2A.
%\end{equation}
\begin{equation}
	g_1(X)\equiv \frac{f_0}{2}(W_0+W_2)|A|^2A=g_1^c|A|^2A.
\end{equation}
At $\mathcal{O}(\epsilon^3)$ the solvability condition only involves terms proportional to $e^{ik_T x}$, and therefore we do not need the expressions for $g_2(X)$ and $g_3(X)$.

The second term on the right leads to 
\begin{equation*}
	(\mathcal{L}_2+\mathcal{N}_2)\phi_1=h_1(X)e^{ik_T x}+h_2(X)e^{2ik_T}+h_3(X)e^{3ik_T}+{\rm c.c.},
\end{equation*}
where 
\begin{equation}
	h_1(X)=h_1^aA(X)+h_1^b\partial_X^2A(X)+h_1^c|A|^2A.
\end{equation}
Here
\begin{subequations}
	%  \begin{equation}
	%h_1^a\equiv2(\alpha-3\beta u_T)u_2= -2\delta(\alpha-3\beta u_T)/\mathcal{M}_T=-\delta\tilde{h}_1^b,
	%  \end{equation} 
	\begin{equation}
		h_1^a\equiv f_0u_2= -\delta f_0/\mathcal{M}_T=-\delta\tilde{h}_1^b,
	\end{equation}   
	\begin{equation}
		h_1^b\equiv 2\nu,
	\end{equation}
	% \begin{equation}
	%     h_1^c\equiv (\alpha-3\beta u_T)(W_0+W_2)-3\beta.
	% \end{equation}
	\begin{equation}
		h_1^c\equiv \frac{f_0}{2}(W_0+W_2)-3\beta.
	\end{equation}
\end{subequations}
Expressions for $h_2(X)$ and $h_3(X)$ are not needed.

The solvability condition at this order,
\begin{equation}
	\langle (\mathcal{L}_1+\mathcal{N}_1)\phi_2,w \rangle+\langle (\mathcal{L}_2+\mathcal{N}_2)\phi_1, w \rangle=0,
\end{equation}
\begin{figure}[!t]
	\centering
	\includegraphics[scale=1]{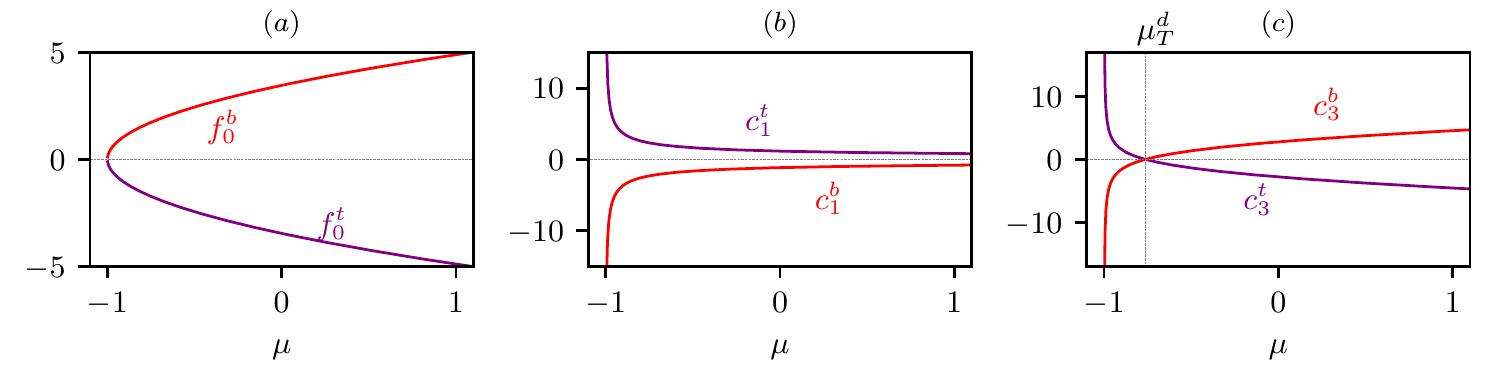}
	\caption{Dependence of (a) $f_0^{b,t}$, (b) $c_1^{b,t}$ and (c) $c_3^{b,t}$ on $\mu$ for $\beta=1$, $\alpha=0$ and $\nu=2$. }
	\label{Coefficietns0}
\end{figure}
\begin{figure}[!t]
	\centering
	\includegraphics[scale=1]{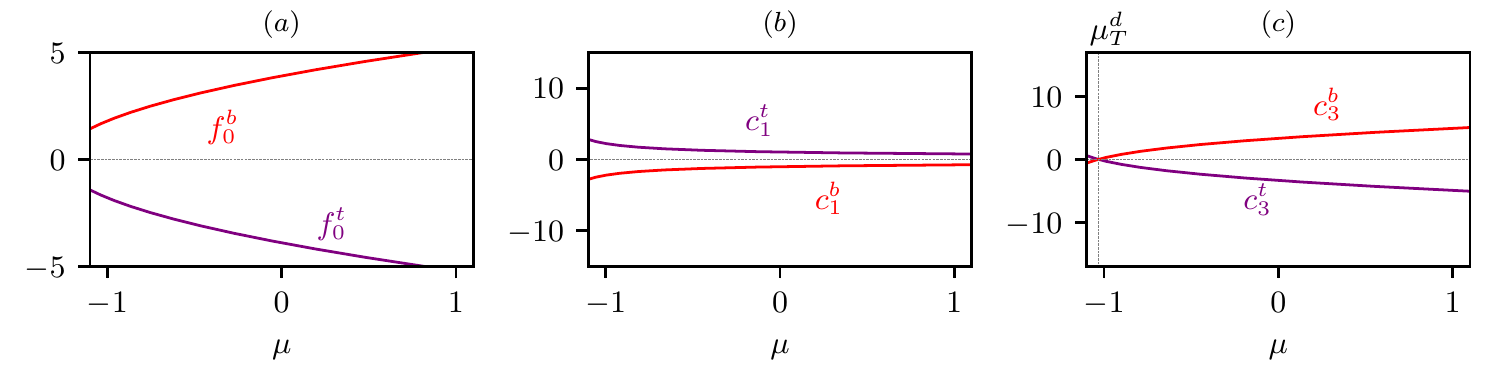}
	\caption{Dependence of (a) $f_0^{b,t}$, (b) $c_1^{b,t}$ and (c) $c_3^{b,t}$ on $\mu$ for $\beta=1$, $\alpha=0.9$ and $\nu=2$. }
	\label{Coefficietns1}
\end{figure}
leads to the amplitude equation 
%\begin{equation}
%    h_1^a A+h_1^b\partial^2_X A+(g_1^c+h_1^c)|A|^2A=0,
%\end{equation}
\begin{equation}
	-\delta\tilde{h}_1^a A+h_1^b\partial^2_X A+(g_1^c+h_1^c)|A|^2A=0,
\end{equation}
which can be further simplified to 
%\begin{equation}\label{amp_eq_HHap}
%{\color{blue}-\delta A+c_1 A_{XX}-c_3|A|^2A=0},
%\end{equation}
\begin{equation}\label{amp_eq_HHap}
	\delta A-c_1 A_{XX}+c_3|A|^2A=0,
\end{equation}
where
\begin{align}
	c_1\equiv \frac{h_1^b}{\tilde{h}_1^a}=\frac{2\nu\mathcal{M}_T}{f_0} && c_3\equiv-\frac{g_1^c+h_1^c}{\tilde{h}_1^a}=-\mathcal{M}_T\left(W_0+W_2-\frac{3\beta}{f_0}\right). 
\end{align}
Using Eqs.~(\ref{wavenumber}) and (\ref{wavenumber2}) the expressions  $\mathcal{M}_T$, $W_0$ and $W_2$ reduce to 
\begin{align}
	\mathcal{M}_T=-\nu^2/4, && W_0=4f_0/\nu^2, && W_2=2f_0/9\nu^2, 
\end{align}
and the coefficients $c_1$ and $c_3$ can be expressed as a function of $f_0$ as
\begin{align}
	c_1=\frac{-\nu^3}{2f_0}, && c_3=\frac{\nu^2}{4}\left(\frac{38f_0}{9\nu^2}-\frac{3\beta}{f_0}\right).
\end{align}
From Eq.~(\ref{uc}) we obtain the dependence of $f_0$ on $\mu$
$$f_0^\pm=\pm2\sqrt{\alpha^2+3\beta\left(\mu+\nu^2/4\right)},$$
which can be used to further simplify the previous coefficients. The $\pm$ signs correspond to the two positions of the Turing instability $u_T^{b,t}$, and hereafter we write $f_0^\pm=f_0^{b,t}$.

Depending on the location of the TI, i.e., on $u_T^{b,t}$, we have two different expressions for the coefficients $c_1$ and $c_3$, hereafter $c_1^{b,t}$ and $c_3^{b,t}$. The dependence of these coefficients on $\mu$ for $(\nu,\beta,\alpha)=(2,1,0)$ and $(\nu,\beta,\alpha)=(2,1,0.9)$ are  shown in Figs.~\ref{Coefficietns0} and \ref{Coefficietns1}, respectively. 

\subsubsection{Solutions of the amplitude equation}
Equation~(\ref{amp_eq_HHap}) admits $X$-independent solutions of the form $A=Be^{i\varphi}$, where $B$ is real and satisfies
\begin{equation}
	B(\delta+c_3B^2)=0,
\end{equation}
with solutions $B=0$ and $B=\sqrt{-\delta/c_3}$. 
According to this equation, a pitchfork bifurcation occurs at $\delta=0$ being supercritical when $c_3>0$ and subcritical otherwise. The transition between these two cases occurs at a degenerate HH bifurcation and corresponds to a codimension-two point at $(\eta,\mu)=(\eta_T^d,\mu_T^d)$, where $\mu_T^d$ satisfies the condition $c_3|_{\mu_T^d}=0$. To determine the location of this bifurcation, we first solve $c_3(f_0)=0$, obtaining
$f^d_0\equiv\pm3\nu\sqrt{3\beta/38}$. Matching $f^d_0$ with $f_0^{t,b}$ we obtain
\begin{equation}
	\mu_T^d\equiv-\frac{1}{3\beta}\left(\frac{87}{152}\beta\nu^2+\alpha^2\right).
\end{equation}
This degenerate value is depicted in Fig.~\ref{Coefficietns0}(c) and Fig.~\ref{Coefficietns1}(c)  using a vertical dashed line. 

The solution $B=\sqrt{-\delta/c_3}$ corresponds to a spatially periodic solution of the form 
\begin{equation}
	u(x)=u_T-\mathcal{M}_T^{-1}(\eta-\eta_T)+2\sqrt{\frac{\eta-\eta_T}{-c_3}}{\rm cos}(k_Tx+\varphi),
\end{equation}
while in the subcritical regime one also finds $X$-dependent solutions $B=B(X)$ of the form 
\begin{equation}
	B(X)=\sqrt{\frac{2\delta}{-c_3}}{\rm sech}\left(\sqrt{\frac{\delta}{c_1}}X\right).
\end{equation}
%\begin{equation}
%B(X)=\sqrt{\frac{2\delta}{c_3}}{\rm %sech}\left(\sqrt{-\frac{\delta}{c_1}}X\right).
%\end{equation}
This solution describes the LS solutions of the form 
\begin{equation}\label{LSpattern}
	u(x)=u_T-\mathcal{M}_T^{-1}(\eta-\eta_T)
	+2\sqrt{\frac{\eta-\eta_T}{-c_3/2}}{\rm sech}\left(\sqrt{\frac{\eta-\eta_T}{c_1}}x\right){\rm cos}(k_Tx+\varphi).
\end{equation}
The coefficients of the previous solution change depending of the position of the TI.  We can write $u_T=u_T^r$, $\eta_T=\eta_T^r$, $c_1=c_1^r$ and $c_3=c_3^r$, where $r=b$ for the TI occurring at the bottom part of the HSS diagram, and $r=t$ for the TI occurring at the top part of the HSS.

\bibliography{codim3Biblio}

%apsrev4-2.bst 2019-01-14 (MD) hand-edited version of apsrev4-1.bst
%Control: key (0)
%Control: author (8) initials jnrlst
%Control: editor formatted (1) identically to author
%Control: production of article title (0) allowed
%Control: page (0) single
%Control: year (1) truncated
%Control: production of eprint (0) enabled
\begin{thebibliography}{64}%
\makeatletter
\providecommand \@ifxundefined [1]{%
 \@ifx{#1\undefined}
}%
\providecommand \@ifnum [1]{%
 \ifnum #1\expandafter \@firstoftwo
 \else \expandafter \@secondoftwo
 \fi
}%
\providecommand \@ifx [1]{%
 \ifx #1\expandafter \@firstoftwo
 \else \expandafter \@secondoftwo
 \fi
}%
\providecommand \natexlab [1]{#1}%
\providecommand \enquote  [1]{``#1''}%
\providecommand \bibnamefont  [1]{#1}%
\providecommand \bibfnamefont [1]{#1}%
\providecommand \citenamefont [1]{#1}%
\providecommand \href@noop [0]{\@secondoftwo}%
\providecommand \href [0]{\begingroup \@sanitize@url \@href}%
\providecommand \@href[1]{\@@startlink{#1}\@@href}%
\providecommand \@@href[1]{\endgroup#1\@@endlink}%
\providecommand \@sanitize@url [0]{\catcode `\\12\catcode `\$12\catcode
  `\&12\catcode `\#12\catcode `\^12\catcode `\_12\catcode `\%12\relax}%
\providecommand \@@startlink[1]{}%
\providecommand \@@endlink[0]{}%
\providecommand \url  [0]{\begingroup\@sanitize@url \@url }%
\providecommand \@url [1]{\endgroup\@href {#1}{\urlprefix }}%
\providecommand \urlprefix  [0]{URL }%
\providecommand \Eprint [0]{\href }%
\providecommand \doibase [0]{https://doi.org/}%
\providecommand \selectlanguage [0]{\@gobble}%
\providecommand \bibinfo  [0]{\@secondoftwo}%
\providecommand \bibfield  [0]{\@secondoftwo}%
\providecommand \translation [1]{[#1]}%
\providecommand \BibitemOpen [0]{}%
\providecommand \bibitemStop [0]{}%
\providecommand \bibitemNoStop [0]{.\EOS\space}%
\providecommand \EOS [0]{\spacefactor3000\relax}%
\providecommand \BibitemShut  [1]{\csname bibitem#1\endcsname}%
\let\auto@bib@innerbib\@empty
%</preamble>
\bibitem [{\citenamefont {Cross}\ and\ \citenamefont
  {Hohenberg}(1993)}]{cross_pattern_1993}%
  \BibitemOpen
  \bibfield  {author} {\bibinfo {author} {\bibfnamefont {M.~C.}\ \bibnamefont
  {Cross}}\ and\ \bibinfo {author} {\bibfnamefont {P.~C.}\ \bibnamefont
  {Hohenberg}},\ }\bibfield  {title} {\bibinfo {title} {Pattern formation
  outside of equilibrium},\ }\href {https://doi.org/10.1103/RevModPhys.65.851}
  {\bibfield  {journal} {\bibinfo  {journal} {Reviews of Modern Physics}\
  }\textbf {\bibinfo {volume} {65}},\ \bibinfo {pages} {851} (\bibinfo {year}
  {1993})}\BibitemShut {NoStop}%
\bibitem [{\citenamefont {Hoyle}(2006)}]{hoyle_pattern_2006}%
  \BibitemOpen
  \bibfield  {author} {\bibinfo {author} {\bibfnamefont {R.}~\bibnamefont
  {Hoyle}},\ }\href {https://doi.org/10.1017/CBO9780511616051} {\emph {\bibinfo
  {title} {Pattern {Formation}: {An} {Introduction} to {Methods}}}}\ (\bibinfo
  {publisher} {Cambridge University Press},\ \bibinfo {address} {Cambridge},\
  \bibinfo {year} {2006})\BibitemShut {NoStop}%
\bibitem [{\citenamefont {Murray}(2003)}]{murray_mathematical_2003}%
  \BibitemOpen
  \bibfield  {author} {\bibinfo {author} {\bibfnamefont {J.~D.}\ \bibnamefont
  {Murray}},\ }\href {https://doi.org/10.1007/b98869} {\emph {\bibinfo {title}
  {Mathematical {Biology} {II}: {Spatial} {Models} and {Biomedical}
  {Applications}}}},\ \bibinfo {edition} {3rd}\ ed.,\ Interdisciplinary
  {Applied} {Mathematics}\ (\bibinfo  {publisher} {Springer-Verlag},\ \bibinfo
  {address} {New York},\ \bibinfo {year} {2003})\BibitemShut {NoStop}%
\bibitem [{\citenamefont {Turing}(1952)}]{turing_chemical_1952}%
  \BibitemOpen
  \bibfield  {author} {\bibinfo {author} {\bibfnamefont {A.~M.}\ \bibnamefont
  {Turing}},\ }\bibfield  {title} {\bibinfo {title} {The chemical basis of
  morphogenesis},\ }\href {https://doi.org/10.1098/rstb.1952.0012} {\bibfield
  {journal} {\bibinfo  {journal} {Philosophical Transactions of the Royal
  Society of London. Series B, Biological Sciences}\ }\textbf {\bibinfo
  {volume} {237}},\ \bibinfo {pages} {37} (\bibinfo {year} {1952})}\BibitemShut
  {NoStop}%
\bibitem [{\citenamefont {Swift}\ and\ \citenamefont
  {Hohenberg}(2008)}]{swift_swift-hohenberg_2008}%
  \BibitemOpen
  \bibfield  {author} {\bibinfo {author} {\bibfnamefont {J.}~\bibnamefont
  {Swift}}\ and\ \bibinfo {author} {\bibfnamefont {P.}~\bibnamefont
  {Hohenberg}},\ }\bibfield  {title} {\bibinfo {title} {Swift-{Hohenberg}
  equation},\ }\href {https://doi.org/10.4249/scholarpedia.6395} {\bibfield
  {journal} {\bibinfo  {journal} {Scholarpedia}\ }\textbf {\bibinfo {volume}
  {3}},\ \bibinfo {pages} {6395} (\bibinfo {year} {2008})}\BibitemShut
  {NoStop}%
\bibitem [{\citenamefont {Knobloch}(2015)}]{knobloch_spatial_2015}%
  \BibitemOpen
  \bibfield  {author} {\bibinfo {author} {\bibfnamefont {E.}~\bibnamefont
  {Knobloch}},\ }\bibfield  {title} {\bibinfo {title} {Spatial {localization}
  in {dissipative} {systems}},\ }\href
  {https://doi.org/10.1146/annurev-conmatphys-031214-014514} {\bibfield
  {journal} {\bibinfo  {journal} {Annual Review of Condensed Matter Physics}\
  }\textbf {\bibinfo {volume} {6}},\ \bibinfo {pages} {325} (\bibinfo {year}
  {2015})}\BibitemShut {NoStop}%
\bibitem [{\citenamefont {Walgraef}(1997)}]{walgraef_book}%
  \BibitemOpen
  \bibfield  {author} {\bibinfo {author} {\bibfnamefont {D.}~\bibnamefont
  {Walgraef}},\ }\href {https://doi.org/10.1007/978-1-4612-1850-0} {\emph
  {\bibinfo {title} {{Spatio-Temporal Pattern Formation: with Examples from
  Physics, Chemistry, and Materials Science.}}}}\ (\bibinfo  {publisher}
  {Springer-Verlag},\ \bibinfo {address} {New York},\ \bibinfo {year} {1997})\
  p.\ \bibinfo {pages} {309}\BibitemShut {NoStop}%
\bibitem [{\citenamefont
  {Benjamin}(1978{\natexlab{a}})}]{benjamin_bifurcation_1978}%
  \BibitemOpen
  \bibfield  {author} {\bibinfo {author} {\bibfnamefont {T.~B.}\ \bibnamefont
  {Benjamin}},\ }\bibfield  {title} {\bibinfo {title} {Bifurcation {Phenomena}
  in {Steady} {Flows} of a {Viscous} {Fluid}. {I}. {Theory}},\ }\href
  {https://www.jstor.org/stable/79505} {\bibfield  {journal} {\bibinfo
  {journal} {Proceedings of the Royal Society of London. Series A, Mathematical
  and Physical Sciences}\ }\textbf {\bibinfo {volume} {359}},\ \bibinfo {pages}
  {1} (\bibinfo {year} {1978}{\natexlab{a}})},\ \bibinfo {note} {publisher: The
  Royal Society}\BibitemShut {NoStop}%
\bibitem [{\citenamefont
  {Benjamin}(1978{\natexlab{b}})}]{benjamin_bifurcation_1978b}%
  \BibitemOpen
  \bibfield  {author} {\bibinfo {author} {\bibfnamefont {T.~B.}\ \bibnamefont
  {Benjamin}},\ }\bibfield  {title} {\bibinfo {title} {Bifurcation {Phenomena}
  in {Steady} {Flows} of a {Viscous} {Fluid}. {II}. {Experiments}},\ }\href
  {https://www.jstor.org/stable/79506} {\bibfield  {journal} {\bibinfo
  {journal} {Proceedings of the Royal Society of London. Series A, Mathematical
  and Physical Sciences}\ }\textbf {\bibinfo {volume} {359}},\ \bibinfo {pages}
  {27} (\bibinfo {year} {1978}{\natexlab{b}})},\ \bibinfo {note} {publisher:
  The Royal Society}\BibitemShut {NoStop}%
\bibitem [{\citenamefont {Zelnik}\ \emph {et~al.}(2018)\citenamefont {Zelnik},
  \citenamefont {Gandhi}, \citenamefont {Knobloch},\ and\ \citenamefont
  {Meron}}]{zelnik_implications_2018}%
  \BibitemOpen
  \bibfield  {author} {\bibinfo {author} {\bibfnamefont {Y.~R.}\ \bibnamefont
  {Zelnik}}, \bibinfo {author} {\bibfnamefont {P.}~\bibnamefont {Gandhi}},
  \bibinfo {author} {\bibfnamefont {E.}~\bibnamefont {Knobloch}},\ and\
  \bibinfo {author} {\bibfnamefont {E.}~\bibnamefont {Meron}},\ }\bibfield
  {title} {\bibinfo {title} {Implications of tristability in pattern-forming
  ecosystems},\ }\href {https://doi.org/10.1063/1.5018925} {\bibfield
  {journal} {\bibinfo  {journal} {Chaos: An Interdisciplinary Journal of
  Nonlinear Science}\ }\textbf {\bibinfo {volume} {28}},\ \bibinfo {pages}
  {033609} (\bibinfo {year} {2018})}\BibitemShut {NoStop}%
\bibitem [{\citenamefont {Kozyreff}\ \emph {et~al.}(2003)\citenamefont
  {Kozyreff}, \citenamefont {Chapman},\ and\ \citenamefont
  {Tlidi}}]{kozyreff_interaction_2003}%
  \BibitemOpen
  \bibfield  {author} {\bibinfo {author} {\bibfnamefont {G.}~\bibnamefont
  {Kozyreff}}, \bibinfo {author} {\bibfnamefont {S.~J.}\ \bibnamefont
  {Chapman}},\ and\ \bibinfo {author} {\bibfnamefont {M.}~\bibnamefont
  {Tlidi}},\ }\bibfield  {title} {\bibinfo {title} {Interaction of two
  modulational instabilities in a semiconductor resonator},\ }\href
  {https://doi.org/10.1103/PhysRevE.68.015201} {\bibfield  {journal} {\bibinfo
  {journal} {Physical Review E}\ }\textbf {\bibinfo {volume} {68}},\ \bibinfo
  {pages} {015201} (\bibinfo {year} {2003})}\BibitemShut {NoStop}%
\bibitem [{\citenamefont {Kozyreff}\ and\ \citenamefont
  {Tlidi}(2007)}]{kozyreff_nonvariational_2007}%
  \BibitemOpen
  \bibfield  {author} {\bibinfo {author} {\bibfnamefont {G.}~\bibnamefont
  {Kozyreff}}\ and\ \bibinfo {author} {\bibfnamefont {M.}~\bibnamefont
  {Tlidi}},\ }\bibfield  {title} {\bibinfo {title} {Nonvariational real
  {Swift}-{Hohenberg} equation for biological, chemical, and optical systems},\
  }\href {https://doi.org/10.1063/1.2759436} {\bibfield  {journal} {\bibinfo
  {journal} {Chaos: An Interdisciplinary Journal of Nonlinear Science}\
  }\textbf {\bibinfo {volume} {17}},\ \bibinfo {pages} {037103} (\bibinfo
  {year} {2007})}\BibitemShut {NoStop}%
\bibitem [{\citenamefont {Hornreich}(1980)}]{hornreich_lifshitz_1980}%
  \BibitemOpen
  \bibfield  {author} {\bibinfo {author} {\bibfnamefont {R.~M.}\ \bibnamefont
  {Hornreich}},\ }\bibfield  {title} {\bibinfo {title} {The {Lifshitz} point:
  {Phase} diagrams and critical behavior},\ }\href
  {https://doi.org/10.1016/0304-8853(80)91100-2} {\bibfield  {journal}
  {\bibinfo  {journal} {Journal of Magnetism and Magnetic Materials}\ }\textbf
  {\bibinfo {volume} {15-18}},\ \bibinfo {pages} {387} (\bibinfo {year}
  {1980})}\BibitemShut {NoStop}%
\bibitem [{\citenamefont {Pisarski}\ \emph {et~al.}(2019)\citenamefont
  {Pisarski}, \citenamefont {Skokov},\ and\ \citenamefont
  {Tsvelik}}]{pisarski_pedagogical_2019}%
  \BibitemOpen
  \bibfield  {author} {\bibinfo {author} {\bibfnamefont {R.~D.}\ \bibnamefont
  {Pisarski}}, \bibinfo {author} {\bibfnamefont {V.~V.}\ \bibnamefont
  {Skokov}},\ and\ \bibinfo {author} {\bibfnamefont {A.}~\bibnamefont
  {Tsvelik}},\ }\bibfield  {title} {\bibinfo {title} {A {Pedagogical}
  {Introduction} to the {Lifshitz} {Regime}},\ }\href
  {https://doi.org/10.3390/universe5020048} {\bibfield  {journal} {\bibinfo
  {journal} {Universe}\ }\textbf {\bibinfo {volume} {5}},\ \bibinfo {pages}
  {48} (\bibinfo {year} {2019})}\BibitemShut {NoStop}%
\bibitem [{\citenamefont {Kozyreff}\ and\ \citenamefont
  {Tlidi}(2004)}]{kozyreff_optical_2004}%
  \BibitemOpen
  \bibfield  {author} {\bibinfo {author} {\bibfnamefont {G.}~\bibnamefont
  {Kozyreff}}\ and\ \bibinfo {author} {\bibfnamefont {M.}~\bibnamefont
  {Tlidi}},\ }\bibfield  {title} {\bibinfo {title} {Optical patterns with
  different wavelengths},\ }\href {https://doi.org/10.1103/PhysRevE.69.066202}
  {\bibfield  {journal} {\bibinfo  {journal} {Physical Review E}\ }\textbf
  {\bibinfo {volume} {69}},\ \bibinfo {pages} {066202} (\bibinfo {year}
  {2004})}\BibitemShut {NoStop}%
\bibitem [{\citenamefont {Bordeu}\ and\ \citenamefont
  {Clerc}(2015)}]{bordeu_rodlike_2015-1}%
  \BibitemOpen
  \bibfield  {author} {\bibinfo {author} {\bibfnamefont {I.}~\bibnamefont
  {Bordeu}}\ and\ \bibinfo {author} {\bibfnamefont {M.~G.}\ \bibnamefont
  {Clerc}},\ }\bibfield  {title} {\bibinfo {title} {Rodlike localized structure
  in isotropic pattern-forming systems},\ }\href
  {https://doi.org/10.1103/PhysRevE.92.042915} {\bibfield  {journal} {\bibinfo
  {journal} {Physical Review E}\ }\textbf {\bibinfo {volume} {92}},\ \bibinfo
  {pages} {042915} (\bibinfo {year} {2015})}\BibitemShut {NoStop}%
\bibitem [{\citenamefont {Vladimirov}\ \emph {et~al.}(2011)\citenamefont
  {Vladimirov}, \citenamefont {Lefever},\ and\ \citenamefont
  {Tlidi}}]{vladimirov_relative_2011-1}%
  \BibitemOpen
  \bibfield  {author} {\bibinfo {author} {\bibfnamefont {A.~G.}\ \bibnamefont
  {Vladimirov}}, \bibinfo {author} {\bibfnamefont {R.}~\bibnamefont
  {Lefever}},\ and\ \bibinfo {author} {\bibfnamefont {M.}~\bibnamefont
  {Tlidi}},\ }\bibfield  {title} {\bibinfo {title} {Relative stability of
  multipeak localized patterns of cavity solitons},\ }\href
  {https://doi.org/10.1103/PhysRevA.84.043848} {\bibfield  {journal} {\bibinfo
  {journal} {Physical Review A}\ }\textbf {\bibinfo {volume} {84}},\ \bibinfo
  {pages} {043848} (\bibinfo {year} {2011})}\BibitemShut {NoStop}%
\bibitem [{\citenamefont {Burke}\ and\ \citenamefont
  {Dawes}(2012)}]{burke_localized_2012}%
  \BibitemOpen
  \bibfield  {author} {\bibinfo {author} {\bibfnamefont {J.}~\bibnamefont
  {Burke}}\ and\ \bibinfo {author} {\bibfnamefont {J.~H.~P.}\ \bibnamefont
  {Dawes}},\ }\bibfield  {title} {\bibinfo {title} {Localized {States} in an
  {Extended} {Swift}–{Hohenberg} {Equation}},\ }\href
  {https://doi.org/10.1137/110843976} {\bibfield  {journal} {\bibinfo
  {journal} {SIAM Journal on Applied Dynamical Systems}\ }\textbf {\bibinfo
  {volume} {11}},\ \bibinfo {pages} {261} (\bibinfo {year} {2012})}\BibitemShut
  {NoStop}%
\bibitem [{\citenamefont {Clerc}\ \emph {et~al.}(2005)\citenamefont {Clerc},
  \citenamefont {Petrossian},\ and\ \citenamefont
  {Residori}}]{clerc_bouncing_2005}%
  \BibitemOpen
  \bibfield  {author} {\bibinfo {author} {\bibfnamefont {M.~G.}\ \bibnamefont
  {Clerc}}, \bibinfo {author} {\bibfnamefont {A.}~\bibnamefont {Petrossian}},\
  and\ \bibinfo {author} {\bibfnamefont {S.}~\bibnamefont {Residori}},\
  }\bibfield  {title} {\bibinfo {title} {Bouncing localized structures in a
  liquid-crystal light-valve experiment},\ }\href
  {https://doi.org/10.1103/PhysRevE.71.015205} {\bibfield  {journal} {\bibinfo
  {journal} {Physical Review E}\ }\textbf {\bibinfo {volume} {71}},\ \bibinfo
  {pages} {015205} (\bibinfo {year} {2005})}\BibitemShut {NoStop}%
\bibitem [{\citenamefont {Clerc}\ and\ \citenamefont
  {Verschueren}(2013)}]{clerc_quasiperiodicity_2013}%
  \BibitemOpen
  \bibfield  {author} {\bibinfo {author} {\bibfnamefont {M.~G.}\ \bibnamefont
  {Clerc}}\ and\ \bibinfo {author} {\bibfnamefont {N.}~\bibnamefont
  {Verschueren}},\ }\bibfield  {title} {\bibinfo {title} {Quasiperiodicity
  route to spatiotemporal chaos in one-dimensional pattern-forming systems},\
  }\href {https://doi.org/10.1103/PhysRevE.88.052916} {\bibfield  {journal}
  {\bibinfo  {journal} {Physical Review E}\ }\textbf {\bibinfo {volume} {88}},\
  \bibinfo {pages} {052916} (\bibinfo {year} {2013})}\BibitemShut {NoStop}%
\bibitem [{\citenamefont {Kawahara}(1972)}]{Kawahara}%
  \BibitemOpen
  \bibfield  {author} {\bibinfo {author} {\bibfnamefont {T.}~\bibnamefont
  {Kawahara}},\ }\bibfield  {title} {\bibinfo {title} {Oscillatory solitary
  waves in dispersive media},\ }\href {https://doi.org/10.1143/JPSJ.33.260}
  {\bibfield  {journal} {\bibinfo  {journal} {J. Phys. Soc. Japan}\ }\textbf
  {\bibinfo {volume} {33}},\ \bibinfo {pages} {260} (\bibinfo {year}
  {1972})}\BibitemShut {NoStop}%
\bibitem [{\citenamefont {Groves}\ and\ \citenamefont
  {Champneys}(1997)}]{Groves}%
  \BibitemOpen
  \bibfield  {author} {\bibinfo {author} {\bibfnamefont {M.}~\bibnamefont
  {Groves}}\ and\ \bibinfo {author} {\bibfnamefont {A.}~\bibnamefont
  {Champneys}},\ }\bibfield  {title} {\bibinfo {title} {A global investigation
  of solitary-wave solutions to a two-parameter model for water waves},\ }\href
  {https://doi.org/10.1017/S0022112097005193} {\bibfield  {journal} {\bibinfo
  {journal} {Journal of Fluid Mechanics}\ }\textbf {\bibinfo {volume} {342}},\
  \bibinfo {pages} {199} (\bibinfo {year} {1997})}\BibitemShut {NoStop}%
\bibitem [{\citenamefont {Haragus}\ and\ \citenamefont
  {Iooss}(2011)}]{haragus_local_2011}%
  \BibitemOpen
  \bibfield  {author} {\bibinfo {author} {\bibfnamefont {M.}~\bibnamefont
  {Haragus}}\ and\ \bibinfo {author} {\bibfnamefont {G.}~\bibnamefont
  {Iooss}},\ }\href {https://doi.org/10.1007/978-0-85729-112-7} {\emph
  {\bibinfo {title} {Local {Bifurcations}, {Center} {Manifolds}, and {Normal}
  {Forms} in {Infinite}-{Dimensional} {Dynamical} {Systems}}}},\ Universitext\
  (\bibinfo  {publisher} {Springer-Verlag},\ \bibinfo {address} {London},\
  \bibinfo {year} {2011})\BibitemShut {NoStop}%
\bibitem [{\citenamefont {Buffoni}\ \emph {et~al.}()\citenamefont {Buffoni},
  \citenamefont {Groves},\ and\ \citenamefont {Toland}}]{Toland}%
  \BibitemOpen
  \bibfield  {author} {\bibinfo {author} {\bibfnamefont {B.}~\bibnamefont
  {Buffoni}}, \bibinfo {author} {\bibfnamefont {M.}~\bibnamefont {Groves}},\
  and\ \bibinfo {author} {\bibfnamefont {J.}~\bibnamefont {Toland}},\
  }\bibfield  {title} {\bibinfo {title} {A plethora of solitary
  gravity-capillary water waves with nearly critical {Bond and Froude}
  numbers},\ }\href@noop {} {\bibfield  {journal} {\bibinfo  {journal} {Phil.
  Trans. R. Soc. Lond. A}\ }\textbf {\bibinfo {volume} {354}},\ \bibinfo
  {pages} {575–607}}\BibitemShut {NoStop}%
\bibitem [{\citenamefont {Iooss}(1995)}]{iooss_codimension_1995}%
  \BibitemOpen
  \bibfield  {author} {\bibinfo {author} {\bibfnamefont {G.}~\bibnamefont
  {Iooss}},\ }\bibfield  {title} {\bibinfo {title} {A codimension 2
  {bifurcation} for reversible {vector} {fields}},\ }\href
  {https://hal.univ-cotedazur.fr/hal-01271013} {\bibfield  {journal} {\bibinfo
  {journal} {Fields Institute Communications}\ }\textbf {\bibinfo {volume}
  {4}},\ \bibinfo {pages} {201} (\bibinfo {year} {1995})}\BibitemShut {NoStop}%
\bibitem [{\citenamefont {Peletier}\ and\ \citenamefont {Troy}(1995)}]{Troy}%
  \BibitemOpen
  \bibfield  {author} {\bibinfo {author} {\bibfnamefont {L.}~\bibnamefont
  {Peletier}}\ and\ \bibinfo {author} {\bibfnamefont {W.}~\bibnamefont
  {Troy}},\ }\bibfield  {title} {\bibinfo {title} {Spatial patterns described
  by the extended {Fisher-Kolmogorov (EFK)} equation: kinks},\ }\href@noop {}
  {\bibfield  {journal} {\bibinfo  {journal} {Differential Integral Equations}\
  }\textbf {\bibinfo {volume} {8}},\ \bibinfo {pages} {1279} (\bibinfo {year}
  {1995})}\BibitemShut {NoStop}%
\bibitem [{\citenamefont {Champneys}\ \emph {et~al.}(2000)\citenamefont
  {Champneys}, \citenamefont {McKenna},\ and\ \citenamefont
  {Zegeling}}]{McKenna}%
  \BibitemOpen
  \bibfield  {author} {\bibinfo {author} {\bibfnamefont {A.}~\bibnamefont
  {Champneys}}, \bibinfo {author} {\bibfnamefont {P.}~\bibnamefont {McKenna}},\
  and\ \bibinfo {author} {\bibfnamefont {P.}~\bibnamefont {Zegeling}},\
  }\bibfield  {title} {\bibinfo {title} {Solitary waves in nonlinear beam
  equations: Stability, fission and fusion},\ }\href
  {https://link.springer.com/article/10.1023/A:1008302207311} {\bibfield
  {journal} {\bibinfo  {journal} {Nonlinear Dynamics}\ }\textbf {\bibinfo
  {volume} {21}},\ \bibinfo {pages} {31–53} (\bibinfo {year}
  {2000})}\BibitemShut {NoStop}%
\bibitem [{\citenamefont {Champneys}\ \emph {et~al.}(2019)\citenamefont
  {Champneys}, \citenamefont {Dodwell}, \citenamefont {Groh}, \citenamefont
  {Hunt}, \citenamefont {Neville}, \citenamefont {Pirrera}, \citenamefont
  {Sakhaei}, \citenamefont {Schenk},\ and\ \citenamefont {Wadee}}]{Happy}%
  \BibitemOpen
  \bibfield  {author} {\bibinfo {author} {\bibfnamefont {A.}~\bibnamefont
  {Champneys}}, \bibinfo {author} {\bibfnamefont {T.}~\bibnamefont {Dodwell}},
  \bibinfo {author} {\bibfnamefont {R.}~\bibnamefont {Groh}}, \bibinfo {author}
  {\bibfnamefont {G.}~\bibnamefont {Hunt}}, \bibinfo {author} {\bibfnamefont
  {R.}~\bibnamefont {Neville}}, \bibinfo {author} {\bibfnamefont
  {A.}~\bibnamefont {Pirrera}}, \bibinfo {author} {\bibfnamefont
  {A.}~\bibnamefont {Sakhaei}}, \bibinfo {author} {\bibfnamefont
  {M.}~\bibnamefont {Schenk}},\ and\ \bibinfo {author} {\bibfnamefont
  {A.}~\bibnamefont {Wadee}},\ }\bibfield  {title} {\bibinfo {title} {Happy
  catastrophe: Recent progress in analysis and exploitation of elastic
  instability},\ }\href
  {https://doi.org/https://doi.org/10.3389/fams.2019.00034} {\bibfield
  {journal} {\bibinfo  {journal} {Frontiers in Applied Mathematics and
  Statistics}\ }\textbf {\bibinfo {volume} {5}} (\bibinfo {year}
  {2019})}\BibitemShut {NoStop}%
\bibitem [{\citenamefont {Champneys}\ \emph {et~al.}(2021)\citenamefont
  {Champneys}, \citenamefont {{Al Saadi}}, \citenamefont {{Bre\~na--Medina}},
  \citenamefont {Grieneisen}, \citenamefont {{Mar\'{e}e}}, \citenamefont
  {Verschueren},\ and\ \citenamefont {Wuyts}}]{MeronIssue}%
  \BibitemOpen
  \bibfield  {author} {\bibinfo {author} {\bibfnamefont {A.}~\bibnamefont
  {Champneys}}, \bibinfo {author} {\bibfnamefont {F.}~\bibnamefont {{Al
  Saadi}}}, \bibinfo {author} {\bibfnamefont {V.}~\bibnamefont
  {{Bre\~na--Medina}}}, \bibinfo {author} {\bibfnamefont {V.}~\bibnamefont
  {Grieneisen}}, \bibinfo {author} {\bibfnamefont {A.}~\bibnamefont
  {{Mar\'{e}e}}}, \bibinfo {author} {\bibfnamefont {N.}~\bibnamefont
  {Verschueren}},\ and\ \bibinfo {author} {\bibfnamefont {B.}~\bibnamefont
  {Wuyts}},\ }\bibfield  {title} {\bibinfo {title} {Bistability, wave pinning
  and localisation in natural reaction-diffusion systems},\ }\href
  {https://doi.org/https://doi.org/10.1016/j.physd.2020.132735} {\bibfield
  {journal} {\bibinfo  {journal} {Physica D.}\ }\textbf {\bibinfo {volume}
  {416}},\ \bibinfo {pages} {132735} (\bibinfo {year} {2021})}\BibitemShut
  {NoStop}%
\bibitem [{\citenamefont {Meron}(2015)}]{Meron}%
  \BibitemOpen
  \bibfield  {author} {\bibinfo {author} {\bibfnamefont {E.}~\bibnamefont
  {Meron}},\ }\href@noop {} {\emph {\bibinfo {title} {Nonlinear Physics of
  Ecosystems}}},\ \bibinfo {edition} {3rd}\ ed.\ (\bibinfo  {publisher} {CRC
  Press},\ \bibinfo {year} {2015})\BibitemShut {NoStop}%
\bibitem [{\citenamefont {Dawes}(2016)}]{Dawes}%
  \BibitemOpen
  \bibfield  {author} {\bibinfo {author} {\bibfnamefont {J.~H.}\ \bibnamefont
  {Dawes}},\ }\bibfield  {title} {\bibinfo {title} {After 1952: The later
  development of alan turing's ideas on the mathematics of pattern formation},\
  }\href {https://doi.org/10.1016/j.hm.2015.03.003} {\bibfield  {journal}
  {\bibinfo  {journal} {Historia Mathematica}\ }\textbf {\bibinfo {volume}
  {43}},\ \bibinfo {pages} {49} (\bibinfo {year} {2016})}\BibitemShut {NoStop}%
\bibitem [{\citenamefont {Al~Saadi}\ and\ \citenamefont
  {Champneys}(2021)}]{al_saadi_unified_2021}%
  \BibitemOpen
  \bibfield  {author} {\bibinfo {author} {\bibfnamefont {F.}~\bibnamefont
  {Al~Saadi}}\ and\ \bibinfo {author} {\bibfnamefont {A.}~\bibnamefont
  {Champneys}},\ }\bibfield  {title} {\bibinfo {title} {Unified framework for
  localized patterns in reaction–diffusion systems; the {Gray}–{Scott} and
  {Gierer}–{Meinhardt} cases},\ }\href
  {https://doi.org/10.1098/rsta.2020.0277} {\bibfield  {journal} {\bibinfo
  {journal} {Philosophical Transactions of the Royal Society A: Mathematical,
  Physical and Engineering Sciences}\ }\textbf {\bibinfo {volume} {379}},\
  \bibinfo {pages} {20200277} (\bibinfo {year} {2021})},\ \bibinfo {note}
  {publisher: Royal Society}\BibitemShut {NoStop}%
\bibitem [{\citenamefont {Tlidi}\ \emph {et~al.}(1994)\citenamefont {Tlidi},
  \citenamefont {Mandel},\ and\ \citenamefont {Lefever}}]{tlidi_1994}%
  \BibitemOpen
  \bibfield  {author} {\bibinfo {author} {\bibfnamefont {M.}~\bibnamefont
  {Tlidi}}, \bibinfo {author} {\bibfnamefont {P.}~\bibnamefont {Mandel}},\ and\
  \bibinfo {author} {\bibfnamefont {R.}~\bibnamefont {Lefever}},\ }\bibfield
  {title} {\bibinfo {title} {Localized structures and localized patterns in
  optical bistability},\ }\href {https://doi.org/10.1103/PhysRevLett.73.640}
  {\bibfield  {journal} {\bibinfo  {journal} {Phys. Rev. Lett.}\ }\textbf
  {\bibinfo {volume} {73}},\ \bibinfo {pages} {640} (\bibinfo {year}
  {1994})}\BibitemShut {NoStop}%
\bibitem [{\citenamefont {Glebsky}\ and\ \citenamefont
  {Lerman}(1995)}]{glebsky_small_1995}%
  \BibitemOpen
  \bibfield  {author} {\bibinfo {author} {\bibfnamefont {L.~Y.}\ \bibnamefont
  {Glebsky}}\ and\ \bibinfo {author} {\bibfnamefont {L.~M.}\ \bibnamefont
  {Lerman}},\ }\bibfield  {title} {\bibinfo {title} {On small stationary
  localized solutions for the generalized 1‐{D} {Swift}–{Hohenberg}
  equation},\ }\href {https://doi.org/10.1063/1.166142} {\bibfield  {journal}
  {\bibinfo  {journal} {Chaos: An Interdisciplinary Journal of Nonlinear
  Science}\ }\textbf {\bibinfo {volume} {5}},\ \bibinfo {pages} {424} (\bibinfo
  {year} {1995})},\ \bibinfo {note} {publisher: American Institute of
  Physics}\BibitemShut {NoStop}%
\bibitem [{\citenamefont {Champneys}\ \emph {et~al.}(2007)\citenamefont
  {Champneys}, \citenamefont {Kirk}, \citenamefont {Knobloch}, \citenamefont
  {Oldeman},\ and\ \citenamefont {Sneyd}}]{champneys_when_2007}%
  \BibitemOpen
  \bibfield  {author} {\bibinfo {author} {\bibfnamefont {A.~R.}\ \bibnamefont
  {Champneys}}, \bibinfo {author} {\bibfnamefont {V.}~\bibnamefont {Kirk}},
  \bibinfo {author} {\bibfnamefont {E.}~\bibnamefont {Knobloch}}, \bibinfo
  {author} {\bibfnamefont {B.~E.}\ \bibnamefont {Oldeman}},\ and\ \bibinfo
  {author} {\bibfnamefont {J.}~\bibnamefont {Sneyd}},\ }\bibfield  {title}
  {\bibinfo {title} {When {Shil}'nikov {meets} {Hopf} in {excitable}
  {systems}},\ }\href {https://doi.org/10.1137/070682654} {\bibfield  {journal}
  {\bibinfo  {journal} {SIAM Journal on Applied Dynamical Systems}\ }\textbf
  {\bibinfo {volume} {6}},\ \bibinfo {pages} {663} (\bibinfo {year}
  {2007})}\BibitemShut {NoStop}%
\bibitem [{\citenamefont {Homburg}\ and\ \citenamefont
  {Sandstede}(2010)}]{homburg_homoclinic_2010}%
  \BibitemOpen
  \bibfield  {author} {\bibinfo {author} {\bibfnamefont {A.~J.}\ \bibnamefont
  {Homburg}}\ and\ \bibinfo {author} {\bibfnamefont {B.}~\bibnamefont
  {Sandstede}},\ }\bibfield  {title} {\bibinfo {title} {Homoclinic and
  heteroclinic bifurcations in vector fields},\ }\href
  {https://doi.org/10.1016/S1874-575X(10)00316-4} {\bibfield  {journal}
  {\bibinfo  {journal} {Handbook of Dynamical Systems: volume 3}\ ,\ \bibinfo
  {pages} {379}} (\bibinfo {year} {2010})}\BibitemShut {NoStop}%
\bibitem [{\citenamefont {Verschueren}\ and\ \citenamefont
  {Champneys}(2017)}]{verschueren_model_2017}%
  \BibitemOpen
  \bibfield  {author} {\bibinfo {author} {\bibfnamefont {N.}~\bibnamefont
  {Verschueren}}\ and\ \bibinfo {author} {\bibfnamefont {A.}~\bibnamefont
  {Champneys}},\ }\bibfield  {title} {\bibinfo {title} {A model for cell
  polarization without mass conservation},\ }\href
  {https://doi.org/10.1137/16M1093847} {\bibfield  {journal} {\bibinfo
  {journal} {SIAM Journal on Applied Dynamical Systems}\ }\textbf {\bibinfo
  {volume} {16}},\ \bibinfo {pages} {1797} (\bibinfo {year}
  {2017})}\BibitemShut {NoStop}%
\bibitem [{\citenamefont {Devaney}(1976)}]{devaney_reversible_1976}%
  \BibitemOpen
  \bibfield  {author} {\bibinfo {author} {\bibfnamefont {R.~L.}\ \bibnamefont
  {Devaney}},\ }\bibfield  {title} {\bibinfo {title} {Reversible
  diffeomorphisms and flows},\ }\href
  {https://doi.org/10.1090/S0002-9947-1976-0402815-3} {\bibfield  {journal}
  {\bibinfo  {journal} {Transactions of the American Mathematical Society}\
  }\textbf {\bibinfo {volume} {218}},\ \bibinfo {pages} {89} (\bibinfo {year}
  {1976})}\BibitemShut {NoStop}%
\bibitem [{\citenamefont {Parra-Rivas}\ \emph
  {et~al.}(2016{\natexlab{a}})\citenamefont {Parra-Rivas}, \citenamefont
  {Gomila}, \citenamefont {Knobloch}, \citenamefont {Coen},\ and\ \citenamefont
  {Gelens}}]{parra-rivas_origin_2016}%
  \BibitemOpen
  \bibfield  {author} {\bibinfo {author} {\bibfnamefont {P.}~\bibnamefont
  {Parra-Rivas}}, \bibinfo {author} {\bibfnamefont {D.}~\bibnamefont {Gomila}},
  \bibinfo {author} {\bibfnamefont {E.}~\bibnamefont {Knobloch}}, \bibinfo
  {author} {\bibfnamefont {S.}~\bibnamefont {Coen}},\ and\ \bibinfo {author}
  {\bibfnamefont {L.}~\bibnamefont {Gelens}},\ }\bibfield  {title} {\bibinfo
  {title} {Origin and stability of dark pulse {Kerr} combs in normal dispersion
  resonators},\ }\href {https://doi.org/10.1364/OL.41.002402} {\bibfield
  {journal} {\bibinfo  {journal} {Optics Letters}\ }\textbf {\bibinfo {volume}
  {41}},\ \bibinfo {pages} {2402} (\bibinfo {year}
  {2016}{\natexlab{a}})}\BibitemShut {NoStop}%
\bibitem [{\citenamefont {Parra-Rivas}\ \emph {et~al.}(2020)\citenamefont
  {Parra-Rivas}, \citenamefont {Mas-Arabí},\ and\ \citenamefont
  {Leo}}]{parra-rivas_parametric_2020}%
  \BibitemOpen
  \bibfield  {author} {\bibinfo {author} {\bibfnamefont {P.}~\bibnamefont
  {Parra-Rivas}}, \bibinfo {author} {\bibfnamefont {C.}~\bibnamefont
  {Mas-Arabí}},\ and\ \bibinfo {author} {\bibfnamefont {F.}~\bibnamefont
  {Leo}},\ }\bibfield  {title} {\bibinfo {title} {Parametric localized patterns
  and breathers in dispersive quadratic cavities},\ }\href
  {https://doi.org/10.1103/PhysRevA.101.063817} {\bibfield  {journal} {\bibinfo
   {journal} {Physical Review A}\ }\textbf {\bibinfo {volume} {101}},\ \bibinfo
  {pages} {063817} (\bibinfo {year} {2020})}\BibitemShut {NoStop}%
\bibitem [{\citenamefont {Champneys}(1998)}]{champneys_homoclinic_1998}%
  \BibitemOpen
  \bibfield  {author} {\bibinfo {author} {\bibfnamefont {A.~R.}\ \bibnamefont
  {Champneys}},\ }\bibfield  {title} {\bibinfo {title} {Homoclinic orbits in
  reversible systems and their applications in mechanics, fluids and optics},\
  }\href {https://doi.org/10.1016/S0167-2789(97)00209-1} {\bibfield  {journal}
  {\bibinfo  {journal} {Physica D: Nonlinear Phenomena}\ }\bibinfo {series}
  {Proceedings of the {Workshop} on {Time}-{Reversal} {Symmetry} in {Dynamical}
  {Systems}},\ \textbf {\bibinfo {volume} {112}},\ \bibinfo {pages} {158}
  (\bibinfo {year} {1998})}\BibitemShut {NoStop}%
\bibitem [{\citenamefont {Iooss}\ and\ \citenamefont
  {Peroueme}(1993)}]{iooss_perturbed_1993}%
  \BibitemOpen
  \bibfield  {author} {\bibinfo {author} {\bibfnamefont {G.}~\bibnamefont
  {Iooss}}\ and\ \bibinfo {author} {\bibfnamefont {M.~C.}\ \bibnamefont
  {Peroueme}},\ }\bibfield  {title} {\bibinfo {title} {Perturbed {Homoclinic}
  {Solutions} in {Reversible} 1:1 {Resonance} {Vector} {Fields}},\ }\href
  {https://doi.org/10.1006/jdeq.1993.1022} {\bibfield  {journal} {\bibinfo
  {journal} {Journal of Differential Equations}\ }\textbf {\bibinfo {volume}
  {102}},\ \bibinfo {pages} {62} (\bibinfo {year} {1993})}\BibitemShut
  {NoStop}%
\bibitem [{\citenamefont {Godey}(2017)}]{godey_bifurcation_2017}%
  \BibitemOpen
  \bibfield  {author} {\bibinfo {author} {\bibfnamefont {C.}~\bibnamefont
  {Godey}},\ }\bibfield  {title} {\bibinfo {title} {A bifurcation analysis for
  the {Lugiato}-{Lefever} equation},\ }\href
  {https://doi.org/10.1140/epjd/e2017-80057-2} {\bibfield  {journal} {\bibinfo
  {journal} {The European Physical Journal D}\ }\textbf {\bibinfo {volume}
  {71}},\ \bibinfo {pages} {131} (\bibinfo {year} {2017})}\BibitemShut
  {NoStop}%
\bibitem [{\citenamefont {Burke}\ \emph {et~al.}(2008)\citenamefont {Burke},
  \citenamefont {Yochelis},\ and\ \citenamefont
  {Knobloch}}]{burke_classification_2008}%
  \BibitemOpen
  \bibfield  {author} {\bibinfo {author} {\bibfnamefont {J.}~\bibnamefont
  {Burke}}, \bibinfo {author} {\bibfnamefont {A.}~\bibnamefont {Yochelis}},\
  and\ \bibinfo {author} {\bibfnamefont {E.}~\bibnamefont {Knobloch}},\
  }\bibfield  {title} {\bibinfo {title} {Classification of {spatially}
  {localized} {oscillations} in {periodically} {forced} {dissipative}
  {systems}},\ }\href {https://doi.org/10.1137/070698191} {\bibfield  {journal}
  {\bibinfo  {journal} {SIAM Journal on Applied Dynamical Systems}\ }\textbf
  {\bibinfo {volume} {7}},\ \bibinfo {pages} {651} (\bibinfo {year}
  {2008})}\BibitemShut {NoStop}%
\bibitem [{\citenamefont {Gandhi}\ \emph {et~al.}(2018)\citenamefont {Gandhi},
  \citenamefont {Zelnik},\ and\ \citenamefont
  {Knobloch}}]{gandhi_spatially_2018}%
  \BibitemOpen
  \bibfield  {author} {\bibinfo {author} {\bibfnamefont {P.}~\bibnamefont
  {Gandhi}}, \bibinfo {author} {\bibfnamefont {Y.~R.}\ \bibnamefont {Zelnik}},\
  and\ \bibinfo {author} {\bibfnamefont {E.}~\bibnamefont {Knobloch}},\
  }\bibfield  {title} {\bibinfo {title} {Spatially localized structures in the
  {Gray}–{Scott} model},\ }\href {https://doi.org/10.1098/rsta.2017.0375}
  {\bibfield  {journal} {\bibinfo  {journal} {Philosophical Transactions of the
  Royal Society A: Mathematical, Physical and Engineering Sciences}\ }\textbf
  {\bibinfo {volume} {376}},\ \bibinfo {pages} {20170375} (\bibinfo {year}
  {2018})}\BibitemShut {NoStop}%
\bibitem [{\citenamefont {Melbourne}(1998)}]{melbourne_derivation_1998}%
  \BibitemOpen
  \bibfield  {author} {\bibinfo {author} {\bibfnamefont {I.}~\bibnamefont
  {Melbourne}},\ }\bibfield  {title} {\bibinfo {title} {Derivation of the
  {time}-{dependent} {Ginzburg}-{Landau} {equation} on the {line}},\ }\href
  {https://doi.org/10.1007/s003329900041} {\bibfield  {journal} {\bibinfo
  {journal} {Journal of Nonlinear Science}\ }\textbf {\bibinfo {volume} {8}},\
  \bibinfo {pages} {1} (\bibinfo {year} {1998})}\BibitemShut {NoStop}%
\bibitem [{\citenamefont {Burke}\ and\ \citenamefont
  {Knobloch}(2006)}]{burkeknobloch2006}%
  \BibitemOpen
  \bibfield  {author} {\bibinfo {author} {\bibfnamefont {J.}~\bibnamefont
  {Burke}}\ and\ \bibinfo {author} {\bibfnamefont {E.}~\bibnamefont
  {Knobloch}},\ }\bibfield  {title} {\bibinfo {title} {Localized states in the
  generalized {Swift}-{Hohenberg} equation},\ }\href
  {https://doi.org/10.1103/PhysRevE.73.056211} {\bibfield  {journal} {\bibinfo
  {journal} {Physical Review E}\ }\textbf {\bibinfo {volume} {73}},\ \bibinfo
  {pages} {056211} (\bibinfo {year} {2006})}\BibitemShut {NoStop}%
\bibitem [{\citenamefont {Kozyreff}\ and\ \citenamefont
  {Chapman}(2006)}]{kozyreff_asymptotics_2006}%
  \BibitemOpen
  \bibfield  {author} {\bibinfo {author} {\bibfnamefont {G.}~\bibnamefont
  {Kozyreff}}\ and\ \bibinfo {author} {\bibfnamefont {S.~J.}\ \bibnamefont
  {Chapman}},\ }\bibfield  {title} {\bibinfo {title} {Asymptotics of {large}
  {bound} {states} of {localized} {structures}},\ }\href
  {https://doi.org/10.1103/PhysRevLett.97.044502} {\bibfield  {journal}
  {\bibinfo  {journal} {Physical Review Letters}\ }\textbf {\bibinfo {volume}
  {97}},\ \bibinfo {pages} {044502} (\bibinfo {year} {2006})}\BibitemShut
  {NoStop}%
\bibitem [{\citenamefont {Chapman}\ and\ \citenamefont
  {Kozyreff}(2009)}]{chapman_exponential_2009}%
  \BibitemOpen
  \bibfield  {author} {\bibinfo {author} {\bibfnamefont {S.~J.}\ \bibnamefont
  {Chapman}}\ and\ \bibinfo {author} {\bibfnamefont {G.}~\bibnamefont
  {Kozyreff}},\ }\bibfield  {title} {\bibinfo {title} {Exponential asymptotics
  of localised patterns and snaking bifurcation diagrams},\ }\href
  {https://doi.org/10.1016/j.physd.2008.10.005} {\bibfield  {journal} {\bibinfo
   {journal} {Physica D: Nonlinear Phenomena}\ }\textbf {\bibinfo {volume}
  {238}},\ \bibinfo {pages} {319} (\bibinfo {year} {2009})}\BibitemShut
  {NoStop}%
\bibitem [{\citenamefont {Kozyreff}(2012)}]{kozyreff_localized_2012}%
  \BibitemOpen
  \bibfield  {author} {\bibinfo {author} {\bibfnamefont {G.}~\bibnamefont
  {Kozyreff}},\ }\bibfield  {title} {\bibinfo {title} {Localized {Turing}
  patterns in nonlinear optical cavities},\ }\href
  {https://doi.org/10.1016/j.physd.2012.02.007} {\bibfield  {journal} {\bibinfo
   {journal} {Physica D: Nonlinear Phenomena}\ }\textbf {\bibinfo {volume}
  {241}},\ \bibinfo {pages} {939} (\bibinfo {year} {2012})}\BibitemShut
  {NoStop}%
\bibitem [{\citenamefont {Doedel}\ \emph
  {et~al.}(1991{\natexlab{a}})\citenamefont {Doedel}, \citenamefont {Keller},\
  and\ \citenamefont {Kernevez}}]{doedel_numerical_1991}%
  \BibitemOpen
  \bibfield  {author} {\bibinfo {author} {\bibfnamefont {E.}~\bibnamefont
  {Doedel}}, \bibinfo {author} {\bibfnamefont {H.~B.}\ \bibnamefont {Keller}},\
  and\ \bibinfo {author} {\bibfnamefont {J.~P.}\ \bibnamefont {Kernevez}},\
  }\bibfield  {title} {\bibinfo {title} {Numerical analysis and control of
  bifurcation problems (i): bifurcation in finite dimensions},\ }\href
  {https://doi.org/10.1142/S0218127491000397} {\bibfield  {journal} {\bibinfo
  {journal} {International Journal of Bifurcation and Chaos}\ }\textbf
  {\bibinfo {volume} {01}},\ \bibinfo {pages} {493} (\bibinfo {year}
  {1991}{\natexlab{a}})}\BibitemShut {NoStop}%
\bibitem [{\citenamefont {Doedel}\ \emph
  {et~al.}(1991{\natexlab{b}})\citenamefont {Doedel}, \citenamefont {Keller},\
  and\ \citenamefont {Kernevez}}]{doedel_numerical_1991-1}%
  \BibitemOpen
  \bibfield  {author} {\bibinfo {author} {\bibfnamefont {E.}~\bibnamefont
  {Doedel}}, \bibinfo {author} {\bibfnamefont {H.~B.}\ \bibnamefont {Keller}},\
  and\ \bibinfo {author} {\bibfnamefont {J.~P.}\ \bibnamefont {Kernevez}},\
  }\bibfield  {title} {\bibinfo {title} {Numerical analysis and control of
  bifurcation problems (ii): bifurcation in infinite dimensions},\ }\href
  {https://doi.org/10.1142/S0218127491000555} {\bibfield  {journal} {\bibinfo
  {journal} {International Journal of Bifurcation and Chaos}\ }\textbf
  {\bibinfo {volume} {01}},\ \bibinfo {pages} {745} (\bibinfo {year}
  {1991}{\natexlab{b}})}\BibitemShut {NoStop}%
\bibitem [{\citenamefont {Burke}\ and\ \citenamefont
  {Knobloch}(2007)}]{burke_snakes_2007}%
  \BibitemOpen
  \bibfield  {author} {\bibinfo {author} {\bibfnamefont {J.}~\bibnamefont
  {Burke}}\ and\ \bibinfo {author} {\bibfnamefont {E.}~\bibnamefont
  {Knobloch}},\ }\bibfield  {title} {\bibinfo {title} {Snakes and ladders:
  {Localized} states in the {Swift}–{Hohenberg} equation},\ }\href
  {https://doi.org/10.1016/j.physleta.2006.08.072} {\bibfield  {journal}
  {\bibinfo  {journal} {Physics Letters A}\ }\textbf {\bibinfo {volume}
  {360}},\ \bibinfo {pages} {681} (\bibinfo {year} {2007})}\BibitemShut
  {NoStop}%
\bibitem [{\citenamefont {Knobloch}\ and\ \citenamefont
  {Wagenknecht}(2005)}]{knobloch_homoclinic_2005}%
  \BibitemOpen
  \bibfield  {author} {\bibinfo {author} {\bibfnamefont {J.}~\bibnamefont
  {Knobloch}}\ and\ \bibinfo {author} {\bibfnamefont {T.}~\bibnamefont
  {Wagenknecht}},\ }\bibfield  {title} {\bibinfo {title} {Homoclinic snaking
  near a heteroclinic cycle in reversible systems},\ }\href
  {https://doi.org/10.1016/j.physd.2005.04.018} {\bibfield  {journal} {\bibinfo
   {journal} {Physica D: Nonlinear Phenomena}\ }\textbf {\bibinfo {volume}
  {206}},\ \bibinfo {pages} {82} (\bibinfo {year} {2005})}\BibitemShut
  {NoStop}%
\bibitem [{\citenamefont {Parra-Rivas}\ \emph
  {et~al.}(2016{\natexlab{b}})\citenamefont {Parra-Rivas}, \citenamefont
  {Knobloch}, \citenamefont {Gomila},\ and\ \citenamefont
  {Gelens}}]{parra-rivas_dark_2016}%
  \BibitemOpen
  \bibfield  {author} {\bibinfo {author} {\bibfnamefont {P.}~\bibnamefont
  {Parra-Rivas}}, \bibinfo {author} {\bibfnamefont {E.}~\bibnamefont
  {Knobloch}}, \bibinfo {author} {\bibfnamefont {D.}~\bibnamefont {Gomila}},\
  and\ \bibinfo {author} {\bibfnamefont {L.}~\bibnamefont {Gelens}},\
  }\bibfield  {title} {\bibinfo {title} {Dark solitons in the
  {Lugiato}-{Lefever} equation with normal dispersion},\ }\href
  {https://doi.org/10.1103/PhysRevA.93.063839} {\bibfield  {journal} {\bibinfo
  {journal} {Physical Review A}\ }\textbf {\bibinfo {volume} {93}},\ \bibinfo
  {pages} {063839} (\bibinfo {year} {2016}{\natexlab{b}})}\BibitemShut
  {NoStop}%
\bibitem [{\citenamefont {Coullet}\ \emph {et~al.}(1987)\citenamefont
  {Coullet}, \citenamefont {Elphick},\ and\ \citenamefont
  {Repaux}}]{coullet_nature_1987}%
  \BibitemOpen
  \bibfield  {author} {\bibinfo {author} {\bibfnamefont {P.}~\bibnamefont
  {Coullet}}, \bibinfo {author} {\bibfnamefont {C.}~\bibnamefont {Elphick}},\
  and\ \bibinfo {author} {\bibfnamefont {D.}~\bibnamefont {Repaux}},\
  }\bibfield  {title} {\bibinfo {title} {Nature of spatial chaos},\ }\href
  {https://doi.org/10.1103/PhysRevLett.58.431} {\bibfield  {journal} {\bibinfo
  {journal} {Physical Review Letters}\ }\textbf {\bibinfo {volume} {58}},\
  \bibinfo {pages} {431} (\bibinfo {year} {1987})}\BibitemShut {NoStop}%
\bibitem [{\citenamefont {Coullet}(2002)}]{coullet_localized_2002}%
  \BibitemOpen
  \bibfield  {author} {\bibinfo {author} {\bibfnamefont {P.}~\bibnamefont
  {Coullet}},\ }\bibfield  {title} {\bibinfo {title} {Localized patterns and
  fronts in nonequilibrium systems},\ }\href
  {https://doi.org/10.1142/S021812740200614X} {\bibfield  {journal} {\bibinfo
  {journal} {International Journal of Bifurcation and Chaos}\ }\textbf
  {\bibinfo {volume} {12}},\ \bibinfo {pages} {2445} (\bibinfo {year}
  {2002})}\BibitemShut {NoStop}%
\bibitem [{\citenamefont {Clerc}\ \emph {et~al.}(2010)\citenamefont {Clerc},
  \citenamefont {Escaff},\ and\ \citenamefont
  {Kenkre}}]{clerc_analytical_2010}%
  \BibitemOpen
  \bibfield  {author} {\bibinfo {author} {\bibfnamefont {M.~G.}\ \bibnamefont
  {Clerc}}, \bibinfo {author} {\bibfnamefont {D.}~\bibnamefont {Escaff}},\ and\
  \bibinfo {author} {\bibfnamefont {V.~M.}\ \bibnamefont {Kenkre}},\ }\bibfield
   {title} {\bibinfo {title} {Analytical studies of fronts, colonies, and
  patterns: {Combination} of the {Allee} effect and nonlocal competition
  interactions},\ }\href {https://doi.org/10.1103/PhysRevE.82.036210}
  {\bibfield  {journal} {\bibinfo  {journal} {Physical Review E}\ }\textbf
  {\bibinfo {volume} {82}},\ \bibinfo {pages} {036210} (\bibinfo {year}
  {2010})}\BibitemShut {NoStop}%
\bibitem [{\citenamefont {Escaff}(2011)}]{escaff_non-local_2011}%
  \BibitemOpen
  \bibfield  {author} {\bibinfo {author} {\bibfnamefont {D.}~\bibnamefont
  {Escaff}},\ }\bibfield  {title} {\bibinfo {title} {Non-local defect
  interaction in one-dimension: weak versus strong non-locality},\ }\href
  {https://doi.org/10.1140/epjd/e2010-10323-8} {\bibfield  {journal} {\bibinfo
  {journal} {The European Physical Journal D}\ }\textbf {\bibinfo {volume}
  {62}},\ \bibinfo {pages} {33} (\bibinfo {year} {2011})}\BibitemShut {NoStop}%
\bibitem [{\citenamefont {Tlidi}\ \emph {et~al.}(2015)\citenamefont {Tlidi},
  \citenamefont {Fernandez-Oto}, \citenamefont {Clerc}, \citenamefont
  {Escaff},\ and\ \citenamefont {Kockaert}}]{tlidi_localized_2015}%
  \BibitemOpen
  \bibfield  {author} {\bibinfo {author} {\bibfnamefont {M.}~\bibnamefont
  {Tlidi}}, \bibinfo {author} {\bibfnamefont {C.}~\bibnamefont
  {Fernandez-Oto}}, \bibinfo {author} {\bibfnamefont {M.~G.}\ \bibnamefont
  {Clerc}}, \bibinfo {author} {\bibfnamefont {D.}~\bibnamefont {Escaff}},\ and\
  \bibinfo {author} {\bibfnamefont {P.}~\bibnamefont {Kockaert}},\ }\bibfield
  {title} {\bibinfo {title} {Localized plateau beam resulting from strong
  nonlocal coupling in a cavity filled by metamaterials and liquid-crystal
  cells},\ }\href {https://doi.org/10.1103/PhysRevA.92.053838} {\bibfield
  {journal} {\bibinfo  {journal} {Physical Review A}\ }\textbf {\bibinfo
  {volume} {92}},\ \bibinfo {pages} {053838} (\bibinfo {year}
  {2015})}\BibitemShut {NoStop}%
\bibitem [{\citenamefont {Parra-Rivas}\ \emph {et~al.}(2017)\citenamefont
  {Parra-Rivas}, \citenamefont {Gomila},\ and\ \citenamefont
  {Gelens}}]{parra-rivas_coexistence_2017}%
  \BibitemOpen
  \bibfield  {author} {\bibinfo {author} {\bibfnamefont {P.}~\bibnamefont
  {Parra-Rivas}}, \bibinfo {author} {\bibfnamefont {D.}~\bibnamefont
  {Gomila}},\ and\ \bibinfo {author} {\bibfnamefont {L.}~\bibnamefont
  {Gelens}},\ }\bibfield  {title} {\bibinfo {title} {Coexistence of stable
  dark- and bright-soliton {Kerr} combs in normal-dispersion resonators},\
  }\href {https://doi.org/10.1103/PhysRevA.95.053863} {\bibfield  {journal}
  {\bibinfo  {journal} {Physical Review A}\ }\textbf {\bibinfo {volume} {95}},\
  \bibinfo {pages} {053863} (\bibinfo {year} {2017})}\BibitemShut {NoStop}%
\bibitem [{\citenamefont {Arabí}\ \emph {et~al.}(2020)\citenamefont {Arabí},
  \citenamefont {Parra-Rivas}, \citenamefont {Parra-Rivas}, \citenamefont
  {Hansson}, \citenamefont {Gelens}, \citenamefont {Wabnitz}, \citenamefont
  {Wabnitz}, \citenamefont {Wabnitz},\ and\ \citenamefont
  {Leo}}]{arabi_localized_2020}%
  \BibitemOpen
  \bibfield  {author} {\bibinfo {author} {\bibfnamefont {C.~M.}\ \bibnamefont
  {Arabí}}, \bibinfo {author} {\bibfnamefont {P.}~\bibnamefont {Parra-Rivas}},
  \bibinfo {author} {\bibfnamefont {P.}~\bibnamefont {Parra-Rivas}}, \bibinfo
  {author} {\bibfnamefont {T.}~\bibnamefont {Hansson}}, \bibinfo {author}
  {\bibfnamefont {L.}~\bibnamefont {Gelens}}, \bibinfo {author} {\bibfnamefont
  {S.}~\bibnamefont {Wabnitz}}, \bibinfo {author} {\bibfnamefont
  {S.}~\bibnamefont {Wabnitz}}, \bibinfo {author} {\bibfnamefont
  {S.}~\bibnamefont {Wabnitz}},\ and\ \bibinfo {author} {\bibfnamefont
  {F.}~\bibnamefont {Leo}},\ }\bibfield  {title} {\bibinfo {title} {Localized
  structures formed through domain wall locking in cavity-enhanced
  second-harmonic generation},\ }\href {https://doi.org/10.1364/OL.399658}
  {\bibfield  {journal} {\bibinfo  {journal} {Optics Letters}\ }\textbf
  {\bibinfo {volume} {45}},\ \bibinfo {pages} {5856} (\bibinfo {year}
  {2020})},\ \bibinfo {note} {publisher: Optica Publishing Group}\BibitemShut
  {NoStop}%
\bibitem [{\citenamefont {Parra-Rivas}\ \emph {et~al.}(2021)\citenamefont
  {Parra-Rivas}, \citenamefont {Arabí},\ and\ \citenamefont
  {Leo}}]{parra-rivas_dark_2021}%
  \BibitemOpen
  \bibfield  {author} {\bibinfo {author} {\bibfnamefont {P.}~\bibnamefont
  {Parra-Rivas}}, \bibinfo {author} {\bibfnamefont {C.~M.}\ \bibnamefont
  {Arabí}},\ and\ \bibinfo {author} {\bibfnamefont {F.}~\bibnamefont {Leo}},\
  }\bibfield  {title} {\bibinfo {title} {Dark quadratic localized states and
  collapsed snaking in doubly resonant dispersive cavity-enhanced
  second-harmonic generation},\ }\href
  {https://doi.org/10.1103/PhysRevA.104.063502} {\bibfield  {journal} {\bibinfo
   {journal} {Physical Review A}\ }\textbf {\bibinfo {volume} {104}},\ \bibinfo
  {pages} {063502} (\bibinfo {year} {2021})},\ \bibinfo {note} {publisher:
  American Physical Society}\BibitemShut {NoStop}%
\bibitem [{\citenamefont {Kolossovski}\ \emph {et~al.}(2002)\citenamefont
  {Kolossovski}, \citenamefont {Champneys}, \citenamefont {Buryak},\ and\
  \citenamefont {Sammut}}]{kolossovski_multi-pulse_2002}%
  \BibitemOpen
  \bibfield  {author} {\bibinfo {author} {\bibfnamefont {K.}~\bibnamefont
  {Kolossovski}}, \bibinfo {author} {\bibfnamefont {A.~R.}\ \bibnamefont
  {Champneys}}, \bibinfo {author} {\bibfnamefont {A.~V.}\ \bibnamefont
  {Buryak}},\ and\ \bibinfo {author} {\bibfnamefont {R.~A.}\ \bibnamefont
  {Sammut}},\ }\bibfield  {title} {\bibinfo {title} {Multi-pulse embedded
  solitons as bound states of quasi-solitons},\ }\href
  {https://doi.org/10.1016/S0167-2789(02)00563-8} {\bibfield  {journal}
  {\bibinfo  {journal} {Physica D: Nonlinear Phenomena}\ }\textbf {\bibinfo
  {volume} {171}},\ \bibinfo {pages} {153} (\bibinfo {year}
  {2002})}\BibitemShut {NoStop}%
\end{thebibliography}%

\end{document}